%% file: dgt-revised3.tex
\documentclass[12pt]{iopart}
\usepackage{iopams, graphicx, rotating}
\usepackage{fleqn}
\usepackage{sidecap}
\usepackage{subfigure}
\usepackage{color}
\usepackage{pdflscape}
\usepackage{dsfont}
\definecolor{Red}{rgb}{1,0,0}

\newcommand{\ket}[1]{|#1\rangle}

\newcommand{\be}{\begin{equation}}
\newcommand{\ee}{\end{equation}}
\newcommand{\beq}{\begin{eqnarray}}
\newcommand{\eeq}{\end{eqnarray}}
\newcommand{\bsw}{\begin{sideways}}
\newcommand{\esw}{\end{sideways}}

\newcommand{\nn}{\nonumber}

\def\dbar{\bar{D_2}}
\def\ebar{\overline{e}}

\def\one{\mathds{1}}
\newcommand{\ZZ}{\mathbb{Z}}
\newcommand{\Z}{\mathbb{Z}}

\newcommand{\ir}[2]{$(#1, #2)$}
\newcommand{\Ir}[2]{(#1, #2)}
\newcommand{\sA}{\mathcal{A}}
\newcommand{\sT}{\mathcal{T}}
\newcommand{\sU}{\mathcal{U}}
\newcommand{\sD}{\mathcal{D}}
\include{Diagrams}

\begin{document}


\title{The modular $S$-matrix as order parameter for topological phase transitions}

\author{F A Bais$^{1,2}$ and J C Romers$^1$}
\address{$^1$ Institute for Theoretical Physics, University of Amsterdam,
Science Park 904, P.O.Box 94485, 1090 GL Amsterdam, The Netherlands}
\ead{j.c.romers@uva.nl}
\address{$^2$ Santa Fe Institute, Santa Fe, NM 87501, USA}

\date{January 15, 2011}

\begin{abstract}
We study topological phase transitions in discrete gauge theories in two spatial dimensions induced by the formation of a Bose condensate.  We analyse a general class of euclidean lattice actions for these theories which contain one coupling constant for each conjugacy class of the gauge group. 
To probe the phase structure we use a complete set of open and closed anyonic string operators. The open strings allow one to determine the particle content of the condensate, whereas the closed strings enable us to determine the matrix elements of the modular $S$-matrix, both in the unbroken and broken phases. From the measured broken $S$-matrix  we may read off the sectors that split or get identified in the broken phase, as well as the sectors that are confined. In this sense the modular $S$-matrix can be employed as a matrix valued non-local order parameter from which the low-energy effective theories that occur in different regions of parameter space can be fully determined. 

To verify our predictions we studied a non-abelian anyon model based on the quaternion  group $H=\dbar$ of order eight by Monte Carlo simulation. We probe part of the phase diagram for the pure gauge theory and find a variety of phases with magnetic condensates leading to various forms of (partial) confinement in complete agreement with the algebraic breaking analysis. Also the order of various transitions is established.
\end{abstract}



\maketitle

\section{Topological order and topological symmetry breaking}

\subsection{Introduction}

The study of  hidden  symmetries and  phase structure of gauge theories has a rich history. A central motivation in the early work was the understanding the  confinement phenomenon in non-abelian gauge theories.  An important step forward were the lattice action formulation by Wilson \cite{Wilson:1974sk} and its Hamiltonian version by Kogut and Susskind \cite{PhysRevD.11.395}, which allowed for  an expansion around the strong coupling limit where confinement is manifest.
The question was whether the confinement regime would extend all the way down to zero coupling. Wilson also introduced his celebrated loop operator as a diagnostic for confinement in the pure gauge theory.

 Not long thereafter 't Hooft and Mandelstam suggested  confinement in 3+1 dimensions to be a consequence of the dual Meissner effect, caused by magnetic disorder, notably a condensate of  topological degrees of freedom, be it monopoles or fluxes \cite{'tHooft:1977hy}.   't Hooft explained the $(2+1)$-dimensional version of confinement  in $SU(N)$ gauge theories as a consequence of the condensation of $\Z_N$ fluxes corresponding to the center elements of the gauge group \cite{'tHooft:1981ht}. Polyakov on the other hand proved confinement of $(2+1)$-dimensional compact $QED$ due to monopoles \cite{Polyakov:1976fu} and showed furthermore that the finite temperature deconfinement  transistion in $d=3+1$ is due to  a Wilson line (string) condensate \cite{Polyakov:1978vu}.  The interpretation of such transitions is  related to the spectrum of  topological defects  in the various  Higgs phases. For example for $U(1)$ in 3+1 dimensions the occurence of a phase transition depends the compactness of the group, i.e. on the presence of monopoles. For non-abelian groups  the situation depends on the representation of the Higgs field. In the case of $SU(2)$ for example one may or may not find a phase transition between the confined and Higgs phases  depending on whether the breaking is achieved with two adjoints or a fundamental representation.  Shenker and Fradkin \cite{Fradkin:1978dv}  constructed the phase diagrams for the $SU(2)$ (lattice) gauge theory for the distinct cases with  Higgs fields in the adjoint and the fundamental representation, and proved  the absence of a transition between confined and Higgs phase in the  latter. The difference can indeed be traced back to the topological  structure: in the broken phase with a fundamental Higgs no defects exist because the group is broken completely. In the broken phase with one isovector, there is a residual $U(1)$ and one has monopoles, and these will be confined if one adds another generic isovector. However with the two vectors there is still the $\Z_2$ center of the group which is unbroken, which implies the existence of $\Z_2$ fluxes in the broken phase. 
 
  Lattice formulations of the $\Z_2$ theories go back till the early seventies \cite{Balian:1974ir, Drouffe:1978py} and were also studied in \cite{Fradkin:1978dv}, where also  the phase diagram of the $\Z_2$ gauge theory with a matter field was computed in detail. A Hamiltonian formulation of $\Z_N$ gauge theories and their duality properties in two and three dimensions was  made in \cite{Horn:1979fy}, where the same algebraic structure (\ref{eq:loopalgebra}) was constructed out of generalized Ising  type lattice models.All authors agree that  the transition to the confined phase is due to  the condensation of magnetic fluxes.  The analysis is facilitated by the introduction of nonlocal order and disorder parameters such as the Wilson and 't Hooft loops and  the  algebra they form, from which the phase structure could be understood qualitatively. As these operators are nonlocal, the algebraic structure  exhibits the nontrivial braid properies that characterize the phase.
For $\Z_N$ theories one obtains an underlying $\Z^{el}_N \otimes \Z^{mag}_N$ 
symmetry of which the electric, magnetic and dyonic (anyonic) charges form irreducible representations $(n,m)$ labeled by a pair of integers mod $N$ ($m,n=0,1,2,\ldots, N-1)$ . If one defines the loop operators $\Delta^{(n,m)}(C)$, then  't  Hooft derived the crucial algebraic relation \cite{'tHooft:1977hy}:
\begin{equation}
\label{eq:loopalgebra}
\Delta^{(n,m)}(C) \; \Delta^{(n',m')}(C') = \Delta^{(n',m')} (C') \; \Delta^{(n,m)} (C) \;\; e^{2\pi i(nm' + mn')k/N}
\end{equation} 
where $k\in \Z$ is the  linking number of the loops $C$ and $C'$. 
These loops can be interpreted as linked timelike loops of worldlines obtained after creating and subsequent annihilation of two particle-antiparticle pairs. But they can also be interpreted as two intersecting spacelike loops on a torus winding around the two different cycles, one being a Wilson line and the other a magnetic Dirac string.  
On the other hand one may think of equation (\ref{eq:loopalgebra}) as  specifying the braiding relations for the representation theory of a $\Z_N \times \Z_N$ algebra endowed with a (unique) nontrivial braiding structure that exactly belongs to the  quantum double algebra  of $\Z_N$ denoted as $D(\Z_N)$. 
This operation on the torus is a topologically nontrivial vacuum to vacuum transition (because no charges or fluxes are left) and that shows that there is a periodic vacuum structure in the theory, leading to a  an $N^2$-fold vacuum degeneracy on the torus. This degeneracy in turn equals the number of particle species, i.e. the total number of unitary irreducible representations  of $D(\Z_N)$, which for the case at hand equals $N^2$.  Similar algebras have surfaced in the study of emergent gauge theories from closely related lattice models like Kitaev's toric code \cite{Kitaev:1997wr}  and the Levin-Wen type spin models \cite{Levin:2004mi}, where furthermore  the vacuum degeneracy (of a gapped phase) on the torus is proposed as a criterion for topological order.

It was also early on pointed out that ``the" Higgs phase structure of a  gauge theory, where no massles gauge degrees of freedom (no continuous groups) survive may  be very rich by itself, allowing for distinct phases which exhibited different spectra of purely topological degrees of freedom \cite{Bais:1980vd}. By breaking with sufficiently high-dimensional representations of  $SU(2)$ for example, one  finds that not just the center, but any discrete subgroup of $SU(2)$ can be selected to survive,  also non-abelian groups like $D_N$, or the tetrahedral group etc. In the corresponding phases one finds a rich variety of non-abelian fluxes that exhibit highly nontrivial braiding and fusion properties with charges and also among them selves.  The effective low energy theory in such a Higgs phase is also called a discrete gauge theory (DGT), these are indeed gapped phases with only topological degrees of freedom, and therefore by definition phases with topological order. The full $(2+1)$-dimensional description of all the anyonic sectors of discrete gauge theories was given in \cite{Bais:1991pe} and revealed an underlying hidden quantum group structure, i.e. the excitations of a discrete gauge theory with gauge group  $H$,  were shown to correspond to the irreducible unitary representations of the quantum double $D(H)$ of $H$ as defined by  by \cite{Roche:1990hs}.  The most interesting aspect of this perspective being that the ordinary  ``electric charge" and topological  ``magnetic flux"  degrees of freedom are both part of the same representation theory, the magnetic degrees of freedom labeled by the conjugacy classes of $H$ and the electric parts by the representations of the centralizer group in $H$ of the given magnetic flux. This leads to a rather intricate interdependence of electric and magnetic sectors in case the goup $H$ is non-abelian and a different way to look at order and disorder operators. We  return to $D(H)$ and its representations in Section \ref{sec:quantumdouble}. An important consequence of this underlying  symmetry is that it suggested that in the {\it pure} DGT a corresponding set of gauge invariant loop operators should exist, thereby generalizing the Wilson/'t Hooft operators to all anyonic species, and indeed such operators have been constructed \cite{Bais:2008xi}. With these operators in hand one can now probe the full phase structure of discrete gauge theories in a lattice formulation and that is what  will be done in this paper.   

The hidden quantum group symmetry  and its representation theory which forms a modular tensor category, is a powerful way to characterize  the distinct topological phases of pure DGTs and forms the natural basis for the  concept of \textit{topological symmetry breaking}, which is defined as the  breaking of quantum group symmetry \cite{Bais2002a,Bais2003,Bais2009}.  A brief review of this symmetry breaking mechanism is given in section \ref{sect.symbreak}. In this paper we study the validity of this mechanism by probing the various groundstates through measuring the  expectation values of single as well as linked anyonic  loop operators, as is  explained in Section \ref{sec:observables}. 
In recent years we have witnessed a growing interest in systems that allow for the realisation of different topological phases, the examples can be found in Levin-Wen models \cite{BurnellSimonSlingerland}, the Kitaev Honeycomb model \cite{Kellsetal2010}, discrete  gauge theories \cite{Bais2002a}, and last but not least in quantum Hall systems \cite{ReadGreen1999,Grosfeld2008,Bais2009a,BarkeshiWen,BarkeshiWen2}.   
Most of these can be understood from the point of view of topological symmetry breaking involving the formation of a Bose condensate. In most instances  a Hamiltonian framework is used, but in this paper we show that to analyse such systems it may be profitable to switch to a euclidean action formulation which allows for the use of  Monte Carlo simulations to  determine the phase structure. In this approach it is easy to directly measure the modular $S$-matrix in any phase of the system which explains why we call this matrix an orderparameter of such a system.
After briefly recalling the basic ingredients of topological order and topological symmetry breaking and settle some notation, we show how open string operators and the modular $S$-matrix can be used as order parameters and phase indicators for topological symmetry breaking. We then introduce a class of  multi-parameter  lattice actions for non abelian discrete gauge theories and verify  the theoretical analysis \cite{Bais2003,Bais2009}  in detail by numerical simulations. 
\subsection{\label{tqft.fusion}TQFT basics}
In this section we set the stage and fix the notation for the rest of this paper. We study phases of  systems that are described by a Topological Quantum Field Theory (TQFT) in $2+1$ dimensions. We label the different sectors or (anyonic) particle species by $a,b,c,\dots$ . The two interactions between two particles in a TQFT are fusion and braiding. 
\paragraph{Fusion}We describe fusion by the rule
\begin{equation} \label{fusrules}
a \times b = \sum_c N^{a b}_c c,
\end{equation}
where the integer multiplicities $N^{a b}_c$ give the number of times $c$ appears in the fusion product of $a$ and $b$. The fusion algebra is associative and commutative, and has a unique identity element denoted as ``1" that represents the vacuum. Each sector $a$ has a unique conjugate $\bar{a}$ (representing the corresponding anti anyon) with the property that their fusion product contains the identity:
\[ a \times \overline{a} = 1 + \sum_{c \neq 1} N^{a \overline{a}}_c c. \]

\paragraph{Braiding}The particles in a $2+1$ dimensional TQFT can have fractional spin and statistics. Rotating a particle $a$ by $2 \pi$ (also called {\em twisting}) multiplies the state vector by a phase equal to the {\em spin factor} $\theta _a$
\[ \ket{a} \stackrel{\rm{twist}}{\rightarrow} \theta _a \ket{a},\]
generalizing the usual $+1$ ($-1$) known from bosons (fermions) in $3+1$ dimensions.
Adiabatically moving a particle $a$ around another particle $b$ in a channel $c$ is called a {\em braiding} and can have a nontrivial effect on the state vector of the system, given by  $\theta_c/\theta_a\theta_b$.
\paragraph{Quantum dimensions}The quantum dimensions $d_a$ of particle species $a$ are another set of important quantities in a TQFT. These numbers satisfy the fusion rules (\ref{fusrules}), i.e. $d_a d_b = \sum_c N^{a b}_c d_c$. The quantum dimension of an anyonic species is a measure for the effective number of degrees of freedom, corresponding to the internal Hilbert space of the corresponding particle type. The Hilbert space dimension of a system with $N$ identical particles of type $a$  grows as $(d_a)^N$ for $N$ large. In general, the quantum dimensions $d_a$ will be real numbers; however for DGTs they are integers. The total quantum dimension $\mathcal{D}$ of the theory is given by
\[ \mathcal{D} = \sqrt{\sum_i d_a^2}, \]
and the topological entanglement entropy of the ground state is proportional to $\log \mathcal{D}$.

\paragraph{Diagrammatics}

There is a powerful diagrammatic language to express the equations describing the TQFT, which we will use to relate the values of observables as they can be measured in the different phases. In this paper we will use the notation and definitions given by Bonderson \cite{Bonderson:2007zz}.  Particle species are represented by lines, fusion and splitting by vertices. A twist is represented by a left or right twist on a particle line:
\begin{equation}
\twistright{a} \quad= \theta_a \quad\anyon{a}\quad,\quad\quad \quad\twistleft{a} \quad = \theta_a^* \quad \anyon{a}.
\end{equation}

The evaluation of simple diagrams is rather straightforward, and complicated diagrams can be simplified using braid relations and the socalled $F$ symbols which follow from associativity of the fusion algebra.  
The simplest examples are the closed loop of type $a$ that evaluates to the quantum dimension $d_a$:
\begin{equation}
\qdim{a}{>} = d_a,
\end{equation}
whereas the twisted loop equals $d_a \theta_a$:
\begin{equation}
\xygraph{
  !{0;/r1.0pc/:}
  [u(0.5)]
  !{\hover}
  !{\hcap}
  [l]!{\vcap[2]}
  [rr]!{\xcapv[1]@(0)<{a}}
  [ll]!{\vcap[-2]} 
} = \theta_a d_a
\end{equation}
Of particular interest are  the generators of the modular group, $S_{ab}$ 
\begin{equation}
\label{eq:sab}
\smatrix{a}{b} = S_{ab} = \frac{1}{\mathcal{D}} \sum_c N^{a\overline{b}}_c \, \frac{\theta_c}{\theta_a \theta_b}\, d_c,
\end{equation}
and $T_{ab} = e^{-2\pi i (\overline{c}/24 )} \theta_a \delta_{a,b}$, where the $\overline{c}$ is the central charge of the theory, not to be confused with a particle type.

As we mentioned before the importance of the rather abstract diagrammatic notation is that the diagrams directly correspond to observables in our euclidean lattice gauge theory formulation. In the euclidean three dimensional formulation of topological theories the values these diagrams have, correspond to the vacuum expectation values of the corresponding anyon loop operators, for example in the unbroken phase one may measure
\begin{equation}
\label{ }
\left< \qdim{a}{>} \right>_0 = d_a,
\end{equation}
where the LHS is now defined as the value of the  path integral with  the nonlocal loop operator for particle species $a$ inserted and the RHS is obtained if we are probing the system in the unbroken phase governed with the groundstate denoted as $0$ and governed by the algebra $\mathcal{A}$. We use the subscript $0$ because the value of the same diagram may be different  if it is evaluated in a different phase with a groundstate that we will denote by $\Phi$;  in the remainder of the paper we will therefore always use brackets with a subscript.

\subsection{\label{sect.symbreak}Topological symmetry breaking}
In this section we briefly recall topological symmetry breaking, the phenomenon that a phase transition to another topological phase occurs due to a Bose condensate \cite{Bais2002a,Bais2003}. The analogy with ordinary symmetry breaking is clear if one thinks of the particle as representations of some quantum group, and assumes that a bosonic degree of freedom i.e. with $\theta_c = 1$ -- fundamental or composite -- condenses. The breaking can then be analyzed, either from the quantum group (Hopf algebra) point of view, or from the dual or representation theory point of view \cite{Bais2009}.

Let us illustrate this by an example of ordinary  group breaking. Suppose we have a gauge group $SU(3)$ and a Higgs triplet  that acquires a vacuum expectation value $\Phi = (1,0,0)$, then the $SU(2)$ subgroup working on the last two entries will leave $\Phi$ invariant. Equivalently this $SU(2)$ subgroup may be characterized by the way the $SU(3)$ triplet decomposes under the $SU(2)$ action as $3 \rightarrow 2+1$ where the singlet on the right corresponds exactly to the new $SU(2)$ invariant groundstate. In that sense one may select a specific  residual gauge symmetry by choosing  an appropriate Higgs representation which has a singlet under that residual group in its branching. For example if we want to break an $SU(3)$ group to the $SO(3)$ subgroup which is characterized by the branching rule $3\; (\mathrm{ as\; well\; as\;} \bar{3}) \rightarrow 3$ then we may choose the Higgs field to be in the $6$-dimensional irrep of $SU(3)$, because then $3\times 3 = \bar{3} + 6 \rightarrow 3\times 3=1+3+5$, from which follows that $6\rightarrow 5+1$ and again the singlet on the right corresponds to the $SO(3)$ invariant vacuum state $\Phi$. 

In the case of general quantum groups it is this branching rule approach which is the most natural and powerful in the context  of  TQFT because the fusion algebra corresponds  to the representation ring of the quantum group.  A general treatment with ample examples can be found in reference \cite{Bais2009}. Let us point out some essential features of this procedure that one has to keep in mind. As the quantum group centralizes the chiral algebra in the operator algebra of a CFT, one  expects that reducing the quantum group will correspond to enlarging the chiral algebra, and this turns out to be the case. In contrast to ordinary group breaking, the topological symmetry breaking procedure involves two steps, firstly the condensate reduces the unbroken fusion algebra (also called a braided modular tensor category) $\mathcal{A}$ to an intermediate algebra denoted by $\mathcal{T}$. This algebra however may contain representations that braid nontrivially with the condensed state, i.e. with the new vacuum and if that is the case, these representation will be confined and will be expelled from the bulk to the boundary of the sample. Confinement implies that in the bulk only the unconfined sectors survive as particles and these are characterized by some subalgebra $\sU\subset \sT$. Let us briefly describe the two steps seperately. 

\paragraph{From $\mathcal{A}$ to $\mathcal{T}$} 
Assuming that a certain bosonic irrep $c$ will condense due to some underlying interaction in the system, implies that $c$ will be identified with the vacuum of $\sT$.  For our purposes, a boson is a sector with trivial (integer) spin, though in fact in the context of $2+1$ dimensions one has to also require that  fusion of this field with itself  has a channel with trivial braiding.

The definition of the  new vacuum requires to a redefinition of fields. Firstly, fields in $\mathcal{A}$ that appear in the orbit under fusion with the condensed field $c$ are identified in $\sT$, so,  if $ c \times a = b$ then $a, \, b \rightarrow a'$.
Secondly, if a field $b$ forms a fixed point under fusion with the condensate $c$, then the field will split at least in two parts: $b\rightarrow \sum_i b_i$. 
The identifications and splittings of representations can be summarized by a rectangular matrix $n_a^t $ that specifies the ``branching" or ``restriction" of fields $a$ from in $\mathcal{A}$ to $\mathcal{T}$ with fields $t,r,s,\dots$:
\[
a \rightarrow  \sum_t n^t_a \; t 
\]
 This {\em branching matrix} is a rectangular matrix (the number of particle types in the $\mathcal{A}$ and $\mathcal{T}$ theories is not equal in general) of positive integers.
 We will also consider the transpose of this matrix denoted as $n_t^a$ which specifies the ``lift" of the fields $t\in \sT$ to fields $a\in \mathcal{A}$:
 \[
 t \rightarrow  \sum_a n^a_t \; a  = \sum_{a\in t} a
 \]

One  may now derive the  fusion rules $\mathcal{T}$ from the fusion algebra (\ref{fusrules}). 
Because  of the identifications, it is often the case that the intermediate algebra $\mathcal{T}$ though being a consistent fusion algebra, is not necessarily braided, in more technical terms, it satisfies the ``pentagon'' equation but not the ``hexagon'' equation.
The physical interpretation of this fact is that the sectors in $\mathcal{T}$ do not yet constitute the low-energy effective theory. This is so because  sectors $t$ that have an ambiguous spin factor, meaning that not all $\theta_a$ of the lift  $a\in t$ are equal,  will be connected to a domain wall and hence are confined in the new vacuum. The confined excitations will be expelled to the edges of the system or have to form hadronic composites that are not confined. Yet the $\mathcal{T}$ algebra plays an important role: in \cite{Bais2009a} for example,  it was shown that the $\mathcal{T}$ algebra governs the edge/interface degrees of freedom in the broken phase.
\paragraph{From $\mathcal{T}$ to $\mathcal{U}$}
Some of the sectors in $\mathcal{T}$ will survive in the bulk, some will be confined. The physical mechanism behind confinement in $2+1$ dimensional topological field theories is nontrivial braiding with the condensate.  The vacuum state or order parameter should be single valued if carried adiabatically around  a localized particle like excitation. If it is not single valued that would  lead to a physical string or ``domain wall" extending from the particle that carries a constant energy per unit length. 
The unconfined algebra $\mathcal{U}$ consists of the representations in $\mathcal{T}$ minus the confined ones, it is this algebra that governs the low energy effective bulk theory. The confined representations can be determined in the following way. First we define the ``lift" of a representation in $\mathcal{T}$ as the set of representations $b \in \mathcal{A}$ that restrict to $t$. Now, if all of the representations in the lift of $t$ braid trivially with the lift of the vacuum, the sector $t$ is part of $\mathcal{U}$. Otherwise, it is confined.
One may prove that the $\mathcal{U}$ algebra closes on itself with consistent fusion rules, while consistent braiding is achieved by assigning the (identical) spin factors of the parent sectors of the unbroken theory to the $\mathcal{U}$ fields.

Let us finally mention a useful quantity, the socalled \textit{quantum embedding index} $q$ defined in \cite{Bais2010}, it is a real number characterizing the topological symmetry breaking. This quantity is defined as
\begin{equation}\label{embindex}
q = \frac{\sum_a n^a_u d_a}{d_u},
\end{equation}
where the index $a$ runs  over the sectors of the unbroken phase $\mathcal{A}$, that correspond to lift of any  sector  $u$ or $t$  of the algebra $\mathcal{U}$ or $\mathcal{T}$; the $n^a_u$ is the lift of sectors $u$ to their parents $a$ and $d_a$ is the quantum dimension of the representation $a$. Observe that this expression is independent of the particular sector $u$, which is a non-trivial result explained in \cite{Bais2010}.

Choosing  for $u$ the new vacuum, we have $d_u=1$ and obtain that $q$ just equals the total quantum dimension of the lift of the $\mathcal{U}$ (or $\mathcal{T}$) vacuum in the unbroken $\mathcal{A}$ theory. 
The quantum embedding index is the analogon for the embedding index defined by Dynkin for the embedding of ordinary groups.  As an aside we mention that the change in topological entanglement entropy of the disk changes also by $\log(D_A/D_U) =  \log{q}$ in a transition from an $\mathcal{A}$ to a $\mathcal{U}$ phase \cite{Bais2010}.

Let us to conclude this subsection on topological symmetry breaking illustrate the procedure with a very straightforward example, namely the breaking of the quantum group $\sA= SU(2)_4$. It has 5 irreps labeled by $\Lambda=0,\ldots,4$ with spinfactors $ \theta_a = 1, \frac{1}{8}, \frac{3}{8}, \frac{5}{8}, 1$. The  $\Lambda = 4$ is the only  boson  and we assume it to condense. The lift of the new vacuum corresponds to the $\Phi=0+4$ of $\sA$, and hence  the embedding index $q= d_0+d_4=1+1=2$. The $1$ and $3$ reps of $\sA$ are identified, but because they have  different spin factors, the corresponding $\sT$ representation will be confined. In $\sU$ we are therefore left with the $\Lambda=2$ rep. which splits because it is a fixed point under fusion with the condensate as $4 \times 2=2$. We write $2 \rightarrow 2_1 + 2_2$. The values for the spin  and the quantum dimensions and the fusion rules  for these representations fully determine the unconfined quantum group to be $\sU= SU(3)_1$. We recall that the nomenclature of the groups is linked to the chiral algebra, it is therefore not surprising that the $SU(2)_4$ quantum group breaks to the \textit{smaller} quantumgroup $SU(3)_1$ which is related to a \textit{larger} chiral algebra. For the chiral algebras one has the conjugate embedding $SU(2)_4 \subset SU(3)_1$ which is a conformal embedding. This conformal embedding in turn is induced by the $SO(3)\subset SU(3)$ embedding mentioned at the beginning of this subsection.

\subsection{\label{sec:observables}Observables}
Our objective is to verify the theoretical predictions of the topological symmetry breaking scheme  in a class of euclidean gauge theories that are expected to exhibit transitions between different topological phases. We will numerically evaluate the expectation values of various topological diagrams using Monte Carlo simulations, and in this section we calculate the predicted outcomes of a variety of possible measurements from  theory. The strategy has  two  steps, (i) the determination of the condensate (including the measurement of the embedding index $q$) by evaluating  the basic nonlocal open string order parameters, given by Eq.~(\ref{eqn.dyonop}), (ii) measuring the socalled broken modular $S$-matrix and from that construct the $S$-matrix of the $\sU$ phase.  We also will see that  the condensate fixes the branching and lift matrices and having determined those we can also predict the outcome of measurements of other topological diagrams corresponding to the lifts of $\mathcal{U}$ fields to $\mathcal{A}$ fields . 
\subsubsection{Determination of the condensate and the embedding index $q$.}
We measure the open string operators in the model. Note that in our pictorial representation time flows upward, so a vertical line physically represents the creation, propagation and annihilation of a {\em single} particle. For the particular case of a DGT, which we study in this work, these lines have a realization as operators on a spacetime lattice, see Eq.~(\ref{eqn.dyonop}).

If the symmetry is unbroken we will have for any nontrivial field $a$ that
\begin{equation}
\label{ }
\left< L_{a\bar{a}}\right>_0 = \left<\quad \anyon{a} \right>_{\Phi=0} = 0.
\end{equation}
because the diagram represents the creation and subsequent annihilation of a single $a$-particle. However in the broken situation the expectation value will be nonzero for all fields $\phi_i \in \mathcal{A}$ in the condensate which we denote by $\Phi$. So writing, 
\begin{equation}
\label{ }
\Phi = 0 + \sum_i \phi_i
\end{equation}
we obtain that in general,
\begin{equation}
\label{ }
\left<\quad \anyon{a} \right>_\Phi = \delta_{a \phi_{i}} d_a.
\end{equation}
This in turn implies that it is simple to measure $q$ as
\begin{equation}
\label{measureq}
\sum_{a \in \sA} \left< \quad \anyon{a} \quad \right>_\Phi = \left< \quad \anyon{0} \quad \right>_\Phi + \sum_i \left< \quad \anyon{\phi_i} \quad \right>_\Phi = d_0 + \sum_i d_{\phi_i} = q
\end{equation}
\subsubsection{Determination of confinement and other topological data of the broken phase.}
Once we have determined the components of the vacuum we can determine the lifts of the $t$ fields simply by studying the fusion rules of $\Phi \times a = \sum t' $, where $t'$ denotes the lifts of those $t$ fields which contain $a$, i.e. for which $n_t^a =1$. Having obtained the lifts of the $t$ fields the next step is to make the measurement determining whether a given t field is confined. This involves the measurement of the index $\eta$, or simply: 
\noindent
\begin{equation}
\label{ }
\left<\quad \sum_{a\in t} \xygraph{
  !{0;/r1.0pc/:}
  [u(0.5)]
  !{\hover}
  !{\hcap}
  [l]!{\vcap[2]}
  [rr]!{\xcapv[1]@(0)<{a}}
  [ll]!{\vcap[-2]} }
 \right>_\Phi = q \sum_{a \in t} \theta_a d_a = q^2 d_t \eta_t = \left\{ \begin{array}{l }
     \mbox{0 if $t \notin \mathcal{U}$ (confined)}   \\
    q^2 \theta_u d_u \mbox{if $t \in \mathcal{U}$ (not confined)}    
\end{array} \right.
\end{equation}
Alternatively one can  measure certain closed $a_i$ loop operators that are also defined for fields $a$ that split under branching and that will be defined later, for which holds that:
\begin{equation}
\label{ }
\left< \qdim{a_i}{>} \right>_\Phi = \left< \qdim{a_i}{>} \right>_0 +\sum_j \; \left<\smatrix{a_i}{\phi_j} \right>_0 =  \left\{ \begin{array}{l }
    \mbox{0 if $t \notin \mathcal{U}$ (confined)}   \\
    q^2 \theta_u d_u \mbox{ if $t \in \mathcal{U}$ (not confined)} 
    \end{array} \right.
\end{equation}
It follows that from these measurements, the fields that are confined can be determined, but also the quantum dimensions $d_u$ and twists $\theta_u$ of the unbroken $\mathcal{U}$ theory are obtained. 

\subsubsection{The broken modular $S$- and $T$-matrices.}
Instead of the fusion coefficients $N^{ab}_c$ an alternative specification of a (modular) topological field theory is by its representation of the modular group $SL(2,\ZZ)$ generated by the $S$ and $T$-matrices
\begin{equation}\label{modgroup}
 S^2 = (ST)^3 = \mathcal{C}, \quad S^* = \mathcal{C}S = S^{-1}, \quad T^* = T^{-1}, \quad \mathcal{C}^2 = 1, \end{equation}
with $\mathcal{C}$ the charge conjugation matrix. The corresponding matrix elements can be expressed in  the fusion coefficients and spin factors:
\begin{equation}
S_{ab} = \frac{1}{\mathcal{D}} \sum_c N^{a\overline{b}}_c \, \frac{\theta_c}{\theta_a \theta_b}\, d_c,
\end{equation}
\begin{equation}
T_{ab} = e^{-2\pi i (\overline{c}/24 )} \theta_a \delta_{a,b}
\end{equation}
where $\mathcal{D}$ is the total quantum dimension and the constant $\overline{c}$ is the conformal central charge of the corresponding conformal field theory. We recall that the central charge of a discrete gauge theory is zero, so in that case the $T$-matrix is just the diagonal matrix containing the spin factors.

The great advantage of switching to the modular data is that unlike the fusion coefficients these generators  can be directly measured using the anyon loop operators that arise naturally  in a  three dimensional euclidean formulation of the theory. We will evaluate the expectation value of these $S$-matrices numerically in our lattice formulation of multiparameter discrete gauge theories later on. The measured $S$- and $T$-matrix elements do not satisfy the relations (\ref{modgroup}) directly; however, using the measurements the full $S$- and $T$-matrices of the $\sU$ theory, which do satisfy the modular group relations, can be constructed, .
In the unbroken theory the measured $S$-matrix elements $\left< S_{ab} \right>$ correspond to  the expectation values of the Hopf link with one loop colored with representation $a$ and the other with representation $b$:
\[
\left< S_{ab} \right>_0 = {1 \over \mathcal{D}}\left<
\xygraph{
  !{0;/r1.0pc/:}
  [u(1.0)]
  !{\vunder}
  !{\vunder-}
  [uur]!{\hcap[2]<{b}}
  [l]!{\hcap[-2]>{a}}
} 
\right>_0 = S_{ab},
\]
where $S_{ab}$ is the $S$-matrix of the unbroken $\sA$ theory.
We can however also determine the modular $S$-matrix of the residual $\sU$ theory $S_{uv}$ directly from measurements if we take the splittings of certain fields $a\Rightarrow \{a_i\}$ in account appropriately. We will show how to do this later for the DGT's in detail and give a more general mathematical treatment of this elsewhere \cite{BER2010}. Then we will arrive at an explicit formula and algorithm to determine $S_{uv}$:
\begin{equation}
\label{eq:Suv}
S_{uv} = {1 \over q} \sum_{a_i,b_j} n^{a_i}_u n^{b_j}_v \left< S_{a{_i}b{_j}} \right>_\Phi.
\end{equation}
This expression involves not only the branching (lift) matrix $n_u^{a_i}$, but also the what we will call the {\it broken $S$-matrix} defined as $\bar{S}_{a{_i}b{_j}} = \langle S_{a{_i}b{_j}} \rangle_\Phi$, which, because of the splitting, clearly involves a larger size matrix then the modular $S$-matrix of the original $\sA$ phase.
From the broken $S$-matrix we may directly read off $S_{uv}$, the $S$-matrix of the effective low energy TQFT governed by $\mathcal{U}$. An important observation is that the values of the $S$-matrix elements in a broken phase will be different from the ones in the unbroken phase, for example because of the contribution of  the {\it vacuum exchange  diagram} $\tilde{S}$ depicted below, in which the condensed particle is exchanged  giving a nonzero contribution in the broken phase while it would give a vanishing contribution in the unbroken phase:
\[
	\tilde{S}_{a{_i}b{_j}} = \frac{1}{q^2}\ \scondensate{a_i}{b_j}
\]
In the explicit calculations later on we show that this \textit{vacuum exchange diagram} leads to a change in the $S$-matrix which depends on the subindices introduced above. It turns out that it is also possible to calculate the broken $S$-matrix from first principles, this will be discussed  in a forthcoming paper \cite{BER2010}.

As to be expected one finds identical rows and columns in the broken $S$-matrix, for components that are identified, whereas the entries for confined fields will be zero. With this prescription the formalism  outlined above is applicable in any phase of the theory including the unbroken one where there is no splitting and the vacuum exchange diagram gives a vanishing contribution. 
The measured $T$-matrix on the other hand is given by
\[
\left< T_{ab} \right>_\Phi =\frac{ \delta_{ab}}{d_a}
\left<
\xygraph{
  !{0;/r1.0pc/:}
  [u(0.5)]
  !{\hover}
  !{\hcap}
  [l]!{\vcap[2]}
  [rr]!{\xcapv[1]@(0)<{a}}
  [ll]!{\vcap[-2]}
}
\;\right>_\Phi
\]
again with $\left< T_{ab} \right>_0 = T_{ab}$.
After measuring or calculating the $S$- and $T$-matrices in a given phase, we can reconstruct the fusion coefficients with the help of the Verlinde formula \cite{Verlinde:1988sn},
\begin{equation}
N_{ab}^c = \sum_x \frac{S_{ax}S_{bx}S_{\overline{c}x}}{S_{1x}}.
\end{equation}
To conclude, we have in this section summarized the basic features of a TQFT and considered some aspects of topological phase transitions induced by a Bose condensate, furthermore we  explained how the measurement of the $L$-, $S$-, and $T$-operators in the broken phase fully determine the quantum group of a (broken) topological phase. The general scheme to analyse the breaking pattern of a some multiparameter TQFT is to first use the open string operators  to probe which fields are condensed in the various regions of parameter space. In a given broken phase we can subsequently  compute/measure what we will call the broken $S$-matrix  $\bar{S}_{a_i b_j}$, where  as mentioned the subindex labels the splitting of the corresponding $\sA$ field. From the broken $S$-matrix we can read off the $S$-matrix of the $\sU$ theory. In the remainder of the paper we will explicitly execute this program for discrete gauge theories.  

\section{A euclidean lattice approach to Discrete Gauge Theories.}
\subsection{$\Z_2$ gauge theory and topological order, a prelude}
Before applying our approach to (non-Abelian) Discrete Gauge Theories (DGTs) in general, let us make some connections to previous work from different perspectives. After Wilson's seminal work on Euclidean lattice gauge theory for non-Abelian Lie groups to study the confinement of quarks \cite{Wilson:1974sk}, a Hamiltonian formalism for the same problem was soon developed \cite{PhysRevD.11.395}. It was clear that because the gauge fields now take values in the {\em group} instead of the Lie {\em algebra}, one could also study models based on a finite group.

These models, which we call DGTs, were mostly studied as approximations to $U(1)$ or $SU(N)$ theories in times when computers were not as powerful as today. They are however also interesting in their own right, since they are purely topological: there are no local degrees of freedom, and only the {\em topological} (generalized Aharonov-Bohm type) interactions survive. 

This does not automatically mean that all observables are topological quantities: the appearance of virtual flux-antiflux pairs gives small size-dependent corrections to the loop-like observables in these theories. However, since these excitations are gapped, these corrections are exponentially small. We will show below by explicit calculation that as long as one stays away from the critical points, it is justified to think of the observables in these theories as topological quantities.

\subsubsection{Hamiltonian formalism}
To connect with work other work on topologically ordered systems, let us first go to a Hamiltonian formalism. This is formally done by taking a timeslice of the spacetime lattice and taking the limit in which the temporal spacing goes to zero \cite{PhysRevD.19.3715}. The Hamiltonian of (2+1)-dimensional $\Z_2$ gauge theory on a square {\em spatial} lattice is
\begin{equation}\label{z2ham}
H = - {1 \over 2} \lambda \sum_l (P_l -1) - \sum_p {1 \over 2} (Q_{p1}Q_{p2}Q_{p3}Q_{p4} -1),
\end{equation}
where the operators $P_l$ and $Q_l$ act on links, the second term is a sum over the elementary plaquettes of the lattice where $p1\dots p4$ are the links of a single plaquette and $\lambda$ is the coupling constant. The operators satisfy
\[
\{ Q_l, P_l \} = 0, \; \; P_l ^2 = Q_l^2 = 1,
\]
which means a possible representation can be given in terms of Pauli matrices $P_l = \sigma_3$, $Q_l= \sigma_1$ acting on spin-$\frac{1}{2}$ bosons living on the links. Note that the algebra above is the same as the $Z_2$ version of (\ref{eq:loopalgebra}), and indeed a closed string of $P_i$ operators generates a Wilson loop, whereas a closed string of $Q_i$  operators creates a closed Dirac-string.   Gauge transformations act on the star of four links $i1\dots i4$ adjacent to a site $i$
\[
G_i = P_{i1}P_{i2}P_{i3}P_{i4},
\]
and build the {\em gauge invariant} Hilbert space, one has to implement a Gauss law for physical states $\ket{\psi}$
\begin{equation}
\label{ham.gaugetr}
(1-G_i)\ket{\psi} = 0 \;\;\; \mbox{for all sites $i$}.
\end{equation}
Now we can make the connection with work by Kitaev \cite{Kitaev20032} and Wen \cite{PhysRevB.72.045141}. Their models (Toric code, $\Z_2$ string nets) correspond to Hamiltonian $\Z_2$ DGT where the coupling $\lambda=0$ and the gauge constraint (\ref{ham.gaugetr}) is not strictly enforced. Setting $\lambda=0$ makes the theory purely topological, the ground state is an equal weight superposition of all states
\begin{equation}\label{ham.loop}
\prod_{l \in C} Q_l \ket{0},
\end{equation}
where $C$ is a closed loop of links and $\ket{0}$ is the state with the property $P_l \ket{0} = \ket{0}$ for all links $l$. Viewing the link variables as spin-$\frac{1}{2}$ bosons, this vacuum state corresponds to all the spins being in the up state. Since the expectation value of any loop operator (\ref{ham.loop}) in the ground state is equal to one, and these loops are Wilson loops in the gauge theory language, the theory is topological.

By not enforcing the gauge constraint (\ref{ham.gaugetr}) strictly but adding it as a term to the Hamiltonian, these models allow for massive open strings. Such open strings are not gauge-invariant at their endpoints and therefore correspond to external charges.
\subsubsection{\label{sec:z2topo}Euclidean formalism}
The $\Z_2$ gauge theory in the Euclidean approach, where we discretize both space and time, is described by the action
\begin{equation}\label{z2action}
S = - \beta \sum_p U_{p1}U_{p2}U_{p3}U_{p4},
\end{equation}where the sum is again over all plaquettes (now both spatial and temporal) and the $U$ variables are numbers $\pm 1$. The partition sum
\[
\mathcal{Z} = \sum_{\{U\}} e^{-S}
\]
and the expectation value of {\em gauge invariant} operators $\mathcal{O}$
\[
\langle \mathcal{O} \rangle = \frac{1}{\mathcal{Z}} \sum_{\{U\}} \mathcal{O} (\{U\})  \,\,e^{-S},
\]
are the quantities of interest here. The gauge invariance, which in the Hamiltonian formulation was enforced by projecting out states from the Hilbert space, is now manifest in the action and the operators. The partition sum is over all gauge field configurations, but since all sums are finite, gauge fixing is not required\footnote{This even holds for continuous groups, since we integrate over the group instead of the algebra.}. 

If the coupling $\beta$ is large, the dominant contribution from the partition sum will be from field configurations where all plaquettes $UUUU=+1$. In the limit $\beta \rightarrow \infty$ this is strictly true, and one is left with a topological quantum field theory, as was the case for the Hamiltonian (\ref{z2ham}) with $\lambda=0$. For $\beta$ small there is a confining phase, the phase transition is at $\beta=0.7613$ \cite{PhysRevD.11.2098}.

In most of this work, we study the topological properties of a DGT, for a general group $H$. To show that for finite coupling constant $\beta$ this is good approximation, let us perturbatively calculate the expectation value of a Wilson loop in this $\Z_2$ theory. The Wilson loop $W(C)$ is the product of $U$ variables around a closed loop $C$
\[
\langle W(C) \rangle = \frac{1}{\mathcal{Z}} \sum_{\{U\}} UU\dots U \,\, e^{-S}.
\]
For large $\beta$, the action is minimized by configurations for wich all plaquettes are $+1$. The first order perturbation comes from those configurations in which one link is $-1$. In three dimensions, this excites 4 plaquettes, so the Boltzmann weight for such configurations is $e^{-4\beta}$ smaller than for those with no excited plaquettes.

If the lattice has size $N \times N \times N$, there are $3N^3$ links. For a contour $C$ of length $L$,
\[
\langle W(C) \rangle \approx \frac{1 - L e^{-4\beta} + (3N^3-L)e^{-4\beta}}{1+3N^3e^{-4\beta}} \approx 1 -2L e^{-4\beta} + O(e^{-8\beta}).
\]
This shows the corrections to the purely topological result $W(C)=1$ are, for $\beta$ several times larger than the critical point, negligible for simulations of reasonable lattice sizes: for a Wilson loop size $10 \times 10$, $\beta=3.0$ yields corrections only in the third digit.

Another gauge-invariant quantity is the 't Hooft loop, which lives on a loop $C'$ of the dual lattice. Such a loop pierces a number of plaquettes $p$, and the 't Hooft operator
\[
H(C') = \prod _{p \in C'} e^{-2 \beta \,\, U_{p1}U_{p2}U_{p3}U_{p4} },
\]
flips the sign of the coupling $\beta \rightarrow -\beta$ for these plaquettes. This forces a $\Z_2$ magnetic flux through these plaquettes. We will define operators generalizing the 't Hooft and Wilson loops for general non-Abelian DGTs shortly.
\subsubsection{Phase structure}
The action (\ref{z2action}) can realize three phases when one also allows for negative coupling. For large positive $\beta$, the phase mentioned before is realized, where almost all plaquettes are $+1$. For large negative $\beta$, almost all plaquettes are $-1$. For small $|\beta|$, a confining phase where the magnetic $\Z_2$ flux has condensed is realized.

To study the phase diagram of a (non-Abelian) DGT in full, we find it convenient to formulate the action in the {\em class basis}, instead of the {\em irrep basis}. This means we do not take the character of the group element of the plaquette product $UUUU$ in the action, but we define delta functions on each class. We will explain in detail how this works in section \ref{sec.actions}. For $\Z_2$ the phase diagram is one-dimensional, but the introduction of a second coupling constant will get rid of the need for negative couplings:
\[
S = - \sum _p \left( \beta_{+1} \delta_{+1} (U_{p1}U_{p2}U_{p3}U_{p4}) + \beta_{-1} \delta_{-1} (U_{p1}U_{p2}U_{p3}U_{p4}) \right),
\]
where $\delta_A(U)$ for a group element $U$ and a conjugacy class $A$ gives $+1$ if $U \in A$ and zero otherwise. For non-Abelian groups this formulation makes the phase diagram much more intuitive, for $\Z_2$ it is rather artificial.
\begin{figure}[t]
\centering
\includegraphics[width=0.4\textwidth]{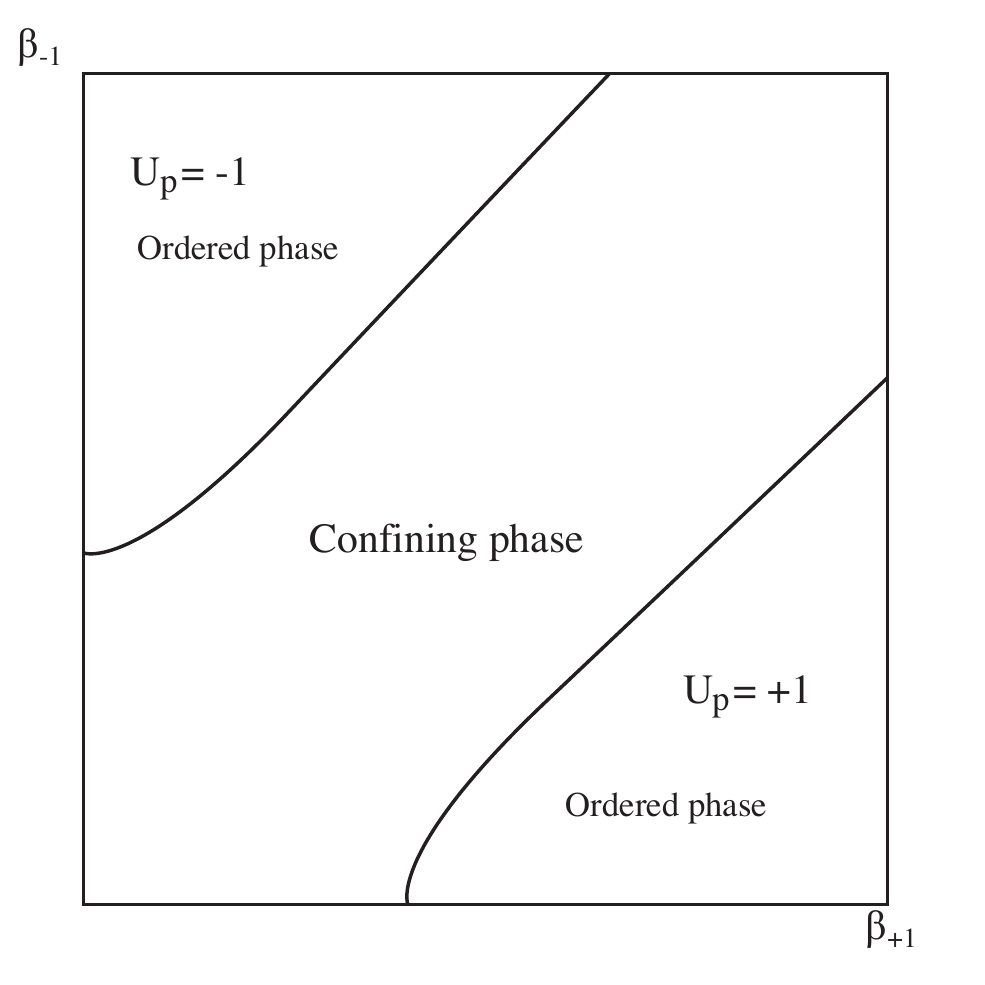}
\caption{\label{fig.pdz2}Phase diagram for a pure $\Z_2$ gauge theory.}
\end{figure}
In Figure \ref{fig.pdz2} the phase diagram of the pure $\Z_2$ gauge theory is shown as a function of the conjugacy class couplings $\beta_{+1}$ and $\beta_{-1}$. Later in this work we present similar phase diagrams for the $\dbar$ gauge theory.

It is well-known that the inclusion of matter coupled to the gauge fields complicates the phase diagram strongly. The question of whether there exist good order parameters to distinguish the phases in coupled gauge-matter systems is interesting in its own right and highly non-trivial \cite{1367-2630-13-2-025009}, but it is not something we will go in to here.

\subsection{DGT and the quantum double of a finite group\label{sec:quantumdouble}}
The particles in a Discrete Gauge Theory, their fusion and braiding properties, spins and so forth are all obtained by working out the representation theory of the underlying quantum group, which is the quantum double of the finite discrete subgroup.
We will not give a detailed account on the emergence of quantum group symmetry in DGT,
this can be found in the literature \cite{Bais:1991pe}, but do present a short summary
of the basics to fix the notation and introduce some key concepts required later on.

Consider the following operators acting on states in the Hilbert space of a DGT.
First there is the {\em flux projection} operator, denoted by $P_h$, which acts on a state $\ket{\psi}$
\begin{displaymath}
	P_h \, \ket{\psi}=\left\{
	\begin{array}{ll}
		\ket{\psi} & \mbox{if the state $\ket{\psi}$ contains flux }h\\
		0 & \mbox{otherwise }
	\end{array}\right.
	.
\end{displaymath}
Secondly, we have the operator $g$, for each group element $g \in H$, 
which realizes a global gauge transformation by the element $g$:
\begin{equation*}
g \, \ket{\psi} = \ket{^g\psi} , \nn
\end{equation*}
where it should be noted that we have not yet modded out by the gauge group to obtain the physical Hilbert space.
These operators do not commute, and realize the algebra
\begin{eqnarray*}
P_h P_{h'} &= \delta_{h,h'} P_h \\
g P_h &= P_{ghg^{-1}} g \, .
\end{eqnarray*}
The set of combined flux projections and gauge transformations $\{ P_h g \}_{h,g \in H} $
generates the quantum double $D(H)$, which is a particular type of algebra called a Hopf algebra. 

The representation theory of the quantum double $D(H)$ of a finite group $H$ was first worked out in 
\cite{Roche:1990hs} but here we follow the discussion presented in \cite{deWildPropitius:1995hk}
and follow the conventions of those lecture notes.

Let $A$ be a conjugacy class in $H$. We will label the elements within $A$ as
\begin{equation*}
\left\{ ^A\!h_1 , ^A\!\!h_2, \dots, ^A\!\!h_k \right\} \in A\, ,
\end{equation*}
for a conjugacy class $A$ of order $k$. In general, the centralizers for the different group elements
within a conjugacy class are different, but they are isomorphic to one another. Let $^A\!N \subset H$
be the centralizer for the first group element in the conjugacy class $A$, denoted by $^A\! h_1$.

The set $^A\!X$ relates the different group elements within a conjugacy class to the first:
\begin{equation}
\label{classcoordinates}
^A\!X = \left\{ ^A\!x_{h_1}, ^A\!\!x_{h_2}, \dots, ^A\!\!x_{h_k} \, \middle\vert \, ^A\!h_i \,=\, ^A\!x_{h_i} \, ^A\!h_1 \, ^A\!x_{h_i}^{-1} \right\}\, .
\end{equation}
This still leaves a lot of freedom, but we fix our convention such that $^A x_{h_1} = e$,
the group identity element. The centralizer $^A\!N$, being a group, will have different irreps,
which we label by $\alpha$. The vector space for a representation $\alpha$ is spanned by a basis $^\alpha v_j$.
The internal Hilbert space corresponding to an irrep of the quantum double that combines magnetic and electric degrees of freedom, $V^{\Ir{A}{\alpha}}$, is then spanned by the set of vectors
\begin{equation*}
\left\{ \ket{^A\!h_i , ^\alpha\!v_j} \right\} \, ,
\end{equation*}
where $i$ runs over the elements of the conjugacy class, $i = 1, 2, \dots, \mathrm{dim}\;A$ and $j$
runs over the basis vectors of the carrier space of $\alpha$, $j=1,2,\dots,\mathrm{dim}\;\alpha$. These irreducible representations correspond precisely to the particle types $a,b,\dots$ in section \ref{tqft.fusion}. They obey a set of fusion rules as in (\ref{fusrules}) and it is possible to calculate the modular $S$-matrix, $F$-symbols and so on.

To see that this basis is a natural one to act on with our flux measurements and gauge transformations,
consider the action of a pure flux projection $P_h e$
\[
P_h e \ket{^A\!h_i , ^\alpha\!v_j} = \delta_{h,^A\!h_i} \ket{^A\!h_i , ^\alpha\!v_j}\, ,
\]
and a pure gauge transformation $\sum_h P_h g$
\[
\sum_h P_h g \ket{^A\!h_i , ^\alpha\!v_j} = \ket{g\,^A\!h_i\,g^{-1} , \sum_m D_\alpha (\tilde{g})_{mj}\,^\alpha v_m}\,.
\]
The matrix action $\Pi^{\Ir{A}{\alpha}}$ of an irreducible representation \ir{A}{\alpha} of some combined projection and gauge transformation $P_h g$:
\begin{equation}
\label{eqn.gaugetr}
\Pi^{\Ir{A}{\alpha}} (  P_h g ) \ket{^A\!h_i , ^\alpha\!v_j}  
= \delta _{h, g\,^A\!h_i\,g^{-1} } \ket{g\,^A\!h_i\,g^{-1} , \sum_m D_\alpha (\tilde{g})_{mj}\,^\alpha v_m}\, ,
\end{equation}
where the element $\tilde{g}$ is the part of the gauge transformation $g$ that commutes with the flux $^A\!h_1$, defined as
\begin{equation}
\label{eqn.centr}
\tilde{g} \left.=\right. ^A\!\!x_{gh_ig^{-1}}^{-1}\,g\,^A\!x_{h_i} \, ,
\end{equation}
and $D_\alpha(\cdot)_{ij}$ is the matrix representation of the centralizer. 

To conclude we give a simple expression for the modular $S$-matrix that can be obtained by calculating the trace of the monodromy matrix
\begin{equation}\label{qdsmat}
S_{\Ir{A}{\alpha}\Ir{B}{\beta}} = {1 \over |H|} \sum _{g \in A, h \in B,[g,h]=e} \mathrm{Tr}_\alpha(x_g^{-1} h x_g)^* \; \mathrm{Tr}_\beta (x_h^{-1} g x_h) ^*, 
\end{equation}
where $[g,h]$ is the group theoretical commutator: $[g,h]=ghg^{-1}h^{-1}$.
\subsection{\label{sec.actions}Lattice actions and observables}
\label{sec:auxiliary}
We discretize three-dimensional spacetime into a set of sites $i,\, j,\, \cdots$ using a rectangular lattice.
The gauge field $U_{ij}$, which takes values in the gauge group $H$, lives on the links $ij, \, jk, \, \dots $
connecting sets of neighboring sites. The links are oriented in the sense that $U_{ij} = U^{-1}_{ji}$ \cite{Wilson:1974sk}.

\begin{figure}
\centering
\includegraphics[width=0.3\textwidth]{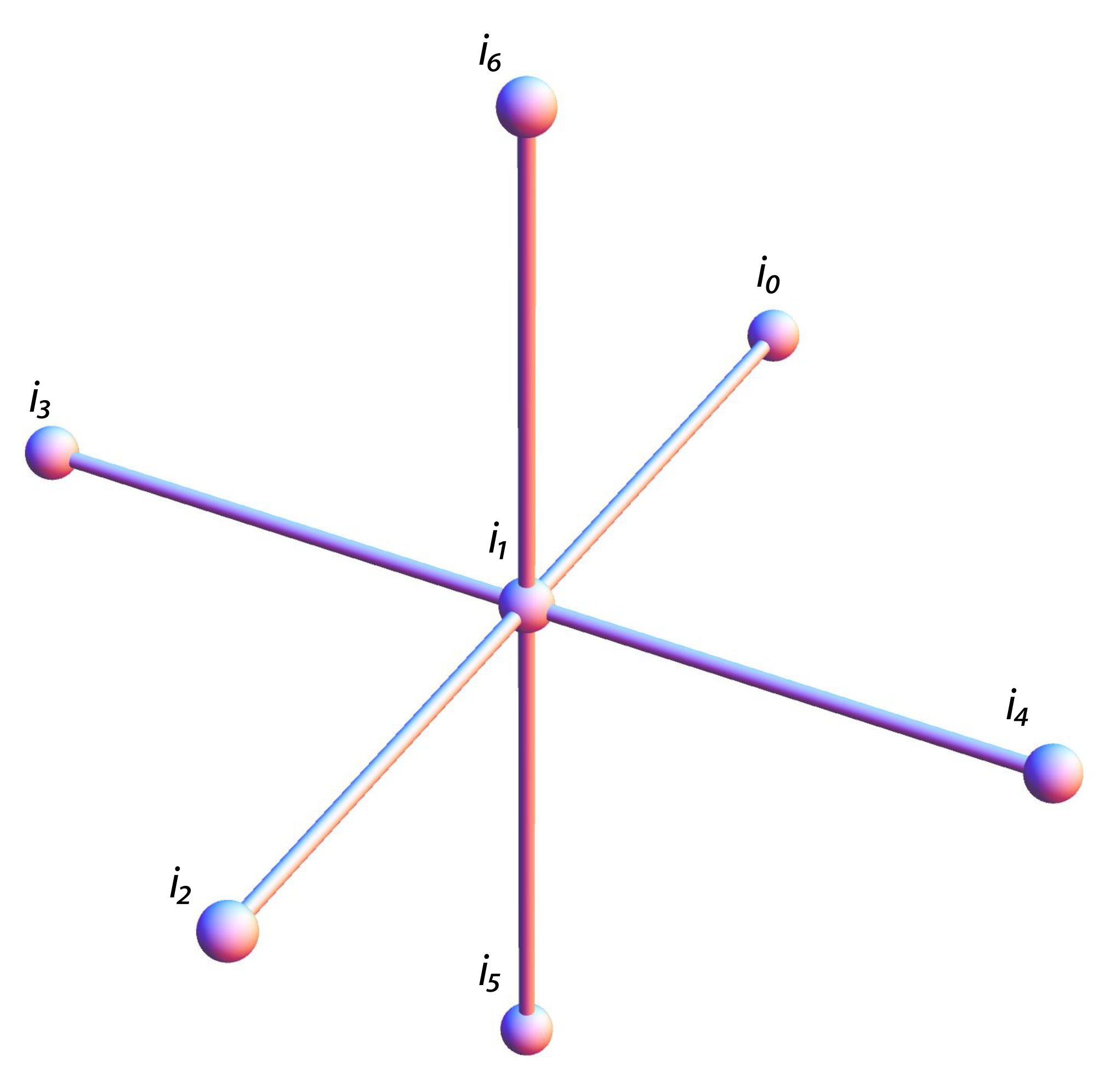}
\caption{In our convention a gauge transformation $g$ at location $i_1$ transforms $U_{i_1i_2}\rightarrow gU_{i_1i_2}$, $U_{i_1i_4} \rightarrow gU_{i_1i_4}$, $U_{i_1i_6} \rightarrow gU_{i_1i_6}$, $U_{i_0i_1} \rightarrow~U_{i_0i_1}g^{-1}$, $U_{i_3i_1} \rightarrow~U_{i_3i_1}g^{-1}$, $U_{i_5i_1} \rightarrow U_{i_5i_1}g^{-1}$.\label{fig.gaugetrans}}
\end{figure}

We note that the gauge field $U_{ij}$ takes care of the parallel transport of matter fields that are charged under the gauge group
from site $i$ to site $j$. An ordered product of links along a closed loop is
gauge invariant up to conjugation by a group element and measures the holonomy of the gauge connection.
Gauge transformations are labeled by a group element $g_i \in H$ and are performed at the sites of the
lattice. The gauge field transforms as
\begin{equation}\label{gaugetransform}
U_{ij} \mapsto g_i \, U_{ij} \, g^{-1}_j\, ,
\end{equation}
where the orientation of the links (incoming or outgoing) has to be taken into account as shown in Figure \ref{fig.gaugetrans}.

The standard form for the lattice gauge field action makes use of the ordered product of links around
a plaquette $ijkl$:
\begin{equation*}
U_p = U_{ijkl} = U_{ij}\,U_{jk}\,U_{kl}\,U_{li} \, ,
\end{equation*}
which transforms under conjugation by the gauge group,
\begin{equation*}
U_p \mapsto g_i\,U_p\,g^{-1}_i\, .
\end{equation*}
The gauge action per plaquette which corresponds to the Yang-Mills form $F_{\mu\nu}^2$ in the
continuum limit for $H=SU(N)$, is given by
\begin{equation}
\label{irrepact}
S_p = - \sum_\alpha \beta_\alpha \chi_\alpha \left(U_p \right) \, ,
\label{gauge1}
\end{equation}
where $\chi_\alpha$ is the group character in irrep $\alpha$ and $\beta_\alpha$ is inversely proportional
to the square of the coupling constant for irrep $\alpha$. This is known, for $H=SU(2)$ and the sum over representations limited to the fundamental one, as the Wilson action \cite{Wilson:1974sk}.
The action (\ref{irrepact}) is the euclidean analog of the Kitaev Hamiltonian introduced in \cite{Kitaev20032} (which in itself is a variation of the Kogut-Susskind Hamiltonian \cite{PhysRevD.11.395}) where the  electric field is represented by the timelike plaquettes and the gauge constraint (\ref{ham.gaugetr}) is not necessary since everything is manifestly gauge invariant.

For $SU(N) $ gauge theories one usually only includes
the fundamental representation and is thus left with only one coupling constant.
This is not necessary however: gauge invariance of the action is ensured by the fact the characters are conjugacy class functions, in
and therefore  we will consider actions where the number of independent couplings equals the number of conjugacy classes i.e. the
number of irreps (for a finite group these numbers are finite and equal).

For our purposes, namely the study of magnetic condensates in DGTs, equation (\ref{gauge1}) is not the most convenient to work with. We perform a change of basis
in the space of coupling constants to write it as a sum over delta functions on conjugacy classes: $\delta_A(h) = 1$ if
$h \in A$, and $0$ otherwise. In this basis the action becomes
\begin{equation*}
S_p = - \sum_A \beta_A \delta_A \left(U_p \right) \, .
\label{gauge2}
\end{equation*}
This formulation allows us in particular to directly control the mass of the different fluxes in the theory, which will ease the search for different vacua in the phase diagram. Increasing the coupling constant for a certain conjugacy class (magnetic flux) $A$ will increase the contribution of configurations carrying many $A$ fluxes to the path integral. Likewise, setting all $\beta_A$ to zero except $\beta_e$, the coupling constant for the trivial conjugacy class, will result in an ``empty'' vacuum and therefore an unbroken phase.

To perform the transformation to the conjugacy class basis, we need to make use of the following orthogonality relations valid
for all finite groups $H$
\begin{eqnarray}
\label{eq.or1}
\int_H dg \; \chi_\alpha (g) \chi^*_\beta (g) &=& \delta_{\alpha, \beta} ,\\
\label{eq.or2}
\sum_{\alpha \in \mathcal{R}} \chi_\alpha (g) \chi^*_\alpha (h) &=& \frac{|H|}{|A|} \; \mathrm{if} \; g,h \in A \\
 &=& 0 \; \mathrm{otherwise} \nn,
\end{eqnarray}
where $|H|$ is the order of the group $H$, $|A|$ is the order of the conjugacy class $A$, $\mathcal{R}$ is the set of
irreps and group integration is defined as
\begin{equation*}
\int_H dg \; f(g) = \frac{1}{|H|} \sum_{g \in H} f(g).
\end{equation*}
Equations (\ref{eq.or1}) and (\ref{eq.or2}) show that the irreducible representations of a group $H$ form an
orthonormal set for functions on conjugacy classes of $H$. We thus expect the conjugacy class delta function to be expressible in terms of characters
\begin{equation*}
\delta_A (g) = \sum_{\alpha \in \mathcal{R}} c_\alpha \chi_\alpha (g),
\end{equation*}
for some set of constants $\{ c_\alpha \}$. We multiply both sides of this expression by a character of the same
group element in another irrep $\beta$ and perform the integrations by use of the orthogonality relations (\ref{eq.or1}) and (\ref{eq.or2})
\begin{eqnarray*}
\int_H dg \; \chi^*_\beta (g) \delta_A(g) &=& \sum_{\alpha \in \mathcal{R}} c_\alpha \int_H dg \; \chi^*_\beta (g) \chi_\alpha (g) ,\\
\frac{|A|}{|H|} \chi^*_\beta (A) &=& \sum_{\alpha \in \mathcal{R}} c_\alpha\delta_{\alpha \beta} = c_\beta,
\end{eqnarray*}
where the slightly abusive notation $\chi_\alpha (A)$ means the character of any group element of $A$ in the
representation $\alpha$. This shows that
\begin{equation}
\label{lgt.deltafn}
\delta_A (g) = \sum_{\alpha \in \mathcal{R}} \frac{|A|}{|H|} \chi^*_\alpha(A) \chi_\alpha(g),
\end{equation}
which in turn implies that the the difference between (\ref{gauge1}) and (\ref{gauge2}) is just a change of basis:
\begin{equation*}
\sum_{A \in \mathcal{C}} \beta_A (\beta_\alpha) \delta_A (g) = \sum_{\alpha \in \mathcal{R}} \beta_\alpha \chi_\alpha (g),
\end{equation*}
where $\mathcal{C}$ is the set of conjugacy classes and $\beta_A ( \beta_\alpha)$ is given by
\begin{equation*}
\beta_A = \sum_\alpha \beta_\alpha \chi_\alpha (A).
\end{equation*}
To probe the physics of the system for a fixed set of values of the coupling constants in the action, we will use
a set of order parameters and phase indicators. These order parameters are in one-to-one with the set of fundamental anyonic excitations of the theory.

\paragraph*{Order parameters and phase indicators.} We distinguish two different sets of order parameters that are closely related to one another. The first is
the set of closed {\em loop operators}, that physically correspond to the creation, propagation and annihilation of an anyon-anti-anyon pair in spacetime. The second is the set of open {\em string operators} that create, propagate and annihilate a single anyon. In the background of a trivial vacuum, only the loops can have nonzero expectation values, since the creation of a single particle would violate the conservation of the quantum numbers of the vacuum in such a background. This means that the open strings tell us something about possible Bose condensates, whereas the closed loops tell us about the behaviour of external particles put into this background. We define the open string operators only for the purely magnetic sectors, since in this work we only study magnetic condensates\footnote{Electric condensates break the gauge group $H$ to some subgroup $K$ by the conventional Higgs breaking, this implies in the present context  that $D(H)$ will be broken to $D(K)$, which in turn means that the fluxes in the coset $G/H$ are confined \cite{Bais2003}.}. 

First we will define the loop operators. This set of nonlocal order parameters was introduced in a previous publication \cite{Bais:2008xi}. For a full discussion, we refer to that work. Here we recall the essentials and fix the notation.
The closed loops are a generalization of the Wilson and 't Hooft loops. They create a particle-antiparticle pair from the vacuum and annihilate them at a later time. These loops allow us to calculate Aharonov-Bohm type phases and determine which anyonic excitations will be confined. In $SU(N)$ gauge theories, the Wilson loop for a free excitation, e.g. in the Higgs phase of $SU(2)$ theory, in general falls off as $e^{-c P}$, with $P$ the perimeter of the loop, whereas a confined excitation, such as the $\mathbf{3}$ charge of an external quark source in pure $SU(3)$ gauge theory describing QCD, falls off as $e^{-c' A}$, with $A$ the area of the loop. 


Because the excitations in a DGT are gapped, numerically we find that the expectation values of loop operators are constant as a function of size. The argument for this behaviour for the $\Z_2$ theory is in Section \ref{sec:z2topo}. Although only strictly true in the limit of infinite coupling constant, the gap supresses the dependence on size so strongly, we will assume that the theory is a purely topological one in the region of coupling constant space in which we are interested. 

\begin{figure}
\centering
\includegraphics[width=0.3\textwidth]{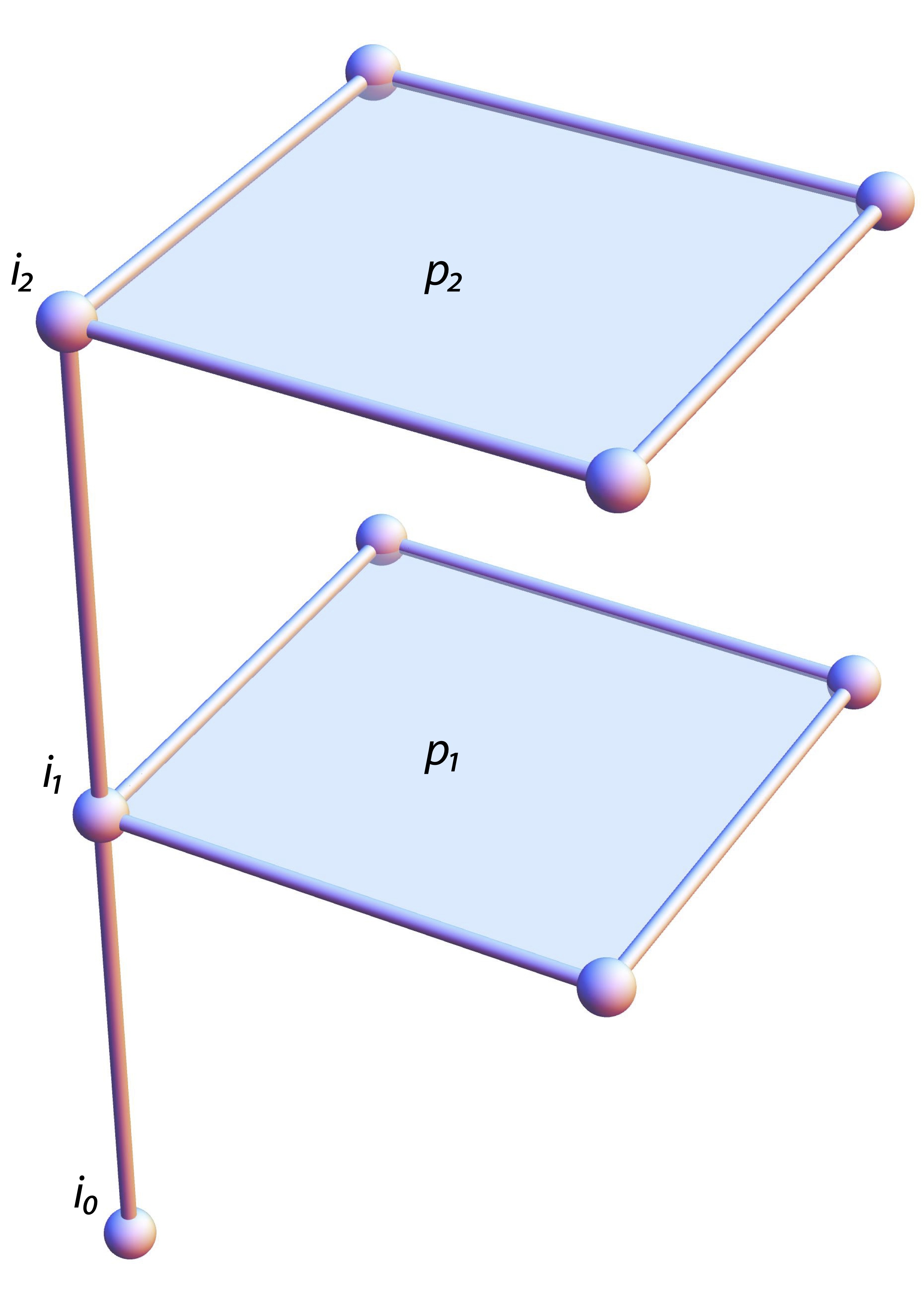}
\caption{The first two plaquettes appearing in expression (\ref{eqn.dyonop}). The ordered product $U_p$ of links around a plaquette $p$ needs to be taken with an orientation that has to be constant throughout the loop.}
\end{figure}

\begin{figure}
\centering
\includegraphics[width=0.5\textwidth]{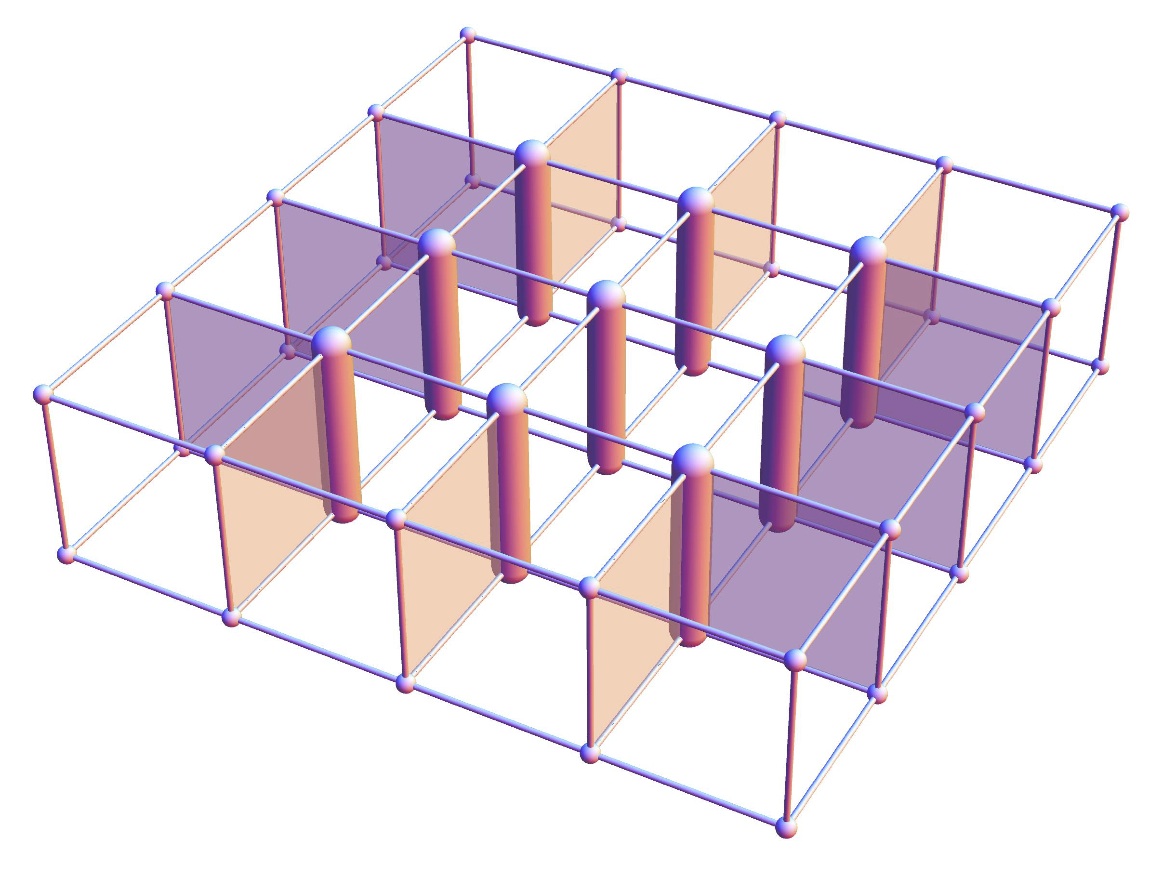}
\caption{A set of plaquettes forming a closed loop on the lattice. The fat links constitute the $h$-forest.\label{fig.gforest}}
\end{figure}
Let us draw a closed loop on the dual lattice, this loop pierces a set of plaquettes $C$ through which we will force magnetic flux.  Now draw another loop , this time on the real lattice, such that (i) each point of this loop lies on the corner of a plaquette in $C$ and (ii) the two loops do not link\footnote{One can also create loops that have a linking number, the expectation values of such loops allow one to calculate the topological spin for a given excitation.}. The combination of the two loops establishes a framing: we have selected a location for the electric charge of a flux-charge composite. This framing also provides us with a point and orientation on each plaquette from which to take the plaquette product $U_p$: for non-Abelian groups the product of the four links depends on which corner you start.

To insert a flux $h$ in a plaquette $p \in C$, we have to ``twist'' the Boltzmann factor of this particular plaquette by locally changing the action from $S(U_p)$ to $S(h^{-1}U_p)$: if the minimum of the action was previously obtained for $U_p=e$, it is now shifted to $U_p=h$. We want to perform this twisting procedure for all plaquettes in $C$.

The notion of a group element as a magnetic flux is not gauge-invariant: under a gauge transformation by $g$, a flux $h$ transforms as $g\,h\,g^{-1}$. Therefore it is necessary to sum over the group elements $h$ in a conjugacy class $A$ in some way. 

One can go about this in two different, and inequivalent, ways.
\begin{itemize}
\item The authors in Ref.~\cite{Alford:1990fc} only studied pure magnetic flux loops (without electric charge) and performed a sum over the conjugacy class for each plaquette in $C$ individually. This leaves a gauge-invariant expression, but the loop loses its framing, since a conjugation by the element $U_{01}$ maps
\[
h^{-1} U_{01}U_{12}U_{23}U_{30} \rightarrow U_{01} h^{-1} U_{12}U_{23}U_{30},
\] 
for a plaquette spanned by group elements $U_{01}\dots U_{30}$.
\item When dealing with nontrivial braiding properties of loop operators  it is necessary to choose  a basepoint $i_0$ in space with respect to which all operators are defined, it provides a calibration that  serves as a ``flux bureau of standards'', borrowing a term from \cite{Preskill2004}. This point can be anywhere in spacetime and does not need to be on the loop. We then define a function $k_{i_p}(h,\{U_{ij}\})$ of $h$ and the gauge field variables $\{U_{ij}\}$ for the twist element that has to be inserted into the plaquette product for plaquette $p$, where $i_p$ is the corner of the plaquette chosen in the framing
\[
k_{i_p} = k_{i_p}(h,\{U_{ij}\}) = U^{-1}_{i_0 i_p}\, h^{-1} U_{i_0 i_p}.
\]
\end{itemize}

With this notation and the above considerations, the anyonic operator $\Delta^{\Ir{A}{\alpha}}$ is given by \footnote{This definition
is different from our original definition  \cite{Bais:2008xi} by a factor of $\frac{1}{|A|}$. This definition gives
the correct $S$-matrix elements directly.}:

\begin{equation}
\label{eqn.dyonop}
\Delta^{\Ir{A}{\alpha}} (C) =  \sum_{h \in A} \prod_{p_j \in C} D_\alpha\left( x^{-1}_{U_{j-1,j}\,k_j \,U^{-1}_{j-1,j}} \; U_{j-1,j} \; x_{k_j} \right) e^{S(U_{p_j}) - S(k_j\,U_{p_j})} .
\end{equation}
Here $p_j$ iterates over the plaquettes in $C$ and $D_\alpha$ is the representation function of the centralizer irrep $\alpha$ of $^AN$. The link $U_{j-1,j}$ neighbours the plaquette $p_j$, and the combination in brackets always takes values in the centralizer subgroup of the conjugacy class $A$. The exponential of the difference of two actions changes the minimal action
configuration to one containing flux $h$ for the plaquette under consideration.

The operator in expression (\ref{eqn.dyonop}) is a generalization of
the Wilson and 't Hooft loops, and by constructing it we have established the desired  one to one correspondence between irreducible  representations of the quantumgroup and loop operators for the pure discrete gauge theory.
 If we fill in for $A$ the trivial conjugacy class, the exponent vanishes and the $x$ group
elements are equal to the group unit, so after we multiply out the $D_\alpha$-matrices we are left with
\[ \Delta^{\Ir{e}{\alpha}} (C) = \chi_\alpha \left(U_{1,2} U_{2,3}\cdots U_{n-1,n} U_{n,1} \right) , \]
where the product of $U$s is an ordered product along the loop on the lattice.

On the other hand, if we replace $\alpha$ by the trivial representation, we are left with
\[
\Delta^{\Ir{A}{1}} (C) = \sum_{h \in A} \prod_{p_j \in C} e^{S(U_{p_j}) - S(h^{p_j}U_{p_j})},
\]
which is comparable to the order parameter proposed in \cite{Alford:1990fc}, but the gauge invariance with respect to the transformations (\ref{gaugetransform}) is ensured
in a different way. We sum over the conjugacy class only once and insert the flux in a gauge invariant way by parallel transporting it along the loop from a fixed basepoint. The operator in \cite{Alford:1990fc} sums over the conjugacy class for each individual plaquette. This way also gauge invariance is achieved, but the loop loses its framing, and therefore is not suitable to describe true anyonic charges.

The open magnetic string operators are a variant of expression (\ref{eqn.dyonop}) where the set of plaquettes $C$ corresponds to an open string on the dual lattice. Looking at the $h$-forest configurations, it can immediately be seen that such a string, corresponding to the creation and subsequent annihilation of a single particle, has zero expectation value in the trivial vacuum. For these strings to acquire a non zero expectation value a {\em vacuum exchange contribution} is required, which we will focus on now.
\paragraph*{The vacuum exchange contribution.} We use the set of operators $\{\Delta^{\Ir{A}{\alpha}}\}$ to measure the elements of the $S$-matrix by picking two loops $C_1$ and $C_2$ that link each other once
\begin{equation}
\label{meassmat}
\left< S_{\Ir{A}{\alpha} \Ir{B}{\beta}} \right> = \left< \Delta^{\Ir{A}{\alpha}} (C_1) \Delta^{\Ir{B}{\beta}} (C_2)\right>.
\end{equation}
In the trivial vacuum the $S_{\Ir{A}{\alpha}\Ir{B}{\beta}}$-matrix elements of fluxes $g\in A$ and $h\in B$ for which $g\,h\,g^{-1}\,h^{-1} = [g,h] \neq e$ evaluate to zero (this is what we measure using the operators (\ref{eqn.dyonop}) and calculate algebraically (\ref{qdsmat}) ). If we however measure the $S$-matrix elements of such noncommuting fluxes in a broken vacuum nonzero matrix elements can appear.

This is most easily explained by considering an example. The main contribution to a single loop of pure magnetic flux is of the form pictured in Figure \ref{fig.gforest}. This configuration is called the $h$-forest state in earlier literature \cite{Alford:1990fc}. Modulo gauge transformations this is the dominant configuration in the trivial vacuum that contributes to a loop of flux labeled by conjugacy class $A$, where $g \in A$. Expression (\ref{eqn.dyonop}) contains a sum over these group elements within a conjugacy class, but let us for now focus on one of the group elements. Each link in this configuration has value $e$, except for the fat links in Figure \ref{fig.gforest}, they have value $h$. That this configuration leads to a loop or tube of flux is easily seen: within the forest each plaquette has a value $e\,h\,e\,h^{-1}=e$, whereas at the edges the value is $e\,h\,e\,e = h$ (depending on the orientation of the plaquette product). This is also the easiest way to see the origin of the Aharonov-Bohm effect on the lattice: an electric charge loop having linking number 1 with the flux loop will have exactly one link with value $h$ in it, therefore its value will be $\chi_\alpha (h)$.

\begin{figure}
\centering
\includegraphics[width=0.5\textwidth]{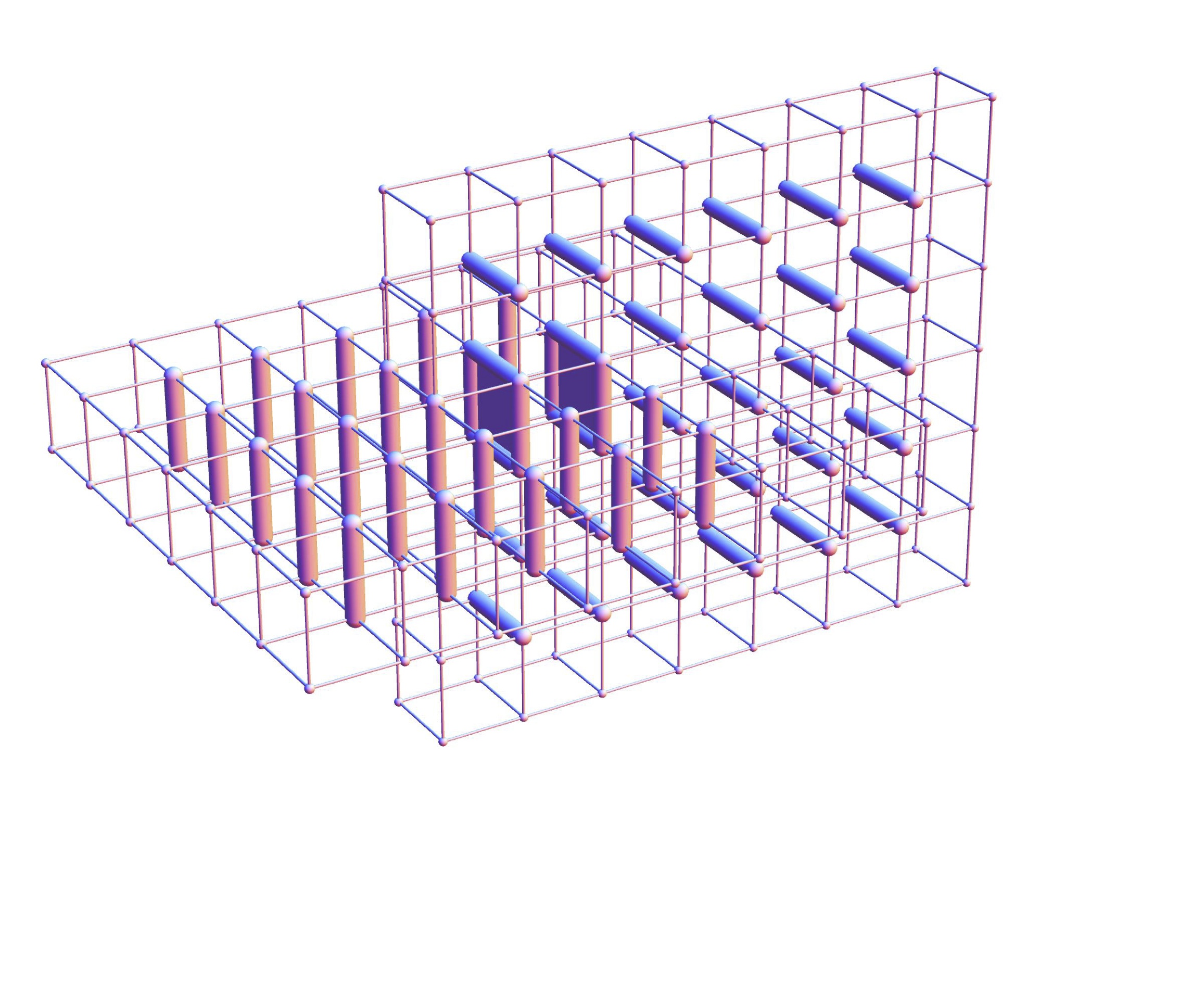}
\caption{\label{fig.smatrixconfig} The vacuum exchange contribution. Double $h$-forest, configuration contributing to the $S$-matrix measurement of two non commuting fluxes. The fat links are the $g$-forest and $h$-forest and the shaded plaquettes are a string of $[g,h]$ flux connecting the the two loops.}
\end{figure}

Consider now the dominant configuration that contributes to the $S$-matrix element $S_{\Ir{A}{1}\Ir{B}{1}}$. We again pick two group elements $g\in A$, $h\in B$ and draw a similar diagram. This is shown in Figure \ref{fig.smatrixconfig}.
By similar logic this causes the plaquettes at the boundary of either forest to have value $g$ respectively $h$. Inside the forests most plaquettes still have value $e$, however there are some plaquettes that are different. There is a tube of plaquettes that have value $[g,h]$, where the two forests intersect. In general this group theoretical commutator is not equal to identity element  for nonabelian groups.
This is the physical reason behind the appearance of zeroes in the $S$-matrix for nonabelian theories. This tube of plaquettes represents a flux $[g,h]$ going from the one loop to the other. In the trivial vacuum this flux will be gapped, so the contribution of this diagram to the path integral expectation value will be negligible.

However, a different situation appears when we are in a vacuum where the flux $[g,h]$ has Bose condensed. We cannot give a single configurations that contributes dominantly to the path integral (there are many), but we can say that configurations like the one in Figure \ref{fig.smatrixconfig} are now contributing since the mass for the flux $[g,h]$ has disappeared.

Thus we expect that in the measurements there will be cases where zeroes in the original $S$-matrix will obtain a nonzero value in the broken phase.
\paragraph*{An  auxiliary $^AN$ gauge symmetry.}
The operators (\ref{eqn.dyonop}) are invariant with respect to the local $H$ gauge transformations (\ref{gaugetransform}). However, in  our formulation of the operators  we have tacitly introduced another, auxiliary $^AN$ gauge symmetry that is less obvious. A crucial property that allows one  to determine the topological symmetry breaking pattern in detail is that the loop operators do transform nontrivially under this symmetry. In a non trivial ground state, these  symmetries may be broken  and will therefore lead to the lifting of certain degeracies related with the splitting of fields in the topological symmetry breaking process. So this hidden symmetry turns out to be a blessing in disguise.

Let us first note that there is no preferred choice for the coordinate system (\ref{classcoordinates}) we define for the conjugacy classes. Once a certain choice $\{x_{h_i}\}$ has been made such that $h_i = x_{h_i} h_1 x_{h_i}^{-1}$, a set $\{x'_{h_i}\}$ with 
\begin{equation}\label{xtrans}
x'_{h_i} = x_{h_i} n_{h_i}, \;\;\;\; [n_{h_i} ,h_i]=e,\;\;\;\; (n_{h_i} \in{^AN}) \end{equation} 
will do just as well. In the trivial vacuum, the $S$-matrix is invariant with respect to this transformation. This is most easily seen by looking at the algebraic expression (\ref{qdsmat}), but it is also confirmed by our measurements of (\ref{meassmat}).

This invariance can be understood on the operator level by multiplying out the representation matrices of the centralizer in equation (\ref{eqn.dyonop}). Generally this will lead to terms of the form
\[
\mathrm{Tr}_\alpha \tilde{g}  =\mathrm{Tr}_\alpha (x_{h_k}^{-1} g x_{h_i} ),
\]
where $g$ is the product of links on the loop and $h_k=g\,h_i\,g^{-1}$, implying that indeed $\tilde{g} \in {^AN}$. When the loop is linked with another loop, the element $g$ will in general be in the conjugacy class of the flux of this other loop.
Under the  transformation (\ref{xtrans}) of the conjugacy class coordinate system, the above expression will transform as
\[
\mathrm{Tr}_\alpha ( n_{h_k}^{-1} x_{h_k}^{-1} g x_{h_i} n_{h_i}) = \mathrm{Tr}_\alpha ( n_{h_i
} n_{h_k}^{-1} x_{h_k}^{-1} g x_{h_i}),
\]
due to the cyclicity of the trace. This elucidates the invariance of the $S$-matrix in the trivial vacuum under the translation of the $x_{h_i}$: non-commuting fluxes never have a non-zero matrix element, and if $[g,h]=e$, we have that $h_i=h_k$ and therefore $n_{h_i
} n_{h_k}^{-1}=e$. In a non-trivial ground state where non-commuting fluxes may have non-zero $S$-matrix elements due to a vacuum interchange contribution, the transformation (\ref{xtrans}) may manifest itself in different measured matrix elements. This means that in such cases the entry $(A,\alpha)$ may split into multiple entries $\{(A_i,\alpha_i)\}$. As we are interested in these multiple entries, we will in our calculations always include the nontrivial behavior of our observables under this auxiliary $^AN$ action. This turns out to be one of two mechanism responsible for the splitting of irreps of $\sA$ into multiple irreps of $\sU$, the other of which we turn to now.

\paragraph*{An auxiliary $H/^AN$ symmetry.} There is another symmetry, but now  on the level of the fusion algebra that turns out to be useful. Suppose in the theory there exists a rule  of the form
\begin{equation}
\label{simplecurrent}
\Ir{A}{\alpha} \times \Ir{e}{\beta} = \Ir{A}{\alpha},
\end{equation}
where \ir{e}{\beta} is some one-dimensional purely electric representation. This turns out to be the case whenever the representation $\Pi^{\Ir{e}{\beta}}(\cdot)$ evaluates to unity for all elements in $^A\!N$, the normalizer of conjugacy class $A$. We can prove this using the explicit expression for the fusion coefficients in terms of the quantum double characters:
\begin{equation}
N^{\Ir{A}{\alpha}\Ir{B}{\beta}}_{\Ir{C}{\gamma}} = {1 \over |H|} \sum_{g,h} \mathrm{Tr} \left[ \Pi^{\Ir{A}{\alpha}} \otimes \Pi^{\Ir{B}{\beta}} \left(\Delta(P_h\,g)\right) \right] \mathrm{Tr} \left[ \Pi^{\Ir{C}{\gamma}} \left( P_h\,g \right) \right]^*.
\end{equation}
Picking $\Ir{B}{\beta} = \Ir{e}{\beta}$ and $\Ir{C}{\gamma} = \Ir{A}{\alpha}$,
\begin{eqnarray*}
N^{\Ir{A}{\alpha}\Ir{e}{\beta}}_{\Ir{A}{\alpha}} &=& {1 \over |H|} \sum_{g,h} \mathrm{Tr} \left[ \Pi^{\Ir{A}{\alpha}} \otimes \Pi^{\Ir{e}{\beta}} \left( \sum_{h_1 h_2 = h} P_{h_1} \,g \otimes P_{h_2}\, g\right) \right] \cdots \\
&&\cdots \mathrm{Tr} \left[ \Pi^{\Ir{A}{\alpha}} (P_h \, g)\right]^* \\
&=& {1 \over |H|} \sum_{g,h} \mathrm{Tr} \left[ \Pi^{\Ir{A}{\alpha}} (P_h \, g) \otimes \Pi^{\Ir{e}{\beta}} (P_e \, g) \right] \mathrm{Tr} \left[ \Pi^{\Ir{A}{\alpha}}(P_h \, g)\right]^* \\
&=&{1 \over |H|} \sum_{g\in ^A\!N,h\in A} \mathrm{Tr} \left[ \Pi^{\Ir{A}{\alpha}} (P_h \, g) \right] \mathrm{Tr} \left[ \Pi^{\Ir{A}{\alpha}} (P_h \, g) \right]^*=1,
\end{eqnarray*}
where in the latter line we have made use of the orthogonality of the characters. We assumed $\Pi^{\Ir{e}{\beta}} (P_e \, g)=1$ for all $g \in ^A\!N$. The sum over $h$ is restricted since if $h \not\in A$ the matrix element will be zero and the sum over $g$ is restricted since if $g \not\in ^A\!N$ the matrix element will be off-diagonal and thus not contribute to the trace. 

So we see that the fusion rule (\ref{simplecurrent})  leads  a degeneracy in the calculation of $S$-matrix elements since by definition
\[
\left<\,\qdsmatrix{\Ir{A}{\alpha}}{\Ir{C}{\gamma}} \,\right>_0=  
\left< \, \qdcloseddoubleloop{\Ir{A}{\alpha}}{\Ir{e}{\beta}}{\Ir{C}{\gamma}} \quad \right>_{\raisebox{22pt}{\footnotesize 0}}.
\]
However,  on the operator level this equality does not hold. Indeed, when we probe the LHS of this equation in a non-trivial vacuum the result will in general differ from the RHS. In particular, it turns out that the different $\sU$ representations that lift to the same $\sA$ representations \ir{A}{\alpha} differ precisely by such a fusion. So this degeneracy may be lifted in the broken phase and give rise to a additional splittings of certain entries $(A,\alpha)$.
Consequently in our numerical calculations we have to explicitly keep track of the presence of such electric representations $(e,\beta)$, that satisfy (\ref{simplecurrent}) and see whether they give rise to additional splittings. 

To conclude this section, we remark that we have very explicitly indicated how one gets from the modular $S$-matrix $S_{ab}$ to the extended or broken $S$-matrix $\bar{S}_{{a_i}{b_j}}$, from which the topological data of the broken $\sU$ phase can be immediately read off.

\section{The $D(\dbar)$ gauge theory}
We turn to the particular example we have chosen to work out in detail: a discrete gauge theory with gauge group $\dbar$, also called the quaternion group. The representation theory was worked out in
\cite{deWildPropitius:1995hk}, here we summarize the results that are required to describe the breaking by a Bose condensate.
\subsection{\label{breakalg}Algebraic analysis}
The group $\dbar$ contains eight elements that can be represented by the set of $2\times 2$-matrices
\begin{equation}\label{d2barrep}
\{ \one, -\one, \pm  i \sigma_1 , \pm i \sigma_2, \pm i \sigma_3 \},
\end{equation}
where the $\sigma_i,\;i=1,2,3$ are the Pauli spin matrices. We denote the conjugacy classes as $e = \{ \one \},\; \ebar = \{ -\one \}, \; X_1 = \{i \sigma_1 , -i\sigma_1 \}, \; X_2 = \{i \sigma_2 , -i\sigma_2 \}, \; X_3 = \{i \sigma_3 , -i\sigma_3 \}$ and the irreducible group representations as $1$, the trivial irrep, $J_1,\;J_2,\;J_3$ three one-dimensional irreps and $\chi$ the two-dimensional irrep given by (\ref{d2barrep}). The character table is given on the left hand side of Table \ref{d2bartab}.
\begin{table}[h]
\begin{center}
\begin{tabular}{c|rrrrr}
$\dbar$& $e$ & $\ebar$ & $X_1$ & $X_2$ & $X_3$ \\
 \hline
$1$ & 1 & 1 & 1 & 1 & 1 \\
$J_1$ & 1 & 1 & 1 & -1 & -1 \\
$J_2$ & 1 & 1 & -1 & 1 & -1 \\
$J_3$ & 1 & 1 & -1 & -1 & 1 \\
$\chi$ &2 & -2 & 0 & 0 & 0 
\end{tabular}
\hspace*{10mm}
\begin{tabular}{c|rrrr}
$\Z_4$& $\one$ & $i \sigma_i$ & $-\one$ & $-i \sigma_i$ \\
 \hline
$\Gamma^0$ & 1 & 1 & 1 & 1 \\
$\Gamma^1$ & 1 & $i$ & -1 & $-i$ \\
$\Gamma^2$ & 1 & -1 & 1 & -1 \\
$\Gamma^3$ & 1 & $-i$ & -1 & $i$ \\
\end{tabular}
\end{center}
\caption{Character table of the group $\dbar$ and of $\Z_4$ as a centralizer of the conjugacy class $X_i$. \label{d2bartab}}
\end{table}
The centralizer groups for the conjugacy classes $e$ and $\ebar$ are both $\dbar$ since the elements in these conjugacy classes constitute the center of the group. The
conjugacy classes $X_i$, $i=1,2,3$ have non-trivial $\Z_4$ centralizer subgroups, of which the character table is given on the right hand side of Table \ref{d2bartab}. 
The irreducible representations of the quantum double are labeled by a combination \ir{A}{\alpha} of a conjugacy class $A$ and
a centralizer irrep $\alpha$. The full set of fusion rules for the $D(\dbar)$ theory is given in \ref{appfusion}.
All in all, there are 22 sectors: the trivial flux paired with the five irreps of $\dbar$, the $\ebar$ flux paired with the five irreps of $\dbar$ and the three $X_i$ fluxes paired with the four $\ZZ_4$ irreps. The sectors that involve an $X_i$ flux or a $\chi$ irrep have quantum dimension 2, the others have unit quantum dimension. One obtains that the total quantum dimension for the theory $D_A = 8$. 

\paragraph{Breaking: \ir{\ebar}{1} condensate.}In this case the lift of the new vacuum is $\phi$ = \ir{e}{1} + \ir{\ebar}{1}, which implies that $q = d_{(e,1)} + d_{(\bar{e},1)} = 2$. To determine the effective
low energy theory we fuse $\phi$ with all particle sectors of the theory and look for the irreducible combinations that appear. As before the notation $\Ir{A}{\alpha}$ stands for a particle with magnetic flux $A$ and electric charge $\alpha$.
\begin{eqnarray*}
\phi \times \Ir{e}{1} 			&=& \Ir{e}{1} + \Ir{\ebar}{1} \\
\phi \times \Ir{e}{J_i} 	 		&=& \Ir{e}{J_i} + \Ir{\ebar}{J_i} \\
\phi \times \Ir{e}{\chi} 		&=& \Ir{e}{\chi} + \Ir{\ebar}{\chi} \;\; (*)\\
\phi \times \Ir{\ebar}{1}	 	&=& \Ir{e}{1} + \Ir{\ebar}{1} \\
\phi \times \Ir{\ebar}{J_i}	 	&=& \Ir{e}{J_i} + \Ir{\ebar}{J_i} \\
\phi \times \Ir{\ebar}{\chi}		&=& \Ir{e}{\chi} + \Ir{\ebar}{\chi} \;\; (*)\\
\phi \times \Ir{X_i}{\Gamma^0}	&=& \Ir{X_i}{\Gamma^0} + \Ir{X_i}{\Gamma^0} \\
\phi \times \Ir{X_i}{\Gamma^1}	&=& \Ir{X_i}{\Gamma^1} + \Ir{X_i}{\Gamma^3} \;\; (*)\\
\phi \times \Ir{X_i}{\Gamma^2}	&=& \Ir{X_i}{\Gamma^2} + \Ir{X_i}{\Gamma^2} \\
\phi \times \Ir{X_i}{\Gamma^3}	&=& \Ir{X_i}{\Gamma^1} + \Ir{X_i}{\Gamma^3} \;\; (*)
\end{eqnarray*}
The lines marked with (*) have components  on the right hand side that carry different spin factors, implying that they are confinement in the broken phase. Studying the fusion rules of the
surviving combinations of irreps leads to the conclusion that the effective $\sU$ theory is $D(\Z_2 \otimes \Z_2)$. We denote the four different irreps and conjugacy classes of the group $\Z_2 \otimes \Z_2$ by the labels $++,+-,-+,--$, the first (second) symbol standing for the first (second) $\Z_2$. This means $\sD^2_\sT = 32$ and $\sD^2_\sU = 16$. 
The branchings of $\sA$ irreps into the unconfined $\sU$ theory are
\begin{eqnarray*}
\Ir{e}{1} + \Ir{\ebar}{1}					&\rightarrow & \Ir{++}{++} \;,\;\;\; d_{\Ir{++}{++}} = 1\\
\Ir{e}{J_1} + \Ir{\ebar}{J_1}				&\rightarrow & \Ir{++}{+-} \;,\;\;\; d_{\Ir{++}{+-}} = 1\\
\Ir{e}{J_2} + \Ir{\ebar}{J_2}				&\rightarrow & \Ir{++}{-+} \;,\;\;\; d_{\Ir{++}{-+}} = 1\\
\Ir{e}{J_3} + \Ir{\ebar}{J_3}				&\rightarrow & \Ir{++}{--} \;,\;\;\; d_{\Ir{++}{--}} = 1\\
\Ir{X_1}{\Gamma^0}_1						&\rightarrow & \Ir{-+}{++} \;,\;\;\; d_{\Ir{-+}{++}} = 1\\
\Ir{X_1}{\Gamma^0}_2						&\rightarrow & \Ir{-+}{+-} \;,\;\;\; d_{\Ir{-+}{+-}} = 1\\
\Ir{X_1}{\Gamma^2}_1						&\rightarrow & \Ir{-+}{-+} \;,\;\;\; d_{\Ir{-+}{-+}} = 1\\
\Ir{X_1}{\Gamma^2}_2						&\rightarrow & \Ir{-+}{--} \;,\;\;\; d_{\Ir{-+}{--}} = 1\\
\Ir{X_2}{\Gamma^0}_1						&\rightarrow & \Ir{+-}{++} \;,\;\;\; d_{\Ir{+-}{++}} = 1\\
\Ir{X_2}{\Gamma^0}_2						&\rightarrow & \Ir{+-}{-+} \;,\;\;\; d_{\Ir{+-}{-+}} = 1\\
\Ir{X_2}{\Gamma^2}_1						&\rightarrow & \Ir{+-}{+-} \;,\;\;\; d_{\Ir{+-}{+-}} = 1\\
\Ir{X_2}{\Gamma^2}_2						&\rightarrow & \Ir{+-}{--} \;,\;\;\; d_{\Ir{+-}{--}} = 1\\
\Ir{X_3}{\Gamma^0}_1						&\rightarrow & \Ir{--}{++} \;,\;\;\; d_{\Ir{--}{++}} = 1\\
\Ir{X_3}{\Gamma^0}_2						&\rightarrow & \Ir{--}{--} \;,\;\;\; d_{\Ir{--}{--}} = 1\\
\Ir{X_3}{\Gamma^2}_1						&\rightarrow & \Ir{--}{+-} \;,\;\;\; d_{\Ir{--}{+-}} = 1\\
\Ir{X_3}{\Gamma^2}_2						&\rightarrow & \Ir{--}{-+} \;,\;\;\; d_{\Ir{--}{-+}} = 1
\end{eqnarray*}
which all have quantum dimension $d_u=1$ , while the confined fields are
\begin{eqnarray*}
\Ir{e}{\chi} + \Ir{\ebar}{\chi}				&\rightarrow & t_1		\;,\;\;\; d_{t_1} = 2\\
\Ir{X_1}{\Gamma^1} + \Ir{X_1}{\Gamma^3}		&\rightarrow & t_2		\;,\;\;\; d_{t_2} = 2\\
\Ir{X_2}{\Gamma^1} + \Ir{X_2}{\Gamma^3}		&\rightarrow & t_3		\;,\;\;\; d_{t_3} = 2\\
\Ir{X_3}{\Gamma^1} + \Ir{X_3}{\Gamma^3}		&\rightarrow & t_4		\;,\;\;\; d_{t_4} = 2\\
\end{eqnarray*}
and have $d_t=2$.

\paragraph{Breaking: \ir{X_1}{\Gamma^0} condensate.} There is an obvious symmetry in the fusion rules between the three \ir{X_i}{\Gamma^0} particle
sectors. We choose to study the case where the \ir{X_1}{\Gamma^0} condenses. This gives for the new vacuum
$\phi$ = \ir{e}{1} + \ir{\ebar}{1} + \ir{X_1}{\Gamma^0}, from which follows that $q=4$ in this case. We now read off the lifts of the $\sT$ fields on the right: 
\begin{eqnarray*}
\phi \times \Ir{e}{1} 			&=& \Ir{e}{1} + \Ir{\ebar}{1} + \Ir{X_1}{\Gamma^0}\\
\phi \times \Ir{e}{J_i} 		&=& \Ir{e}{J_i} + \Ir{\ebar}{J_i} + \delta_{1i} \Ir{X_1}{\Gamma^0} + \eta_{1i} \Ir{X_1}{\Gamma^2} \\
\phi \times \Ir{e}{\chi} 		&=& \Ir{e}{\chi} + \Ir{\ebar}{\chi} + \Ir{X_1}{\Gamma^1} + \Ir{X_1}{\Gamma^3}\\
\phi \times \Ir{\ebar}{1}	 	&=& \Ir{e}{1} + \Ir{\ebar}{1} + \Ir{X_1}{\Gamma^0} \\
\phi \times \Ir{\ebar}{J_i} 	&=& \Ir{e}{J_i} + \Ir{\ebar}{J_i} + \delta_{1i} \Ir{X_1}{\Gamma^0} + \eta_{1i} \Ir{X_1}{\Gamma^2} \\
\phi \times \Ir{\ebar}{\chi}	&=& \Ir{e}{\chi} + \Ir{\ebar}{\chi} + \Ir{X_1}{\Gamma^1} + \Ir{X_1}{\Gamma^3} \\
\phi \times \Ir{X_1}{\Gamma^0}	&=& \Ir{X_1}{\Gamma^0} + \Ir{X_1}{\Gamma^0} + \Ir{e}{1} + \Ir{\ebar}{1} + \Ir{e}{J_1} + \Ir{\ebar}{J_1} \\
\phi \times \Ir{X_1}{\Gamma^1}	&=& \Ir{X_1}{\Gamma^1} + \Ir{X_1}{\Gamma^3} + \Ir{e}{\chi} + \Ir{\ebar}{\chi} \\
\phi \times \Ir{X_1}{\Gamma^2}	&=& \Ir{X_1}{\Gamma^2} + \Ir{X_1}{\Gamma^2} + \Ir{e}{J_2} + \Ir{\ebar}{J_2} + \Ir{e}{J_3} + \Ir{\ebar}{J_3} \\
\phi \times \Ir{X_1}{\Gamma^3}	&=& \Ir{X_1}{\Gamma^1} + \Ir{X_1}{\Gamma^3} + \Ir{e}{\chi} + \Ir{\ebar}{\chi} \\
\phi \times \Ir{X_i}{\Gamma^0}	&=& \Ir{X_i}{\Gamma^0} + \Ir{X_i}{\Gamma^0} + \Ir{X_k}{\Gamma^0} + \Ir{X_k}{\Gamma^2} \;\;\;\; (i \neq k \neq 1)\\
\phi \times \Ir{X_i}{\Gamma^1}	&=& \Ir{X_i}{\Gamma^1} + \Ir{X_i}{\Gamma^3} + \Ir{X_k}{\Gamma^1} + \Ir{X_k}{\Gamma^3} \\
\phi \times \Ir{X_i}{\Gamma^2}	&=& \Ir{X_i}{\Gamma^2} + \Ir{X_i}{\Gamma^2} + \Ir{X_k}{\Gamma^0} + \Ir{X_k}{\Gamma^2}\\
\phi \times \Ir{X_i}{\Gamma^3}	&=& \Ir{X_i}{\Gamma^1} + \Ir{X_i}{\Gamma^3} + \Ir{X_k}{\Gamma^1} + \Ir{X_k}{\Gamma^3} 
\end{eqnarray*}
 We have used the symbol $\delta_{ij}$ which is $1$ when $i$ and $j$ are equal and is zero otherwise, and $\eta_{ij}$ which is $1$ when $i$ and $j$ are
not equal and is zero when $i$ and $j$ are. The $\sU$ theory is $D(\Z_2) \simeq \Z_2\otimes \Z_2$. This means $\sD^2_\sT = 16$ and $\sD^2_\sU = 4$. The lifts of the unconfined   fields are:
\begin{eqnarray*}
\Ir{e}{1} + \Ir{\ebar}{1} + \Ir{X_1}{\Gamma^0}_1			&\rightarrow & \Ir{+}{+} \;,\;\;\; d_{\Ir{+}{+}} = 1\\
\Ir{e}{J_1} + \Ir{\ebar}{J_1} + \Ir{X_1}{\Gamma^0}_2		&\rightarrow & \Ir{+}{-} \;,\;\;\; d_{\Ir{+}{-}} = 1\\
\Ir{X_2}{\Gamma^0}_1 + \Ir{X_3}{\Gamma^0}_1				&\rightarrow & \Ir{-}{+} \;,\;\;\; d_{\Ir{-}{+}} = 1\\
\Ir{X_2}{\Gamma^2}_1 + \Ir{X_3}{\Gamma^2}_1				&\rightarrow & \Ir{-}{-} \;,\;\;\; d_{\Ir{-}{-}} = 1
\end{eqnarray*}
and of the confined fields:
\begin{eqnarray*}
\Ir{e}{J_2} + \Ir{\ebar}{J_2} + \Ir{X_1}{\Gamma^2}_1				&\rightarrow & t_1		\;,\;\;\; d_{t_1} = 1\\
\Ir{e}{J_3} + \Ir{\ebar}{J_3} + \Ir{X_1}{\Gamma^2}_2				&\rightarrow & t_2		\;,\;\;\; d_{t_2} = 1\\
\Ir{e}{\chi} + \Ir{\ebar}{\chi} + \Ir{X_1}{\Gamma^1} + \Ir{X_1}{\Gamma^3} &\rightarrow & t_3		\;,\;\;\; d_{t_3} = 2\\
\Ir{X_2}{\Gamma^0}_2 + \Ir{X_3}{\Gamma^2}_2						&\rightarrow & t_4		\;,\;\;\; d_{t_4} = 1\\
\Ir{X_3}{\Gamma^0}_2 + \Ir{X_2}{\Gamma^2}_2						&\rightarrow & t_5		\;,\;\;\; d_{t_5} = 1\\
\Ir{X_2}{\Gamma^1} + \Ir{X_2}{\Gamma^3} + \Ir{X_3}{\Gamma^1} + \Ir{X_3}{\Gamma^3} &\rightarrow & t_6		\;,\;\;\; d_{t_6} = 2
\end{eqnarray*}

\subsection{\label{sec.measurements}Measurements by  lattice Monte Carlo simulations}
The five couplings $\{\beta_A\}$ for conjugacy class $A$ that appear in the action of the $D(\dbar)$ theory
\begin{equation}\label{d2act}
S_{p}= \sum_p - \left\{ \beta_e \delta_e(U_p) + \beta_{\ebar} \delta_{\ebar}(U_p)+\beta_{X_1} \delta_{X_1}(U_p)+\beta_{X_2} \delta_{X_2}(U_p)+\beta_{X_3} \delta_{X_3}(U_p)\right\},
\end{equation}
are inversely proportional to the masses of the fluxes $A$. For example if we put all couplings to zero except for $\beta_e$, which we make large (at least as large as $2.0$ as we will see shortly), the trivial vacuum is realized: this is the configuration where for all plaquettes $U_p=e$. Deviations from this configuration occur because of quantum fluctuations, but since all excitations are gapped they will be exponentially suppressed.
The gap in this vacuum is easily calculated to be of the order of $4 \beta_e$, since the smallest excitation above the configuration in which all plaquettes are $e$ is one in which one link has a value $h \, \neq \, e$. This excites four plaquettes and changes the action (\ref{d2act}) by a value of $4 \beta_e$. 

\paragraph{Monte Carlo considerations.} For the other, nontrivial phases in this theory, the dominant configurations contributing to the path integral are not so readily identified. To gain insight into what configurations contribute we use a Monte Carlo simulation, in particular a modified {\em heat bath} algorithm. Bluntly applying this algorithm to our problem leads to various complications, therefore we briefly point out the method, the complications and how we have resolved them. 

The procedure starts with some initial configuration of link variables $\{U\}_{1}$. We then update all links in lexicographic order, a process called a sweep, and arrive at a new configuration $\{U\}_{2}$. The updating process for each link proceeds as follows. Consider the link $U_{ij}$. We identify which plaquettes contain this link: in three dimensions, there are four such plaquettes. Now we calculate, for each element $g \, \in \, H$, what the sum of the plaquette actions for each of these four plaquettes would be if $U_{ij}$ were to have the value $g$. This gives a set of numbers
\begin{equation*}
\{ S_{g_1} , S_{g_2}, \cdots , S_{g_{|H|}} \},
\end{equation*}
where $S_{g_k}$ is the sum of the four plaquette actions with $U_{ij}$ equal to $g_k$. We now calculate a localized partition sum $Z_{U_{ij}}$:
\begin{equation*}
Z_{U_{ij}} = \sum_{g \in H } e^{-S_g},
\end{equation*}
which can be used to calculate a set of probabilities $\{ p(g) \} _{g \in H}$ for each group element $g$
\begin{equation*}
\label{pg}
p(g) = \frac{e^{-S_g}}{Z_{U_{ij}}}.
\end{equation*}
After a given number of sweeps $n_0$, the Monte Carlo algorithm arrives at the minimum of the action and the path integral expectation value of the operator $O$
\begin{equation} \label{pathint}
\left< O \right> = \frac{\int DU \, O[U]\, e^{-S[U]}}{\int DU \, e^{-S[U]}}
\end{equation}
is given by taking the average of $O[\{U\}_n]$, the value of $O$ at gauge field configuration $\{U\}_n$:
\begin{equation}\label{MCavg}
\left< O \right>_{\textrm{MC estimate}} = {1 \over m} \sum_{n=n_0+1}^{n_0+m} O[\{U\}_n].
\end{equation}

\begin{figure}[h!]
\centering
\includegraphics[width=0.4\textwidth]{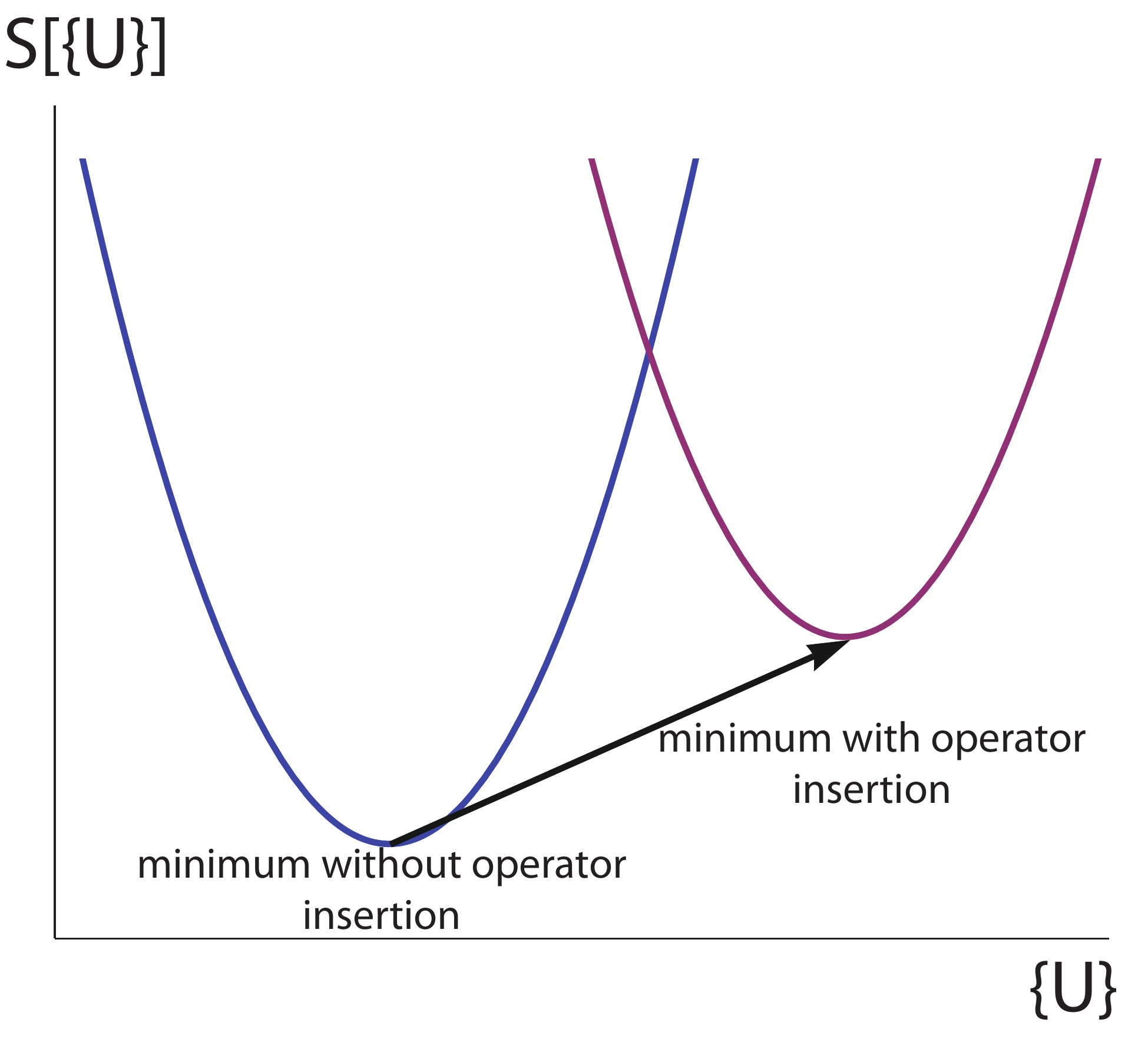}
\caption{\label{actionshift}Schematic drawing of the action as a functional of the gauge field configuration $\{U\}$. The insertion of non-commuting fluxes shifts the minimum of the action to a different location in the configuration space (due to the non-local nature of the excitations) and to a different value (due to the presence of a string).}
\end{figure}

However, for our purposes this scheme is troublesome for two reasons: it is tacitly assumed that the presence of the operator $O$ in (\ref{pathint}) does not change the value of the minimum of the action $S$ and furthermore the loops of magnetic flux are very non-local objects and therefore highly unlikely to appear when using a local updating algorithm. This is illustrated in Figure \ref{actionshift}. The shift upward of the functional $S[{U}]$ is due to the presence of a magnetic flux string and the shift to the left is due to the non-locality of the magnetic excitations. The latter shift also occurs when a single loop of flux is inserted.

The minimum of the action in the calculation of an $S$-matrix element (\ref{meassmat}) is altered by the insertions of the loop operators: the configuration for two non-commuting fluxes carries a string (see the discussion around Figure \ref{fig.smatrixconfig}) that is massive and thus costs a finite amount of action. There is no way to get rid of this string and therefore the minimum value of the action in the presence of the two loops is shifted. We therefore have to amend the standard MC algorithm. Defining
\begin{eqnarray}
S\,&=&\,S_{\mathrm{min}} + \delta S \quad  \textrm{without operator insertion,} \nonumber \\
\tilde{S}\,&=&\tilde{S}_{\mathrm{min}} + \delta \tilde{S}  \quad \, \textrm{with operator insertion,} \nonumber
\end{eqnarray}
and noticing that around the minimum the actions behave identically, implying that $\delta S$ and $\delta \tilde{S}$ are the same functions, expression (\ref{pathint}) becomes
\begin{equation} \label{pathintshift}
\left< O \right> = \frac{\int DU \, e^{-(\tilde{S}_{\mathrm{min}} - S_{\mathrm{min}})} O[U]\, e^{-\delta S[U]}}{\int DU \, e^{-\delta S[U]}}.
\end{equation}
This leads to a modified Monte Carlo average
\begin{equation}\label{MCavg}
\left< O \right>_{\textrm{MC estimate}} = {1 \over m} e^{-(\tilde{S}_{\mathrm{min}} - S_{\mathrm{min}})} \sum_{n=n_0+1}^{n_0+m} O[\{U\}_n].
\end{equation}

We now describe two approaches to the second problem in our MC measurements: the low probability that the local updating algorithm will converge to a gauge field configuration containing a (set of) magnetic flux loop(s). We will assume a single loop of pure magnetic flux is inserted, as nothing substantial will change in the case of multiple loops or the addition of dyonic charge. 

The first approach is based on the observation, illustrated in Figure \ref{fig.gforest}, that we know  the gauge field configuration (up to gauge transformations) that extremizes the action in the trivial vacuum with the insertion of a loop of magnetic flux: the $h$-forest. We can therefore use this configuration as an {\em ansatz} in the MC algorithm. We start with a ``cold lattice", all links $U_{ij}=e$, except for the $h$-forest, for these links we set $U_{ij}=h$. This is an extremum of the action for the action if we set all $\beta_{A\neq e}=0$ and $\beta_e \gg 1$. To perform a measurement at some other value of the coupling constants, we can slowly change the coupling constants towards the desired values, performing a few MC updates after each step.
\begin{figure}[t]
\centering
\includegraphics[width=0.7\textwidth]{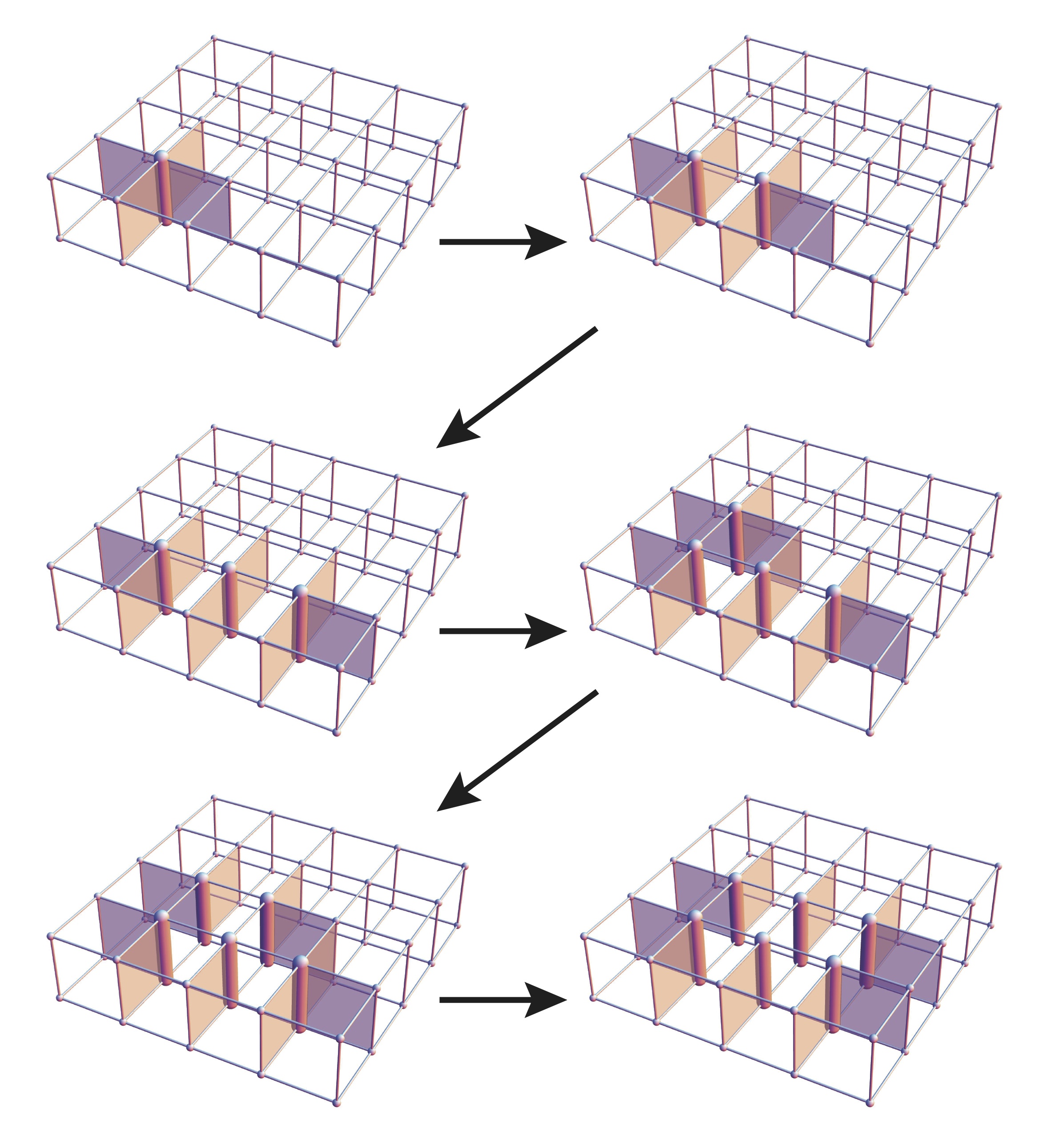}
\caption{\label{fig.growloop}Growing a flux loop in multiple steps. The shaded plaquettes have a twisted action, and the fat links show the convergence towards a $h$-forest state.}
\end{figure}
The second approach is  a more physical one. We initialize the lattice directly at the desired point in coupling constant space. The trick is then not to insert the loop all at once, but to slowly grow it, as illustrated in Figure \ref{fig.growloop}.
We start by twisting the action for four plaquettes around one link, as shown by the shaded plaquettes in the top left of Figure \ref{fig.growloop}. After this, a number of MC updates are performed. Then the set of plaquettes that have a twisted action is changed as in the top right corner of the Figure. Again a number of MC updates is performed and so on. We have checked that in the trivial vacuum one obtains the $h$-forest configuration using this procedure.
Both of  methods to insert flux loops have been used by us and we have verified that they lead to completely equivalent results.
\subsection{Results}
In this subsection we present the results of our Monte Carlo simulations. The first quantity we measured was the free energy as a means to  map out a suitable subspace of the parameter space. It gives us an indication of the validity of our naive intuition about  where nontrivial condensates should occur. 

Once we have found some region where symmetry breaking occurs we measure the open string expectation values to determine the respective condensates. After that we measure the unbroken and broken $S$-matrix elements.  Using the straightforward algorithm involving the auxiliary symmetries of our loop operators discussed in section \ref{sec:auxiliary},  allows us to find the branching matrix $n^{a_1}_u$ as well as the $S$-matrix of the effective $\sU$ theory in the broken phase.

\paragraph{Mapping out the phase diagram.}The space of coupling constants in our theory is five-dimensional but it is not our goal to analyse it completely.  We have restricted our search to some representative regions where nontrivial condensates do indeed occur. To study the location of the corresponding phase transitions we measured the free energy $F$, which we define as the expectation value of the plaquette action, averaged over the spacetime lattice.
\begin{figure}[h]
\centering
\includegraphics[width=0.45\textwidth]{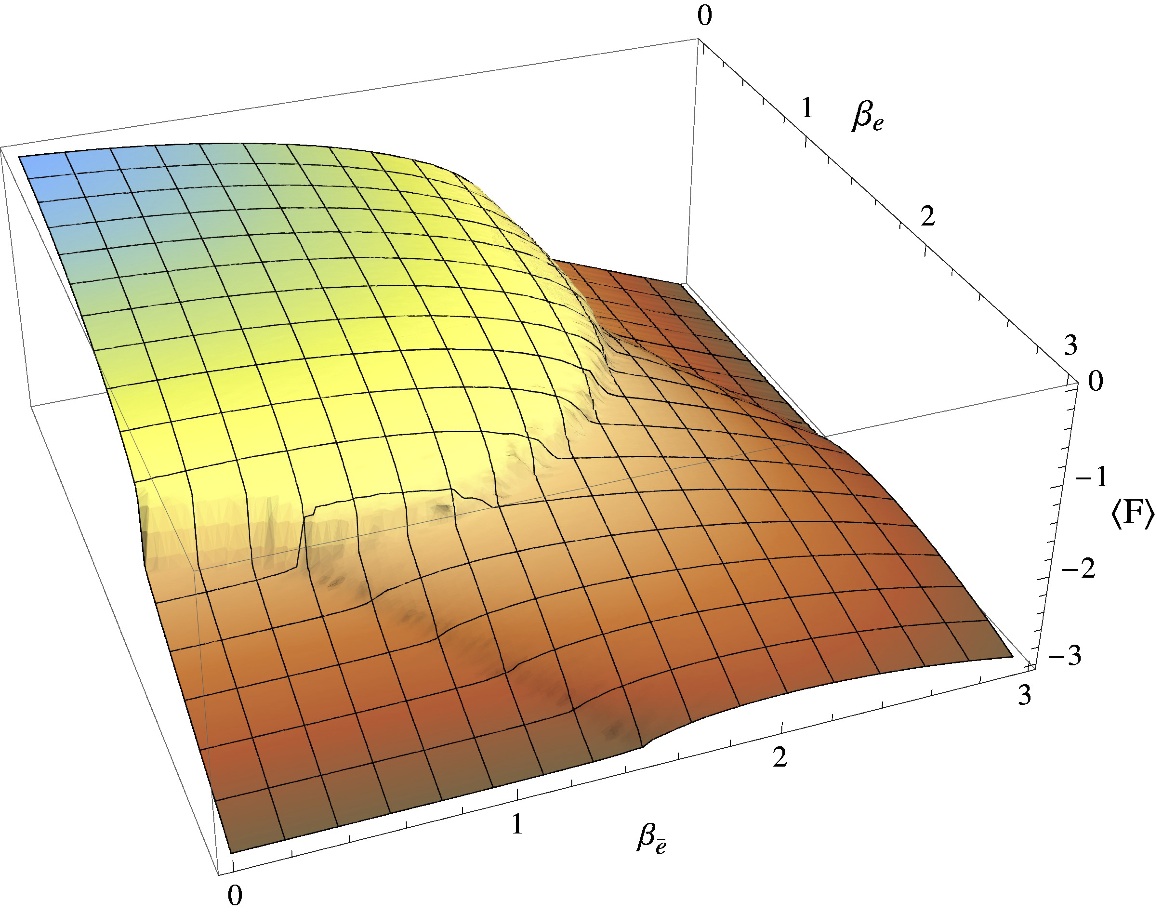}\vspace*{5mm}
\includegraphics[width=0.45\textwidth]{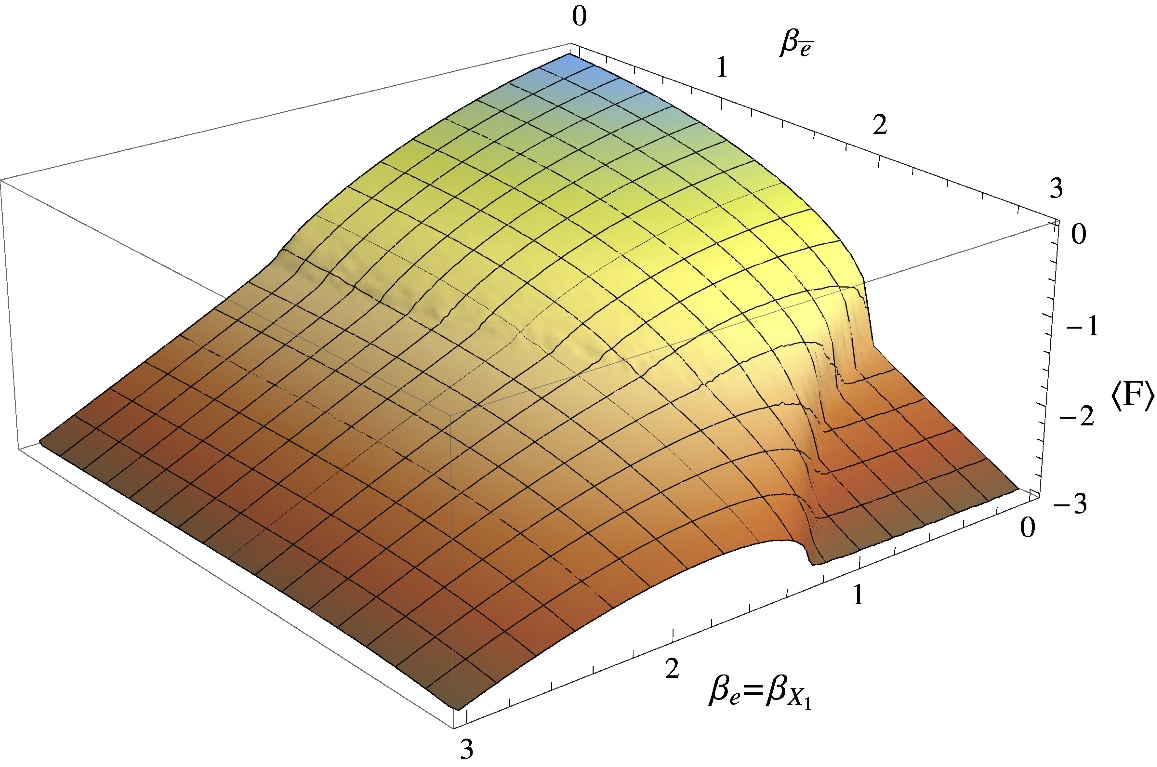}
\caption{\label{free}Plots of the free energy $F$ for two-dimensional planes through the origin of the parameter space of the lattice model. In the left figure we have the $(\beta_e,\beta_{\bar{e}})$ plane and in the right figure we have the $(\beta_e=\beta_{X_1}, \beta_{\bar{e}})$ plane. See text for further comments. }
\end{figure}
The left plot of Figure \ref{free} shows  $F$ as a function of $(\beta_e$ and $\beta_{\bar{e}})$ and all other couplings equal zero. For small values of all the couplings appearing in the action (\ref{d2act}), we are in the completely  confining phase of the gauge theory, where all the open string operators of magnetic flux have a non-zero expectation value, and all loop operators carrying electric charge are confined. This corresponds to the plateau in the graph where $F$ is maximal and tends to zero.

\begin{figure}[h]
\centering
\subfigure[$\langle \delta_e(U_p)\rangle$]{
\includegraphics[scale=0.46]{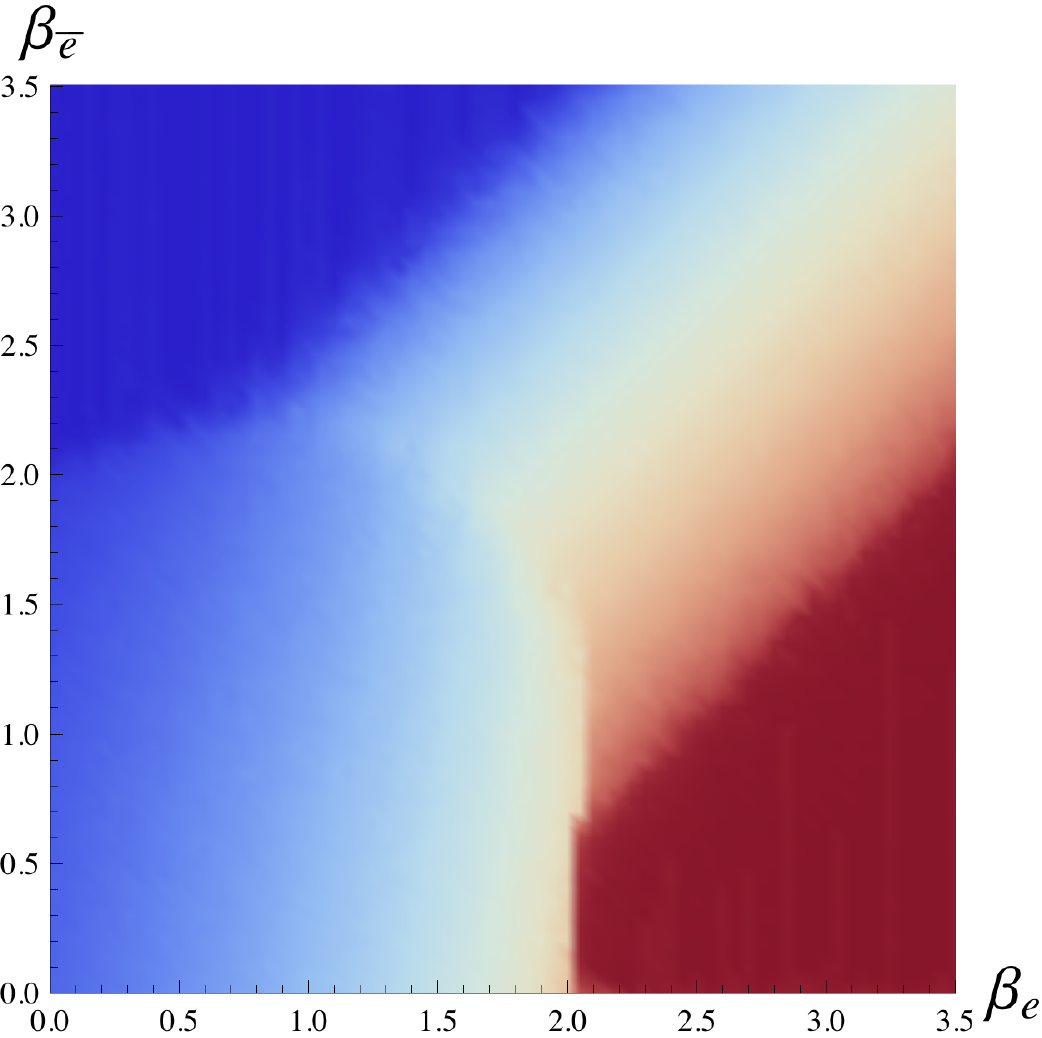}
}
\subfigure[$\langle \delta_{\ebar}(U_p)\rangle$]{
\includegraphics[scale=0.46]{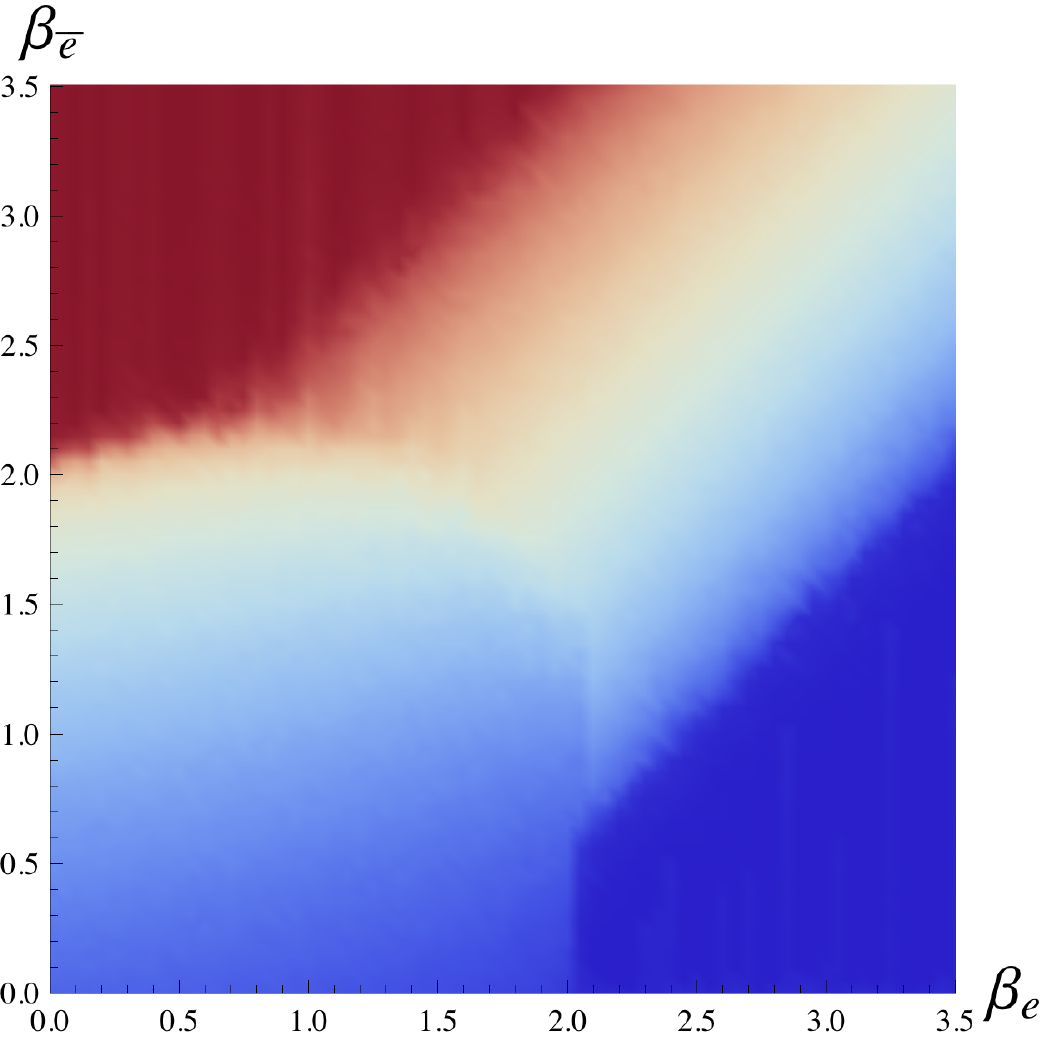}
}
\subfigure[$\langle \delta_{X_1}(U_p)\rangle$]{
\includegraphics[scale=0.46]{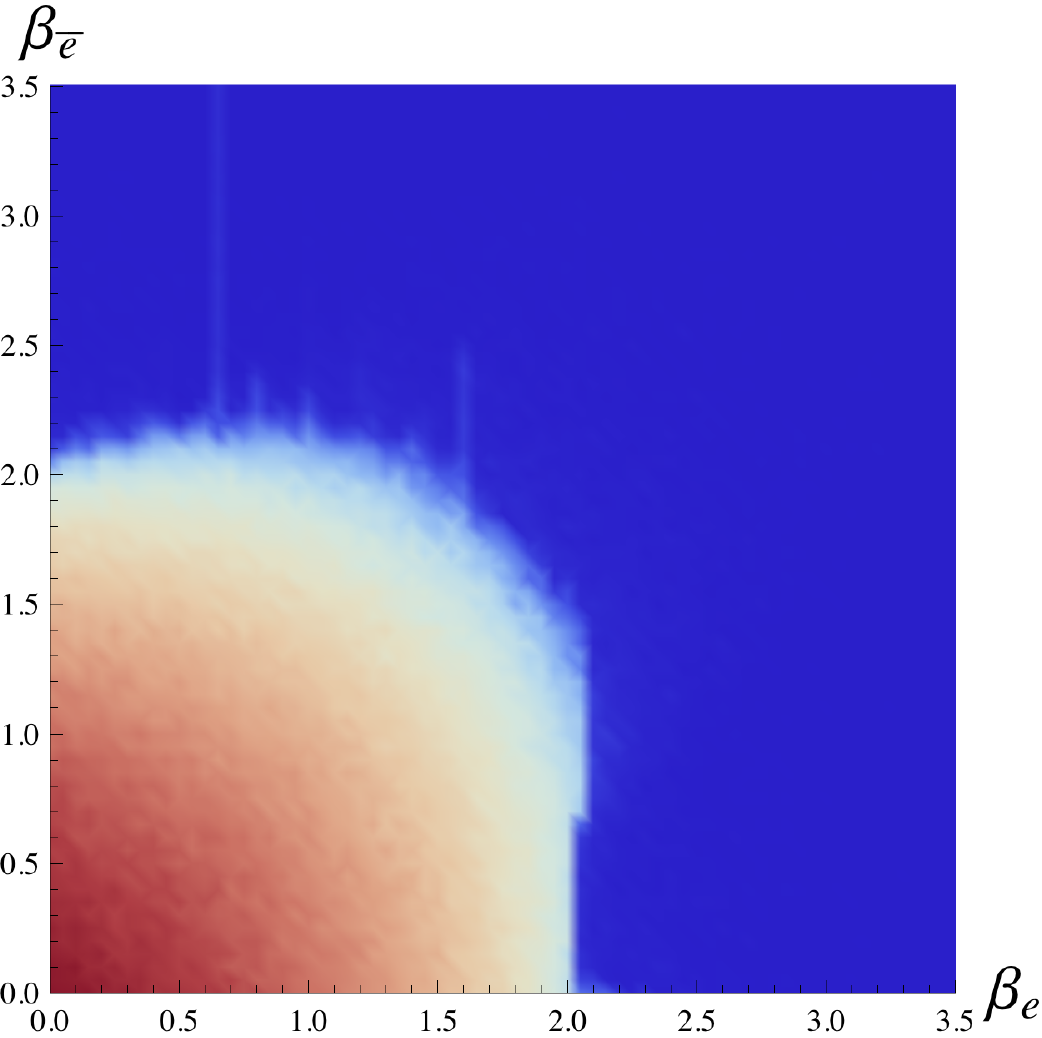}
}\\
\subfigure[$\langle \delta_{X_2}(U_p)\rangle$]{
\includegraphics[scale=0.46]{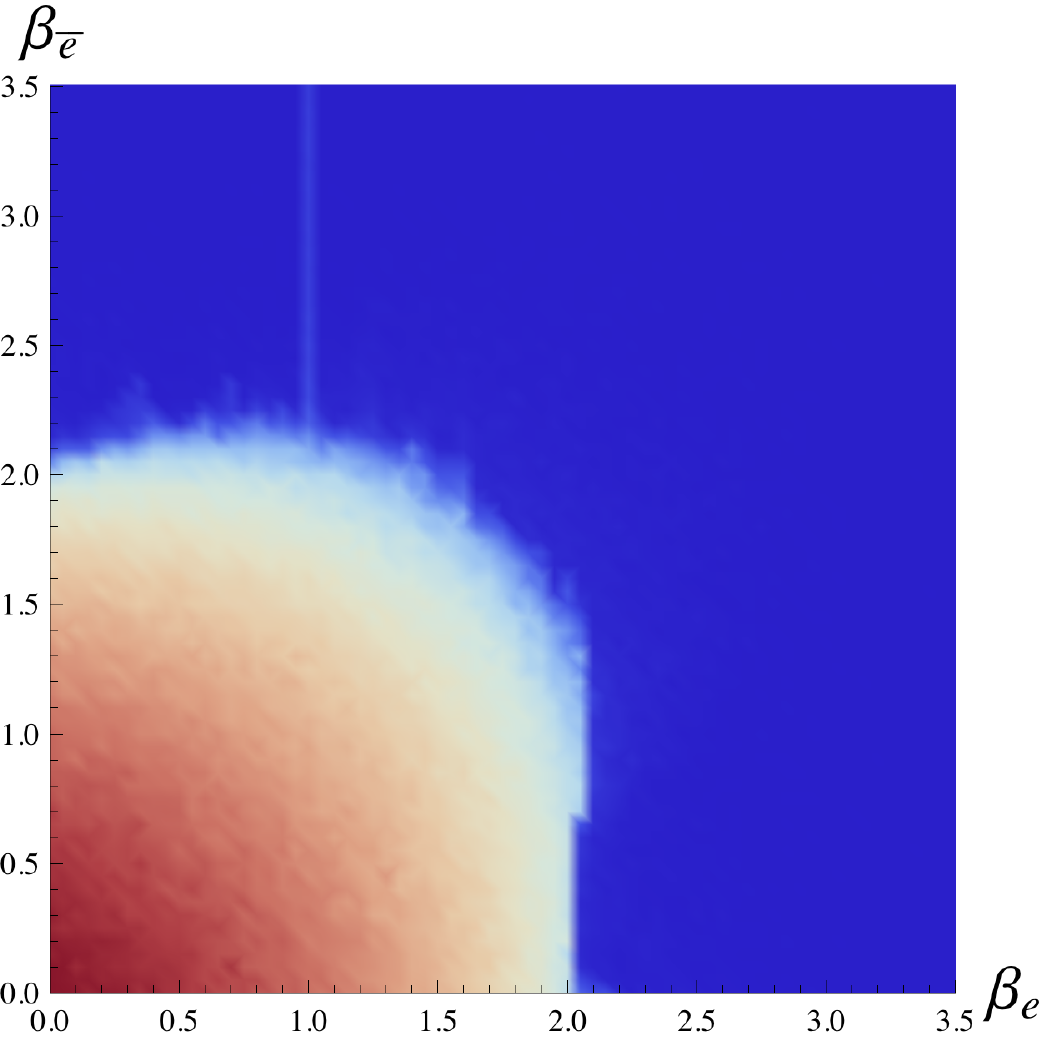}
}
\subfigure[$\langle \delta_{X_3}(U_p)\rangle$]{
\includegraphics[scale=0.46]{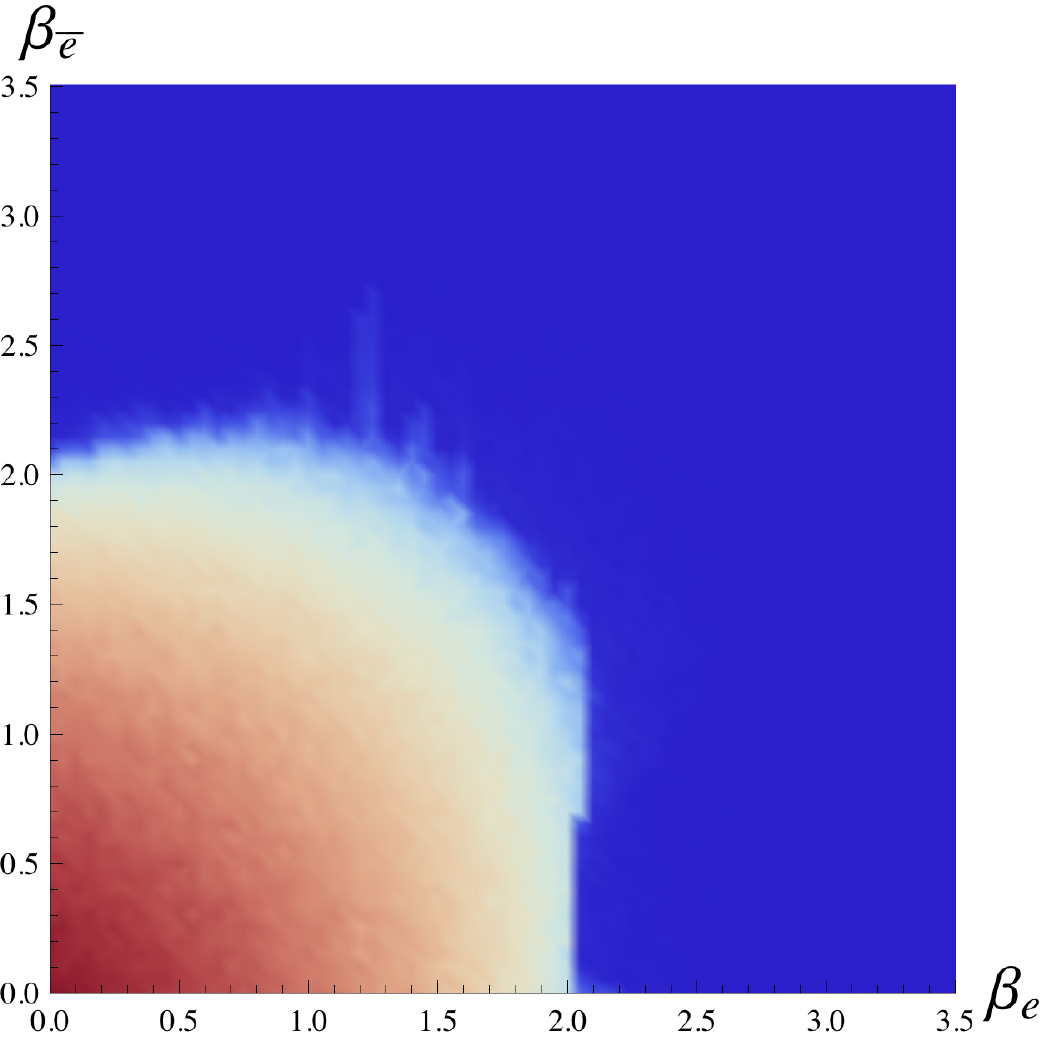}
}
\subfigure[Schematic of diagram]{
\includegraphics[scale=0.52]{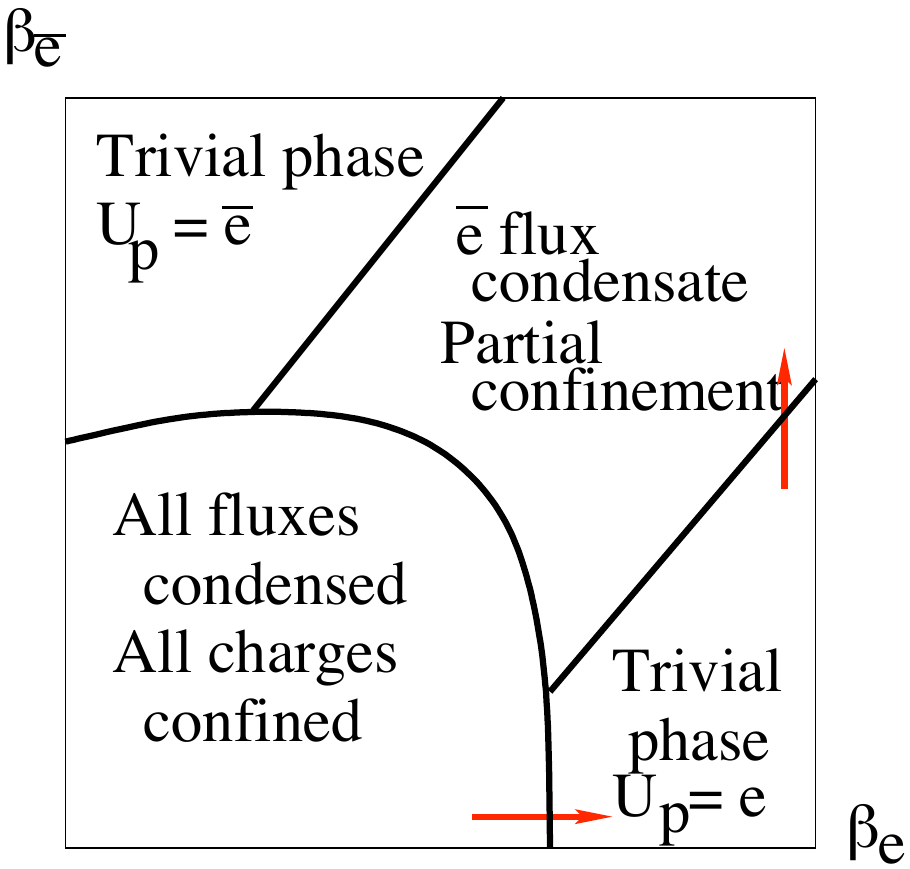}
}
\caption{Space-time avaraged expectation value of $\delta_A(U_p)$ for each conjugacy class $A$. Shown is a $(\beta_e, \beta_{\ebar})$-plane in coupling constant space where the other three couplings are zero. The color coding is such that red is the highest and blue the lowest value in each figure. In (f) we have identified the  meaning of the various regions and the transition lines, where the red arrows indicate the trajectories used to determine whether  the transitions   are first or second order (see figure \ref{fig:pt1}f and \ref{fig:pt2}f). \label{fig:phase1}}
\end{figure}
The regions where the magnetic flux \ir{\ebar}{1} and \ir{X_1}{\Gamma^0} have condensed can be anticipated on theoretical grounds by realizing that the coupling $\beta_A$ is inversely proportional to the mass of flux $A$. In fact, when we look at the subgroup $K_A$ generated by the elements in conjugacy class $A$, in particular
\begin{eqnarray*}
K_{\ebar} &=& \{ \one, -\one \} \\
K_{X_1}  &=& \{ \one, -\one, i\sigma_1, -i\sigma_1 \},
\end{eqnarray*}
and set the couplings for the conjugacy classes containing the elements in $K_A$ equal to one another, there is an extra gauge invariance $U_p \rightarrow k\, U_p$ for an element $k \in A$ in the plaquette action (\ref{d2act}). In particular
\[
S = \beta \left( \delta_e(U_p) + \delta_{\ebar} (U_p) \right)
\]
is invariant with respect to $U_p \rightarrow -\one U_p$ and
\[
S = \beta \left( \delta_e(U_p) + \delta_{\ebar} (U_p) +\delta_{X_1} (U_p) \right)
\]
is invariant with respect to $U_p \rightarrow k\, U_p$, where $k \in \{ -\one, i\sigma_1, -i\sigma_1 \}$.

These left multiplications are exactly the kind appearing in the definition of the (loop) order parameters (\ref{eqn.dyonop}). Therefore one can establish, even without reverting to MC measurements, that the above actions, for large values of $\beta$, produce the desired flux condensates.

\begin{figure}[h]
\centering
\subfigure[$\langle \delta_e(U_p)\rangle$]{
\includegraphics[scale=0.46]{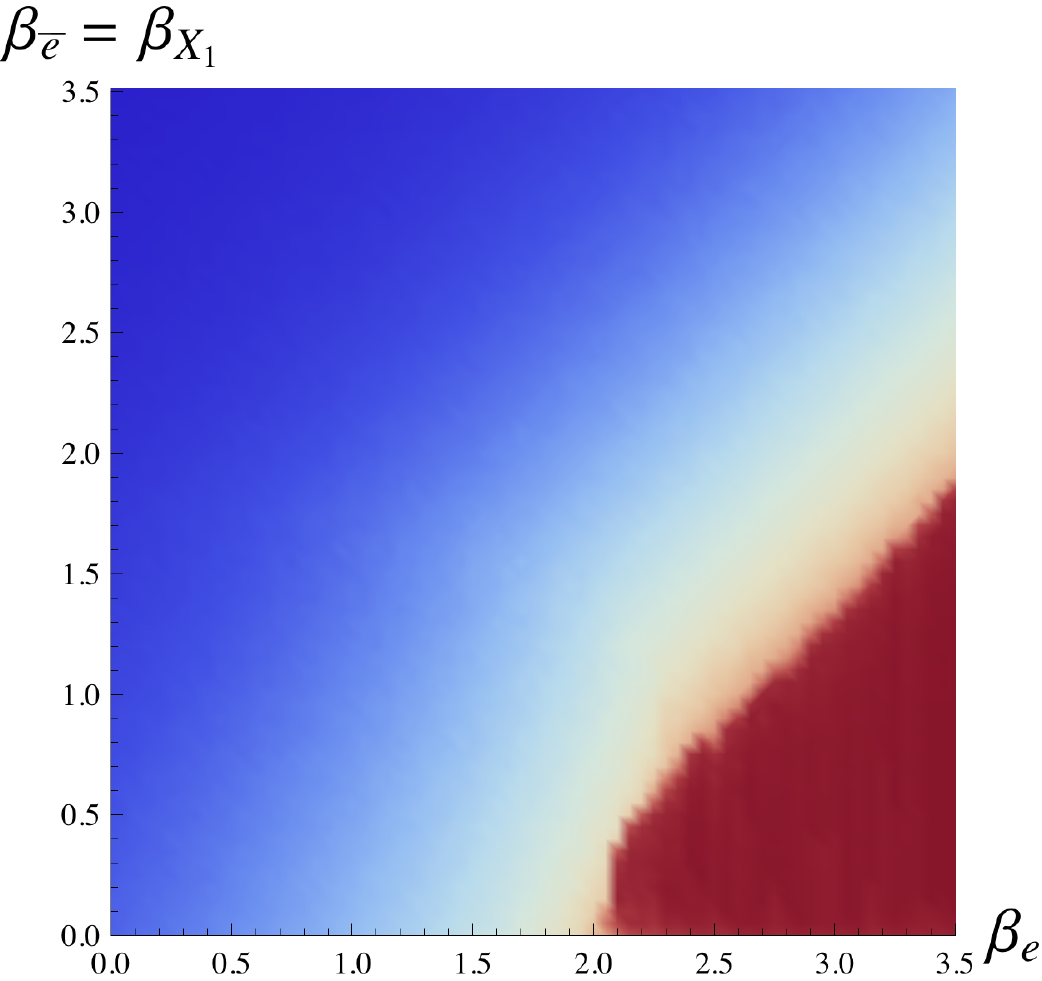}
}
\subfigure[$\langle \delta_{\ebar}(U_p)\rangle$]{
\includegraphics[scale=0.46]{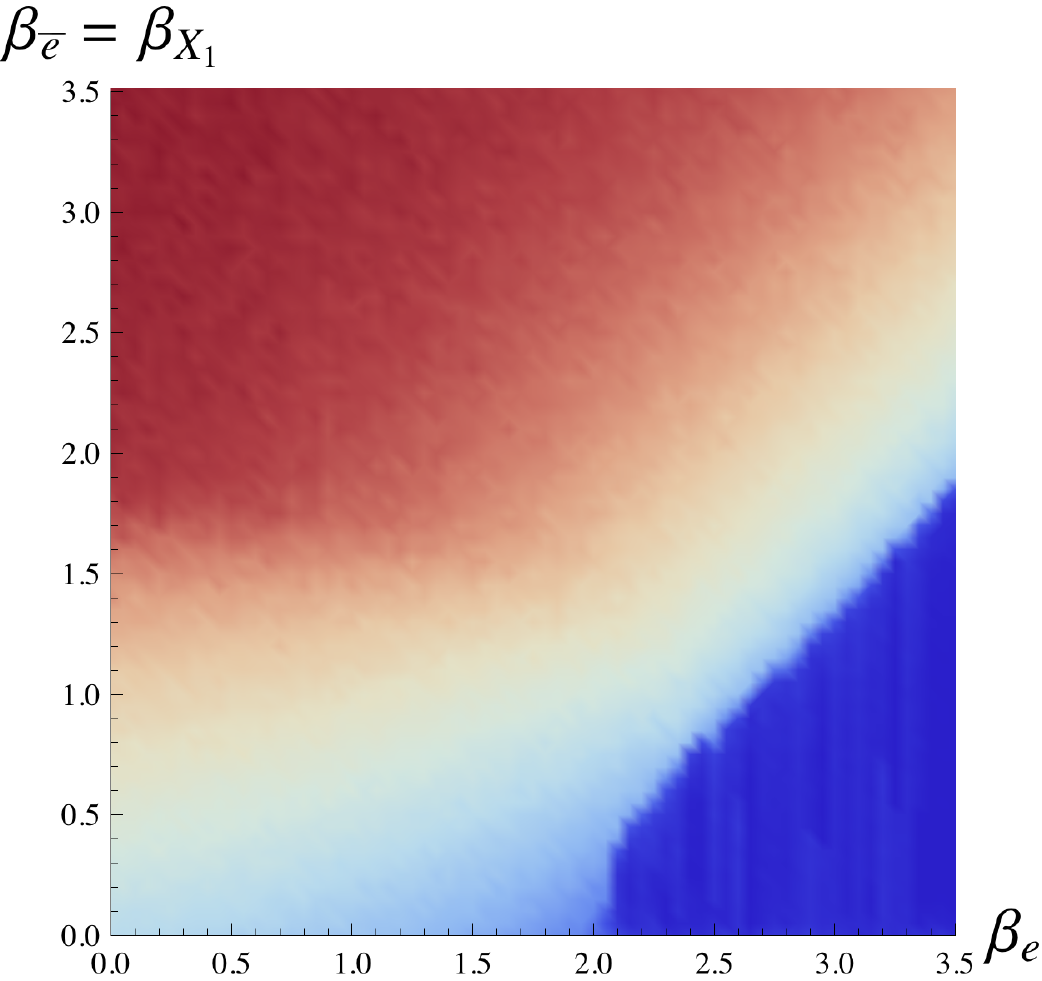}
}
\subfigure[$\langle \delta_{X_1}(U_p)\rangle$]{
\includegraphics[scale=0.46]{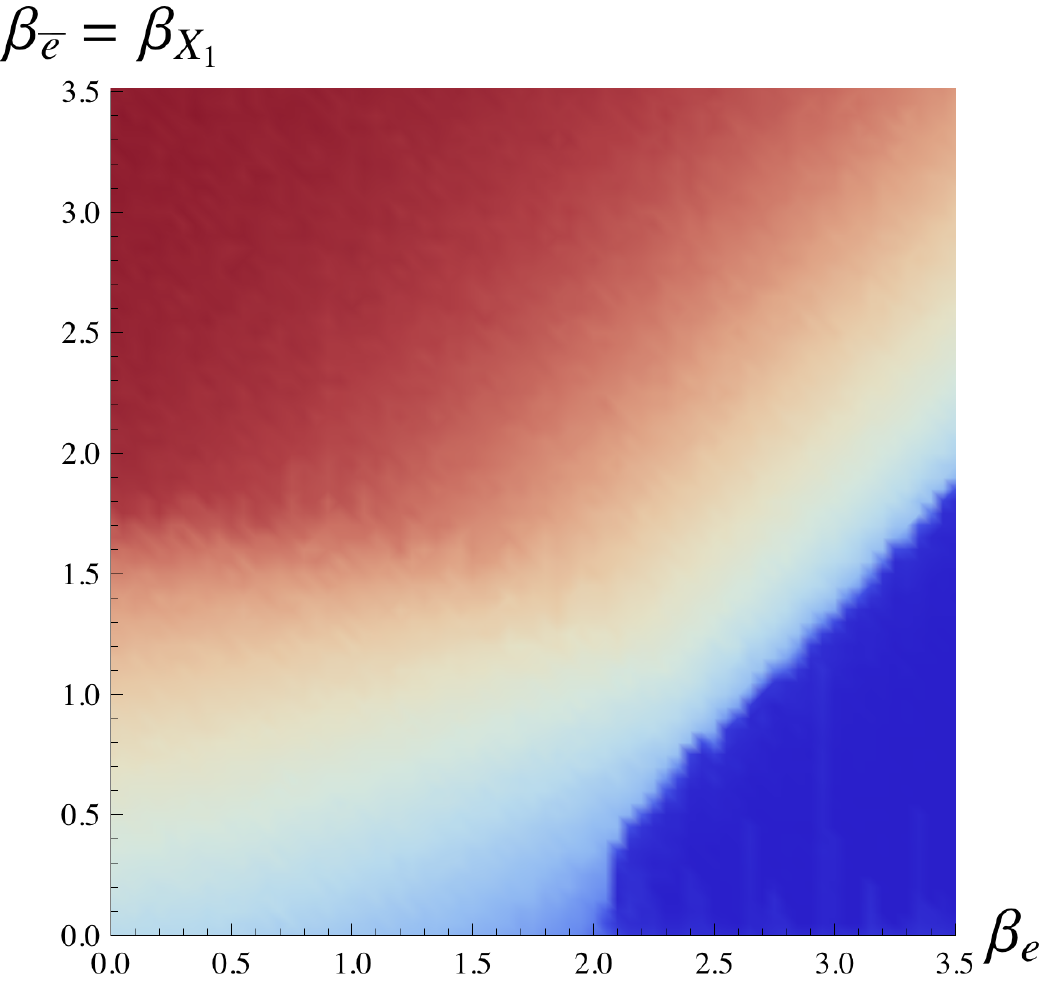}
}\\
\subfigure[$\langle \delta_{X_2}(U_p)\rangle$]{
\includegraphics[scale=0.46]{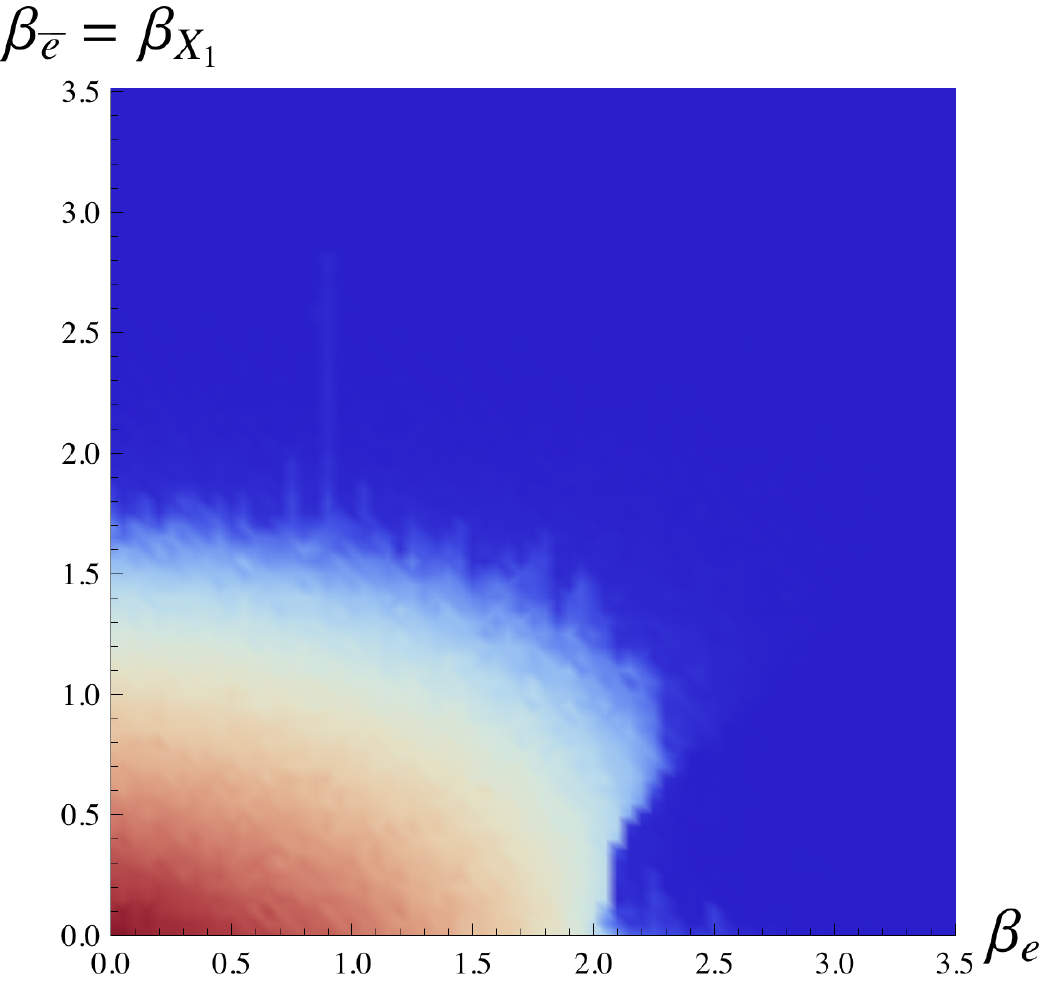}
}
\subfigure[$\langle \delta_{X_3}(U_p)\rangle$]{
\includegraphics[scale=0.46]{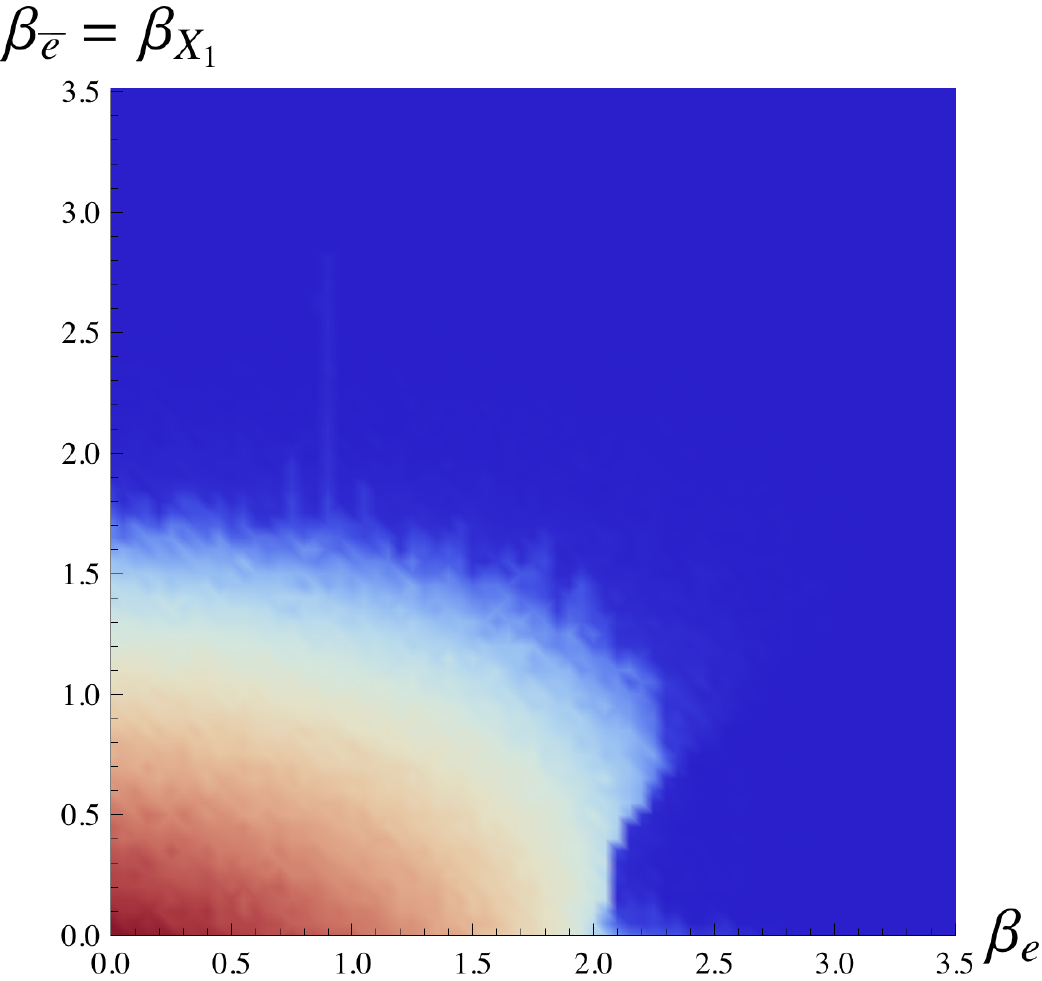}
}
\subfigure[Schematic of diagram]{
\includegraphics[scale=0.51]{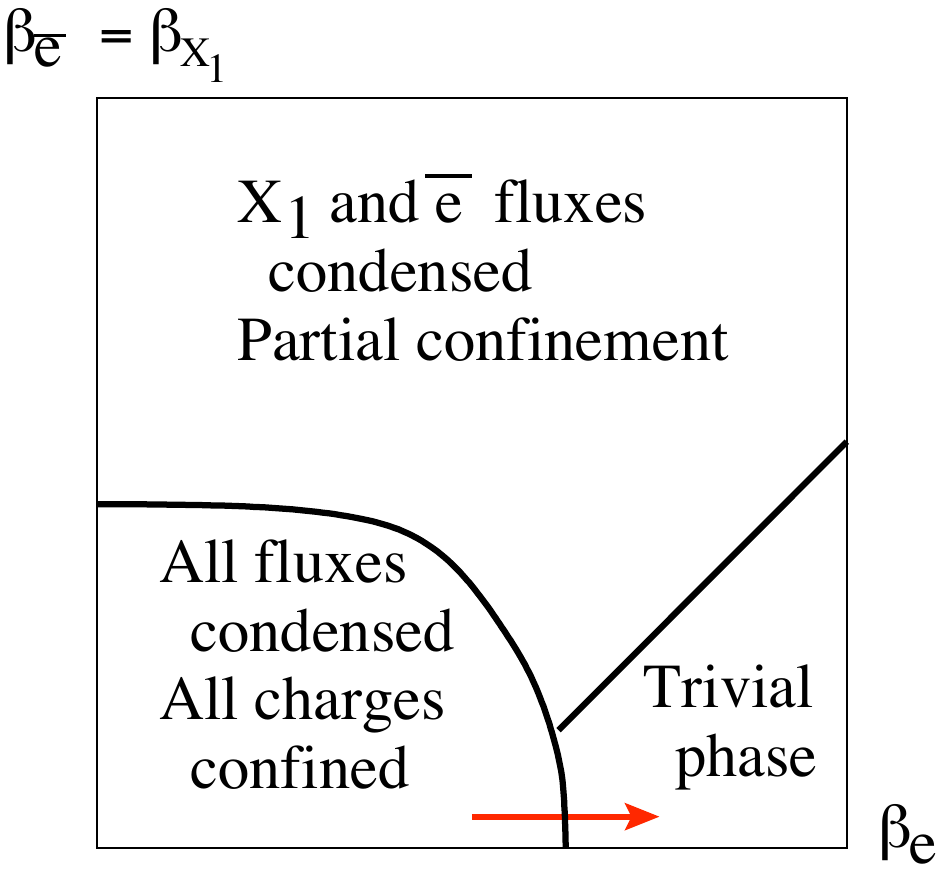}
}
\caption{Space-time avaraged expectation value of $\delta_A(U_p)$ for each conjugacy class $A$. Shown is a $(\beta_e, \beta_{\ebar}=\beta_{X_1})$-plane in coupling constant space where the other two couplings are zero. The color coding is such that red is the highest and blue the lowest value in each figure. In (f) we have identified the  meaning of the various regions and the transition lines, where the red arrow indicates the trajectory used to determine whether  the transition is first or second order (see figure \ref{fig:pt1}f). \label{fig:phase2}}
\end{figure}

One may verify this reasoning in the Figures \ref{fig:phase1} and \ref{fig:phase2} where we have probed the phase diagram more in detail by measuring the spacetime averaged expectation value of  $\delta_A(U_p)$ for all conjugacy classes $A$ as a function of the relevant coupling parameters $\beta$. The red color  indicates high values for the expectation value and we see that for all coupling parameters near zero all fluxes are condensed and thus all charges will be confined. This is what traditionally is called the ``strong coupling phase ($g\sim 1/\beta \gg 1$). Looking at the colorings for the various operators one readily identifies the various phases as indicated in the schematics of the subfigures (f). For example the symmetry with respect to the diagonal of the Figures \ref{fig:phase1}a and \ref{fig:phase1}b, shows that there are ``Ising" like ordered phases, one with all plaquette values $U_p=e$ and the other with all $U_p=\bar{e}$. The in-between region is the region with the $\bar{e}$ flux condensate. Note that if the $\bar{e}$ flux would be the only one that phase would continue all the way to the origin, end we would exactly end up with the $Z_2$ pure gauge theopry phse diagram.

In the region with $\beta_e$ larger than approximately $2.0$ and all other  couplings near zero, the trivial vacuum is realized. All string operators with nontrivial magnetic flux have zero expectation value there. 

\begin{figure}[t]
\begingroup
  \makeatletter
  \providecommand\color[2][]{%
    \GenericError{(gnuplot) \space\space\space\@spaces}{%
      Package color not loaded in conjunction with
      terminal option `colourtext'%
    }{See the gnuplot documentation for explanation.%
    }{Either use 'blacktext' in gnuplot or load the package
      color.sty in LaTeX.}%
    \renewcommand\color[2][]{}%
  }%
  \providecommand\includegraphics[2][]{%
    \GenericError{(gnuplot) \space\space\space\@spaces}{%
      Package graphicx or graphics not loaded%
    }{See the gnuplot documentation for explanation.%
    }{The gnuplot epslatex terminal needs graphicx.sty or graphics.sty.}%
    \renewcommand\includegraphics[2][]{}%
  }%
  \providecommand\rotatebox[2]{#2}%
  \@ifundefined{ifGPcolor}{%
    \newif\ifGPcolor
    \GPcolorfalse
  }{}%
  \@ifundefined{ifGPblacktext}{%
    \newif\ifGPblacktext
    \GPblacktexttrue
  }{}%
  \let\gplgaddtomacro\g@addto@macro
  \gdef\gplbacktext{}%
  \gdef\gplfronttext{}%
  \makeatother
  \ifGPblacktext
    \def\colorrgb#1{}%
    \def\colorgray#1{}%
  \else
    \ifGPcolor
      \def\colorrgb#1{\color[rgb]{#1}}%
      \def\colorgray#1{\color[gray]{#1}}%
      \expandafter\def\csname LTw\endcsname{\color{white}}%
      \expandafter\def\csname LTb\endcsname{\color{black}}%
      \expandafter\def\csname LTa\endcsname{\color{black}}%
      \expandafter\def\csname LT0\endcsname{\color[rgb]{1,0,0}}%
      \expandafter\def\csname LT1\endcsname{\color[rgb]{0,1,0}}%
      \expandafter\def\csname LT2\endcsname{\color[rgb]{0,0,1}}%
      \expandafter\def\csname LT3\endcsname{\color[rgb]{1,0,1}}%
      \expandafter\def\csname LT4\endcsname{\color[rgb]{0,1,1}}%
      \expandafter\def\csname LT5\endcsname{\color[rgb]{1,1,0}}%
      \expandafter\def\csname LT6\endcsname{\color[rgb]{0,0,0}}%
      \expandafter\def\csname LT7\endcsname{\color[rgb]{1,0.3,0}}%
      \expandafter\def\csname LT8\endcsname{\color[rgb]{0.5,0.5,0.5}}%
    \else
      \def\colorrgb#1{\color{black}}%
      \def\colorgray#1{\color[gray]{#1}}%
      \expandafter\def\csname LTw\endcsname{\color{white}}%
      \expandafter\def\csname LTb\endcsname{\color{black}}%
      \expandafter\def\csname LTa\endcsname{\color{black}}%
      \expandafter\def\csname LT0\endcsname{\color{black}}%
      \expandafter\def\csname LT1\endcsname{\color{black}}%
      \expandafter\def\csname LT2\endcsname{\color{black}}%
      \expandafter\def\csname LT3\endcsname{\color{black}}%
      \expandafter\def\csname LT4\endcsname{\color{black}}%
      \expandafter\def\csname LT5\endcsname{\color{black}}%
      \expandafter\def\csname LT6\endcsname{\color{black}}%
      \expandafter\def\csname LT7\endcsname{\color{black}}%
      \expandafter\def\csname LT8\endcsname{\color{black}}%
    \fi
  \fi
  \setlength{\unitlength}{0.0500bp}%
  \begin{picture}(7200.00,5040.00)%
    \gplgaddtomacro\gplbacktext{%
      \csname LTb\endcsname%
      \put(1254,704){\makebox(0,0)[r]{\strut{} 0}}%
      \put(1254,1383){\makebox(0,0)[r]{\strut{} 0.2}}%
      \put(1254,2061){\makebox(0,0)[r]{\strut{} 0.4}}%
      \put(1254,2740){\makebox(0,0)[r]{\strut{} 0.6}}%
      \put(1254,3419){\makebox(0,0)[r]{\strut{} 0.8}}%
      \put(1254,4097){\makebox(0,0)[r]{\strut{} 1}}%
      \put(1254,4776){\makebox(0,0)[r]{\strut{} 1.2}}%
      \put(1386,484){\makebox(0,0){\strut{} 0}}%
      \put(2182,484){\makebox(0,0){\strut{} 0.5}}%
      \put(2978,484){\makebox(0,0){\strut{} 1}}%
      \put(3774,484){\makebox(0,0){\strut{} 1.5}}%
      \put(4570,484){\makebox(0,0){\strut{} 2}}%
      \put(5366,484){\makebox(0,0){\strut{} 2.5}}%
      \put(6162,484){\makebox(0,0){\strut{} 3}}%
      \put(6958,484){\makebox(0,0){\strut{} 3.5}}%
      \put(484,2740){\rotatebox{90}{\makebox(0,0){\strut{}$\langle \Delta^{\Ir{\ebar}{1}} \rangle$}}}%
      \put(4172,154){\makebox(0,0){\strut{}$\beta_{\ebar}$}}%
    }%
    \gplgaddtomacro\gplfronttext{%
    }%
    \gplbacktext
    \put(0,0){\includegraphics{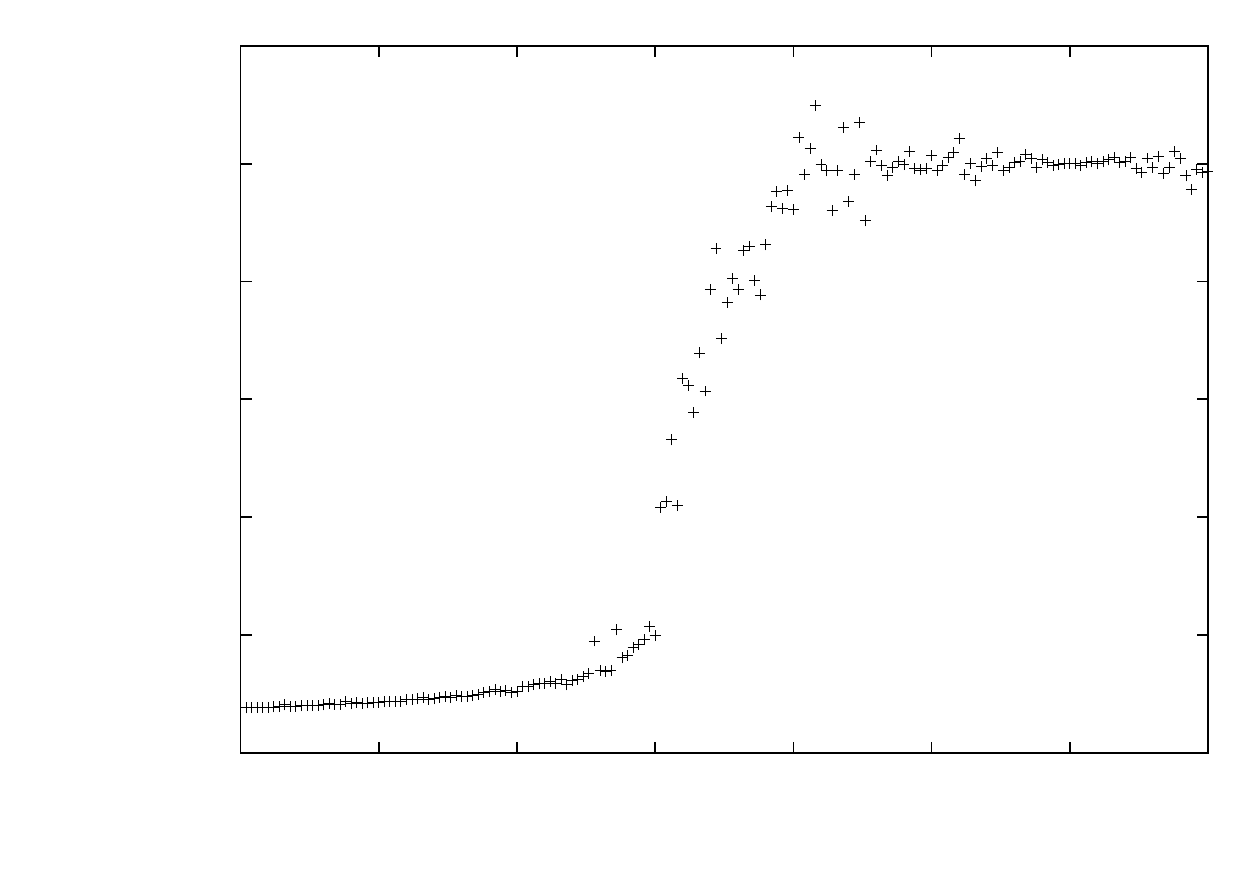}}%
    \gplfronttext
  \end{picture}%
\endgroup
\caption{\label{fig.ebar}Vacuum expectation value of the \ir{\ebar}{1} open string as a function of the coupling constant $\beta_{\bar{e}}$, showing that the nonlocal open string operators are good order parameters  to characterize topological phase transitions. Length of the string: 4 plaquettes, measurements on a $16^3$ lattice, $\beta_e=3.0$, other couplings zero.\label{fig:openstring}}
\end{figure}

As pointed out in previous sections for example in relation (\ref{measureq}), there is a very direct way to determine the condensate as well as the quantum embedding index $q$. This is by measuring the expectation value of the open string for each pure flux $A$ and then summing over all fluxes. In the \ir{\ebar}{1} vacuum we obtain
\[
\left<\quad \qdanyon{\Ir{e}{1}} \right>_\Phi = \left<\quad \qdanyon{\Ir{\ebar}{1}} \right>_\Phi = 1.0,
\]
so $q=2$, whereas in the \ir{X_1}{\Gamma^0} vacuum
\[
\left<\quad \qdanyon{\Ir{e}{1}} \right>_\Phi = \left<\quad \qdanyon{\Ir{\ebar}{1}} \right>_\Phi = 1.0, \quad\left<\quad \qdanyon{\Ir{X_1}{\Gamma^0}} \right>_\Phi =2.0,
\]
so in this case $q=4$. In Figure \ref{fig.ebar} we show the measurement of the vacuum expectation value for the \ir{\ebar}{1} open string as a function of the coupling constant $\beta_{\bar{e}}$, which demonstrates that such measurements  clearly indicate  where the transition takes place.

There is one more issue we like to address in our simulations, that is to determine the order of the transitions we have identified. A conventional approach is to search for a hysteresis effect across a first order transition, but because of the relative modest size of the lattices used this is not an optimal approach. A method that is working much better is to directly probe the system at a given sequence of coupling constants around the transition and to see whether there is a coexistence region where both phases occur in the sampling\footnote{We would like to thank Simon Trebst for pointing this out to us.}. To perform these measurements we use the parallel tempering method \cite{JPSJ.65.1604} to overcome local minima in the action landscape. The idea behind this method is to initialize a range of lattices simultaneously, all at different couplings along a trajectory in coupling constant space starting in phase one and ending in phase two. The updates of this ensemble then consist of the updates of each of the individual lattices and, occasionaly, a swap of two adjacent lattices. The swap between lattices ${\bf 1}$ and ${\bf 2}$ is accepted with a probability
\[
p({\bf 1} \leftrightarrow {\bf 2})= {\rm min} \left\{ 1,{ {\rm exp}\left( S_{\bf 1}({\bf 1}) + S_{\bf 2}({\bf 2})   \right) \over {\rm exp}\left( S_{\bf 2}({\bf 1}) + S_{\bf 1}({\bf 2})   \right) }    \right\},
\]
where $S_{\bf 1} ({\bf 2})$ means using the action (in particular, the set of couplings) of lattice ${\bf 1}$ to evaluate the field configuration of lattice ${\bf 2}$ et cetera.
One can prove that this satisfies detailed balance. In effect, each lattice will perform a random walk through coupling constant space along the chosen trajectory, allowing a ``cold" lattice to thermalize in the ``high temperature" region, thus overcoming the local minima of the action.

\begin{figure}[t]
\centering
\subfigure[$\beta_e=1.906$]{
\includegraphics[scale=0.46]{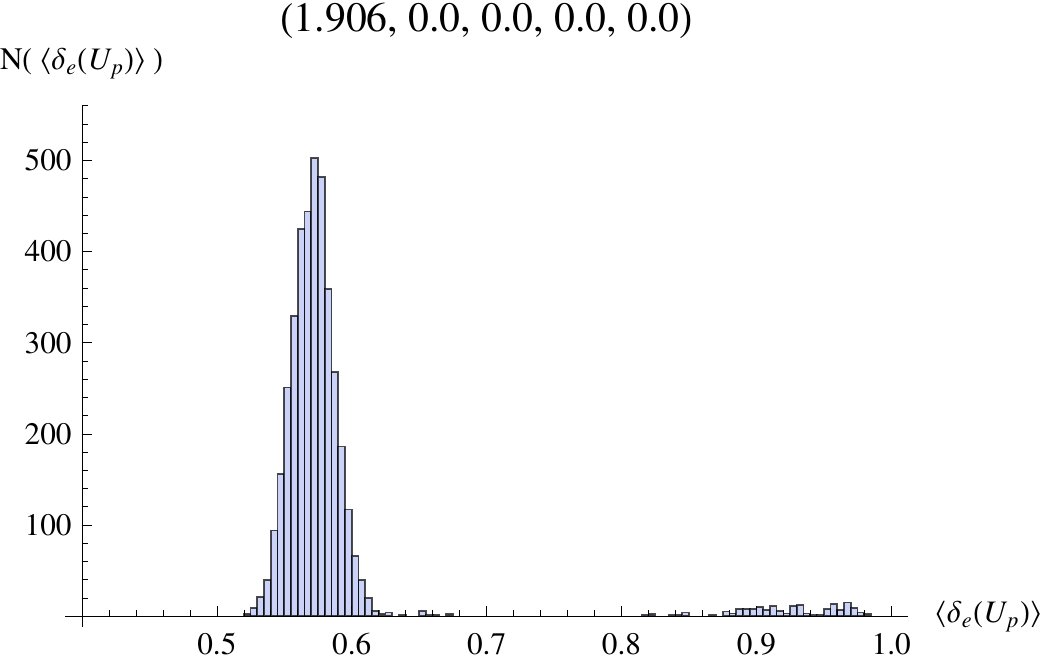}
}
\subfigure[$\beta_e=1.916$]{
\includegraphics[scale=0.46]{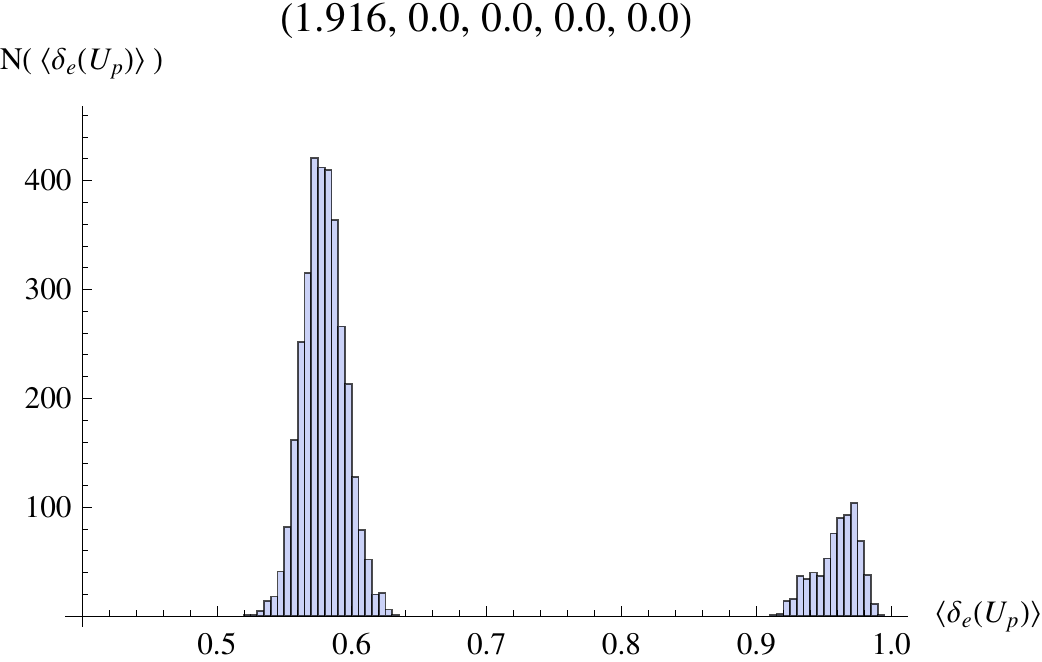}
}
\subfigure[$\beta_e=1.926$]{
\includegraphics[scale=0.46]{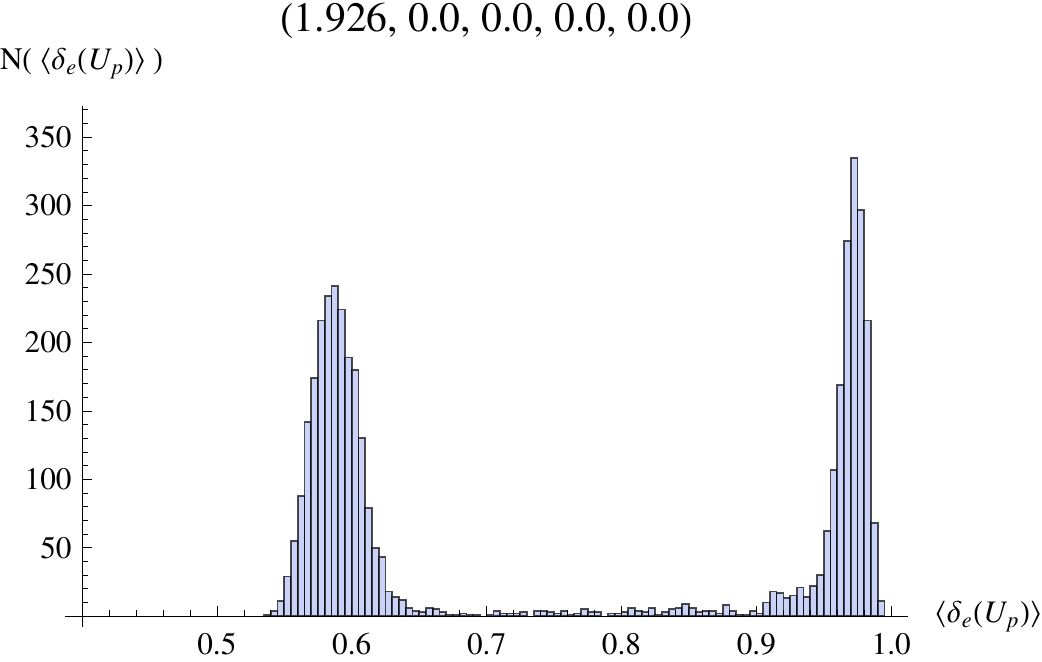}
}\\
\subfigure[$\beta_e=1.936$]{
\includegraphics[scale=0.46]{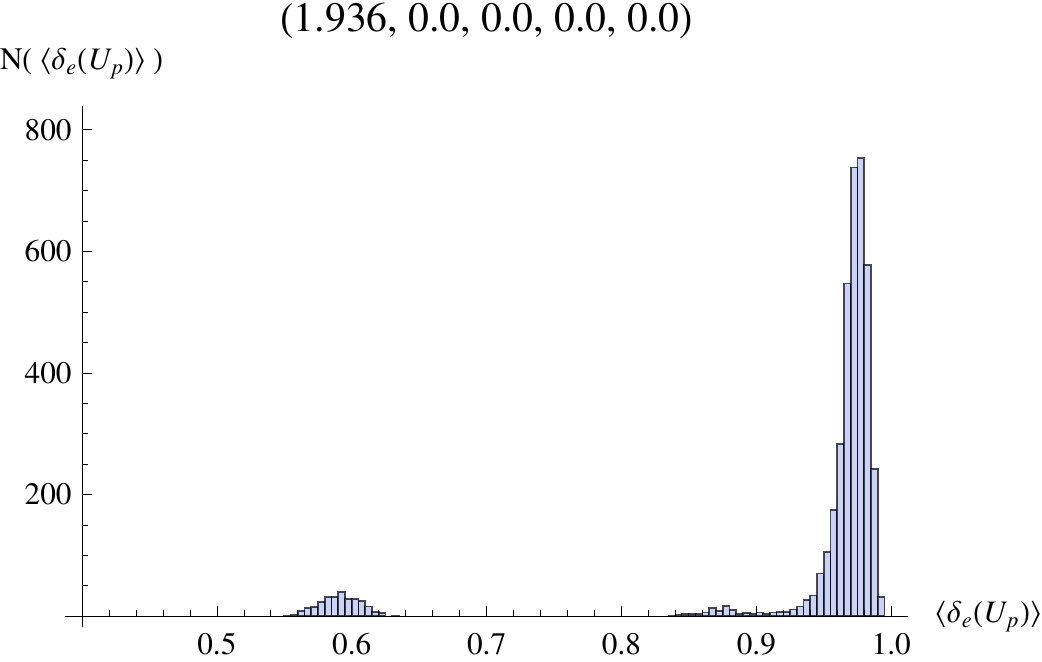}
}

\caption{The sequence of plots is across the transition from the strongly coupled phase with all fluxes condensed and all charges coinfined, to the trivial phase. This trajectory corresponds to the horizontal arrow in figures \ref{fig:phase1}f  and \ref{fig:phase2}f, where $1.906~\le~\beta_e~\le~1.936$ and all other couplings equal zero. Plotted along the $x$-axis is  average expectation value of the percentage of trivial plaquettes with $U_p=e$ and along the $y$-axis we plot the number of times that that percentage is measured in a simulation of $4000$ runs on a $10^3$ lattice. The figures clearly show a shift from peak on the left to on the right, with a double peak in between, this is the signature of region where both phases coexist, i.e. of a first order transition.\label{fig:pt1}}
\end{figure}

\begin{figure}[h!]
\centering
\subfigure[$\beta_{\bar{e}}=1.42$]{
\includegraphics[scale=0.46]{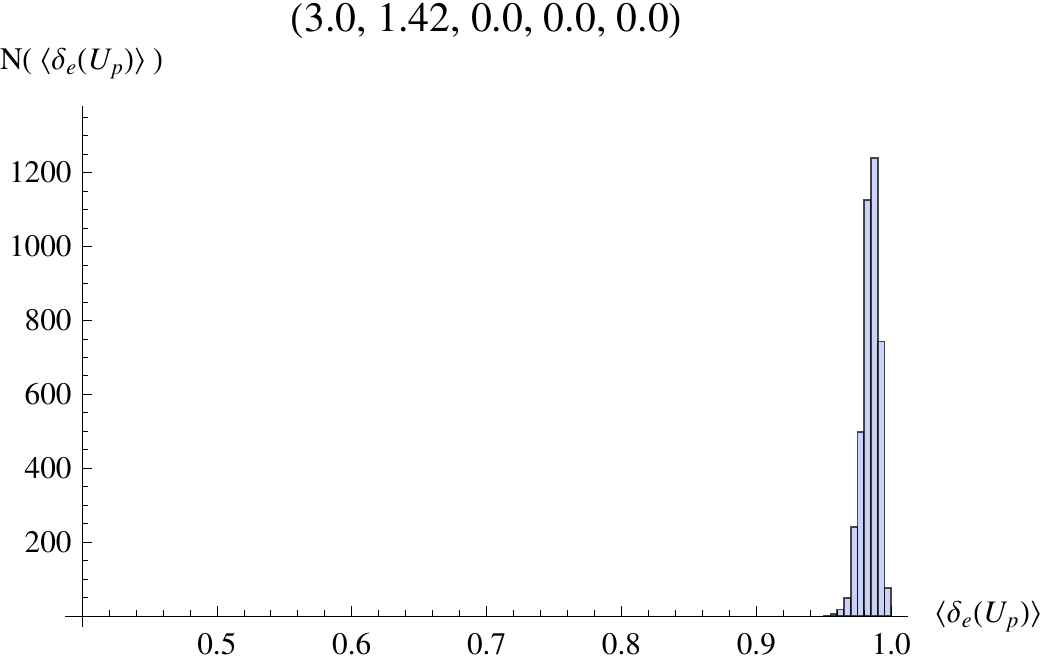}
}
\subfigure[$\beta_{\bar{e}}=1.52$]{
\includegraphics[scale=0.46]{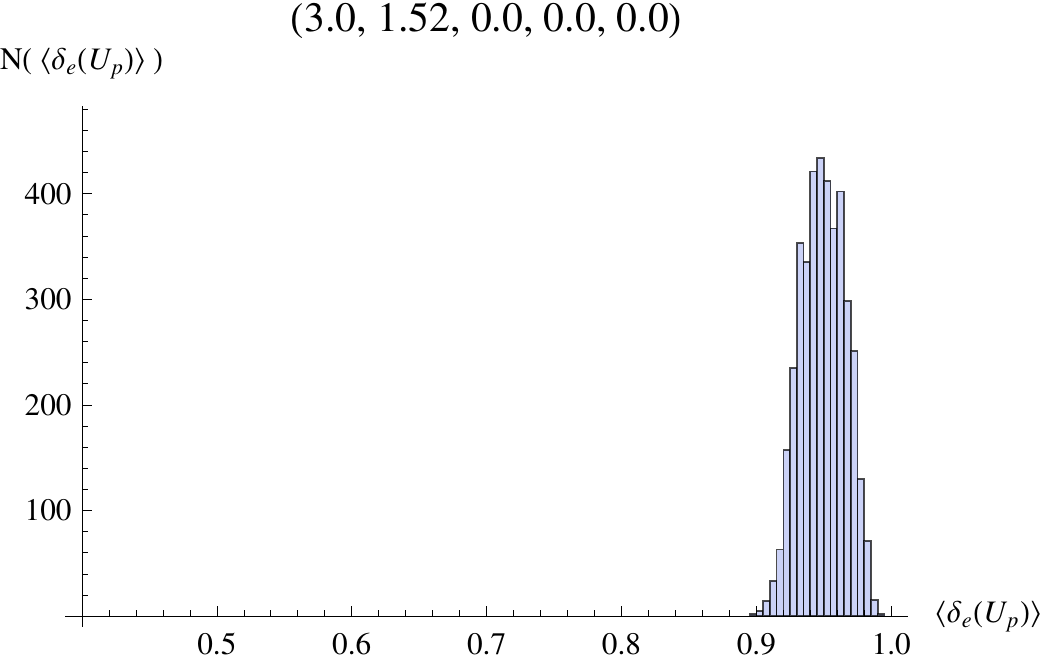}
}
\subfigure[$\beta_{\bar{e}}=1.62$]{
\includegraphics[scale=0.46]{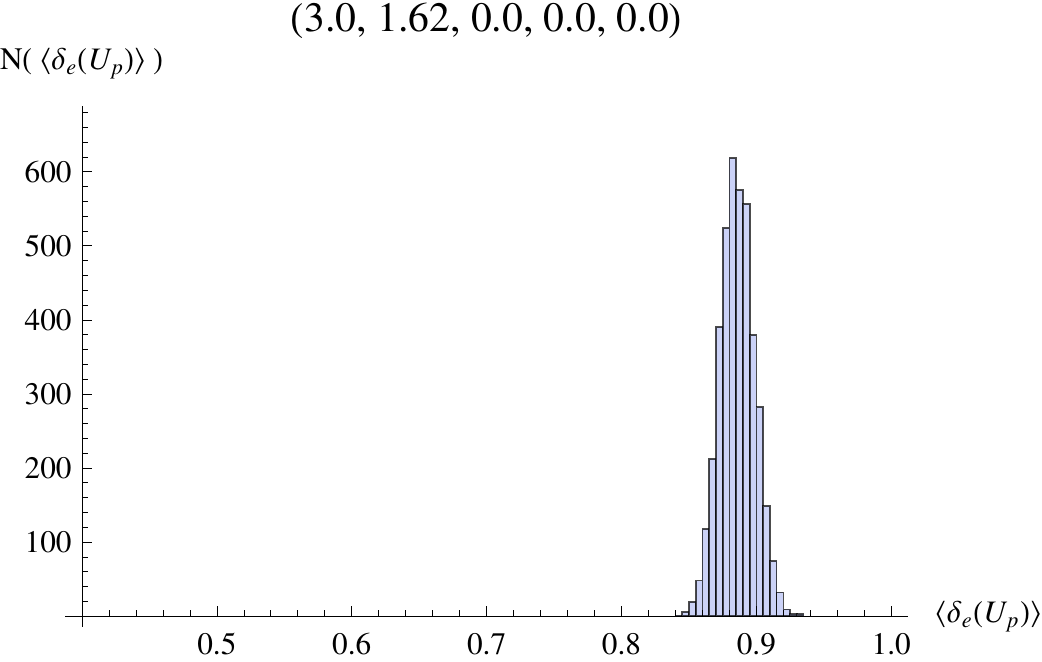}
}\\
\subfigure[$\beta_{\bar{e}}=1.72$]{
\includegraphics[scale=0.46]{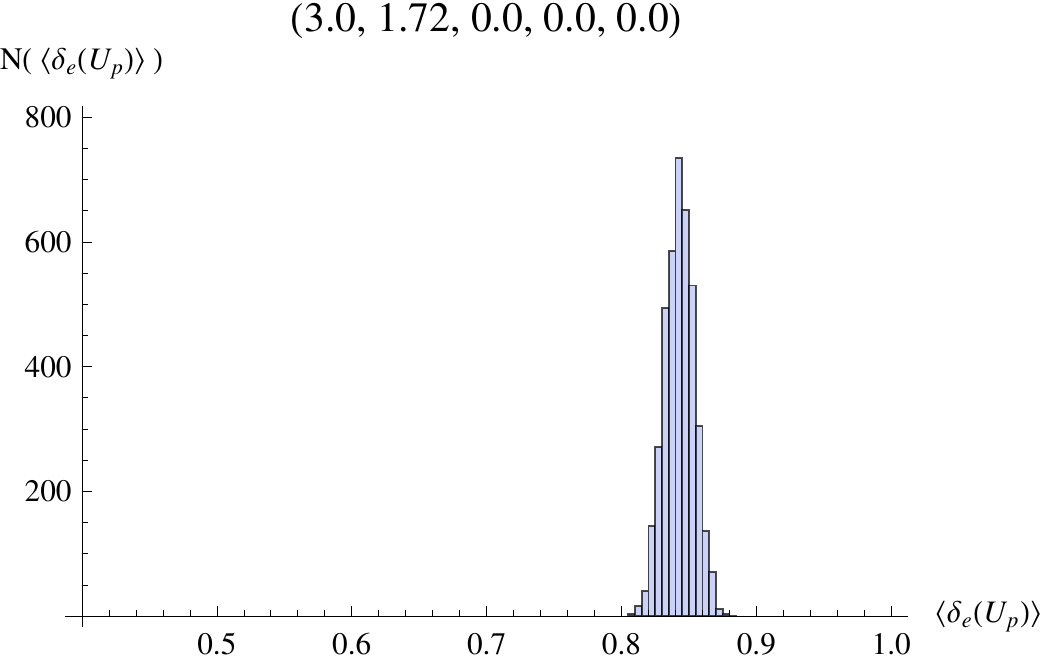}
}
\caption{The sequence of plots is across the transition from the trivial  phase to the  phase with the $\bar{e}$ condensate, the trajectory corresponds to the vertical arrow in figure \ref{fig:phase1}f, where $1.42~\le~\beta_{\bar{e}}~\le~1.72$, $\beta_e=3.0$ and all other couplings equal zero. Plotted along the x-axis is  average expectation value of the percentage of trivial plaquettes with $U_p=e$ and along the $y$-axis we plot the number of times that that percentage is measured in a simulation of $4000$ runs on a $10^3$ lattice. 
The figures only feature only a single peak that smoothly moves from one phase to the other, indicating a smooth second order transition, presumably corresponding to the 3-d ising model transition.  \label{fig:pt2} }
\end{figure}

We have made measurements for the trajectories indicated by the arrows in the Figures \ref{fig:phase1}f  and \ref{fig:phase2}f. The results of these measurements for the horizontal arrow is given in Figure \ref{fig:pt1}  and for the vertical arrow in Figure \ref{fig:pt2}. We find that in that the horizontal trajectory the transition from the strongly coupled phase corresponding to the left peak in Figure \ref{fig:pt1}  to the trivial phase corresponding to the right peak  indeed goes through a coexistence region corresponding to the values of the coupling parameter where both peaks are present as in subfigures \ref{fig:pt1}b and \ref{fig:pt1}c.  

The result for the vertical trajectory corresponding to the transition from the trivial phase to the broken ($X_1,\Gamma_0$) phase is  given in figure \ref{fig:pt2}, where we see that the peak shifts continuously implying that the transition is second order. We can understand this transition as follows. In this region of coupling constant space, all fluxes except the $\ebar$ flux are very heavy. This means the ground state is essentially that of a $\Z_2$ gauge theory. Since $\Z_2$ gauge theory in 3 dimensions is Kramers-Wannier dual to the 3 dimensional Ising model, it has the same phase structure \cite{PhysRevD.11.2098}. We therefore expect this phase transition to lie in the same universality class.

\paragraph*{Measuring the (broken)  modular S-matrices.}

We have  measured  the (broken) $S$-matrix elements using the simple algorithm involving the auxiliary symmetries of our loop operators. This allows us to obtain the unbroken $S$-matrix as well as the branching matrix $n^{a_1}_u$ and the 
 $S$-matrix of the effective $\sU$ theory in the various broken phases.  Here we exploit the relation (\ref{qdsmat}) for the measurement, and  relation (\ref{eq:Suv}) :
\begin{equation*}
S_{uv} = {1 \over q} \sum_{a_i,b_j} n^{a_i}_u n^{b_j}_v \left< S_{a{_i}b{_j}} \right>_\Phi,
\label{eq:suv}
\end{equation*}relating $S_{uv}$ to the measured  $S$-matrix in the broken phase.
We first measured the unbroken  $S$-matrix  in the $D(\bar{D}_2)$ phase and obtain results identical to the matrix calculated using defining formula (\ref{eq:sab}), the result is given in Table \ref{ap.smattriv} and is of course also consistent to the matrix obtained  from the relation (\ref{eq:Suv}) with $\Phi=0$.  The accuracy of the measured matrix elements in represented in the table as integers is smaller than 5\%.

\begin{table}[h!]
\begin{center}
\tiny
\tabcolsep 3.8pt
\begin{tabular}{@{}r|rrrrrrrrrrrrrrrrrrrrrr@{}}

   & \bsw \ir{e}{1}\esw & \bsw\ir{e}{J_1}\esw & \bsw\ir{e}{J_2}\esw & \bsw\ir{e}{J_3}\esw & \bsw\ir{e}{\chi}\esw & \bsw \ir{\ebar}{1}\esw & \bsw\ir{\ebar}{J_1}\esw & \bsw\ir{\ebar}{J_2}\esw & \bsw\ir{\ebar}{J_3}\esw & \bsw\ir{\ebar}{\chi}\esw & \bsw\ir{X_1}{\Gamma^0}\esw & \bsw\ir{X_1}{\Gamma^1}\esw & \bsw\ir{X_1}{\Gamma^2}\esw & \bsw\ir{X_1}{\Gamma^3}\esw & \bsw\ir{X_2}{\Gamma^0}\esw & \bsw\ir{X_2}{\Gamma^1}\esw & \bsw\ir{X_2}{\Gamma^2}\esw & \bsw\ir{X_2}{\Gamma^3}\esw & \bsw\ir{X_3}{\Gamma^0}\esw & \bsw\ir{X_3}{\Gamma^1}\esw & \bsw\ir{X_3}{\Gamma^2}\esw & \bsw\ir{X_3}{\Gamma^3}\esw  \\
 \hline
\ir{e}{1}			& 1 & 1 & 1 & 1 & 2 & 1 & 1 & 1 & 1 & 2 & 2 & 2 & 2 & 2 & 2 & 2 & 2 & 2 & 2 & 2 & 2 & 2 \\
\ir{e}{J_1}			& 1 & 1 & 1 & 1 & 2 & 1 & 1 & 1 & 1 & 2 & 2 & 2 & 2 & 2 &-2 &-2 &-2 &-2 &-2 &-2 &-2 &-2 \\
\ir{e}{J_2}			& 1 & 1 & 1 & 1 & 2 & 1 & 1 & 1 & 1 & 2 &-2 &-2 &-2 &-2 & 2 & 2 & 2 & 2 &-2 &-2 &-2 &-2 \\
\ir{e}{J_3}			& 1 & 1 & 1 & 1 & 2 & 1 & 1 & 1 & 1 & 2 &-2 &-2 &-2 &-2 &-2 &-2 &-2 &-2 & 2 & 2 & 2 & 2 \\
\ir{e}{\chi}			& 2 & 2 & 2 & 2 & 4 &-2 &-2 &-2 &-2 &-4 & 0 & 0 & 0 & 0 & 0 & 0 & 0 & 0 & 0 & 0 & 0 & 0 \\
\ir{\ebar}{1}		& 1 & 1 & 1 & 1 &-2 & 1 & 1 & 1 & 1 &-2 & 2 & 2 & 2 & 2 & 2 & 2 & 2 & 2 & 2 & 2 & 2 & 2 \\
\ir{\ebar}{J_1}		& 1 & 1 & 1 & 1 &-2 & 1 & 1 & 1 & 1 &-2 & 2 & 2 & 2 & 2 &-2 &-2 &-2 &-2 &-2 &-2 &-2 &-2 \\
\ir{\ebar}{J_2}		& 1 & 1 & 1 & 1 &-2 & 1 & 1 & 1 & 1 &-2 &-2 &-2 &-2 &-2 & 2 & 2 & 2 & 2 &-2 &-2 &-2 &-2 \\
\ir{\ebar}{J_3}		& 1 & 1 & 1 & 1 &-2 & 1 & 1 & 1 & 1 &-2 &-2 &-2 &-2 &-2 &-2 &-2 &-2 &-2 & 2 & 2 & 2 & 2 \\
\ir{\ebar}{\chi}		& 2 & 2 & 2 & 2 &-4 &-2 &-2 &-2 &-2 & 4 & 0 & 0 & 0 & 0 & 0 & 0 & 0 & 0 & 0 & 0 & 0 & 0 \\
\ir{X_1}{\Gamma^0}	& 2 & 2 &-2 &-2 & 0 & 2 & 2 &-2 &-2 & 0 & 4 & 0 &-4 & 0 & 0 & 0 & 0 & 0 & 0 & 0 & 0 & 0 \\
\ir{X_1}{\Gamma^1}	& 2 & 2 &-2 &-2 & 0 & 2 & 2 &-2 &-2 & 0 & 0 &-4 & 0 & 4 & 0 & 0 & 0 & 0 & 0 & 0 & 0 & 0 \\
\ir{X_1}{\Gamma^2}	& 2 & 2 &-2 &-2 & 0 & 2 & 2 &-2 &-2 & 0 &-4 & 0 & 4 & 0 & 0 & 0 & 0 & 0 & 0 & 0 & 0 & 0 \\
\ir{X_1}{\Gamma^3}	& 2 & 2 &-2 &-2 & 0 & 2 & 2 &-2 &-2 & 0 & 0 & 4 & 0 &-4 & 0 & 0 & 0 & 0 & 0 & 0 & 0 & 0 \\
\ir{X_2}{\Gamma^0}	& 2 &-2 & 2 &-2 & 0 & 2 &-2 & 2 &-2 & 0 & 0 & 0 & 0 & 0 & 4 & 0 &-4 & 0 & 0 & 0 & 0 & 0 \\
\ir{X_2}{\Gamma^1}	& 2 &-2 & 2 &-2 & 0 & 2 &-2 & 2 &-2 & 0 & 0 & 0 & 0 & 0 & 0 &-4 & 0 & 4 & 0 & 0 & 0 & 0 \\
\ir{X_2}{\Gamma^2}	& 2 &-2 & 2 &-2 & 0 & 2 &-2 & 2 &-2 & 0 & 0 & 0 & 0 & 0 &-4 & 0 & 4 & 0 & 0 & 0 & 0 & 0 \\
\ir{X_2}{\Gamma^3}	& 2 &-2 & 2 &-2 & 0 & 2 &-2 & 2 &-2 & 0 & 0 & 0 & 0 & 0 & 0 & 4 & 0 &-4 & 0 & 0 & 0 & 0 \\
\ir{X_3}{\Gamma^0}	& 2 &-2 &-2 & 2 & 0 & 2 &-2 &-2 & 2 & 0 & 0 & 0 & 0 & 0 & 0 & 0 & 0 & 0 & 4 & 0 &-4 & 0 \\
\ir{X_3}{\Gamma^1}	& 2 &-2 &-2 & 2 & 0 & 2 &-2 &-2 & 2 & 0 & 0 & 0 & 0 & 0 & 0 & 0 & 0 & 0 & 0 &-4 & 0 & 4 \\
\ir{X_3}{\Gamma^2}	& 2 &-2 &-2 & 2 & 0 & 2 &-2 &-2 & 2 & 0 & 0 & 0 & 0 & 0 & 0 & 0 & 0 & 0 &-4 & 0 & 4 & 0 \\
\ir{X_3}{\Gamma^3}	& 2 &-2 &-2 & 2 & 0 & 2 &-2 &-2 & 2 & 0 & 0 & 0 & 0 & 0 & 0 & 0 & 0 & 0 & 0 & 4 & 0 &-4
\end{tabular}
\end{center}
\caption{The $S$-matrix for the (unbroken) $D(\dbar)$ theory (up to the normalisation factor $1/D_A=1/8$) as measured in the trivial vacuum. We put  integers in the table as the accuracy is below the 5\%,  i.e. 1 actually stands for read as $1. \pm 0.05$. \label{ap.smattriv}}
\end{table}

The branching matrices $n^a_u$ can be obtained from measuring the broken $S$-matrices. The columns in these matrices correspond to the different $\sU$ sectors. If we see two rows or columns with different parents $a$, $b$ in the $\sA$ theory that are proportional to each other, $a$ and $b$ branch to the same $\sU$ sector $u$. Conversely, if different $u$ fields correspond to the same $a$ field that means that the $a$ splits in the broken phase.  We have listed the results for the broken $S$-matrix in  the \ir{\ebar}{1} vacuum in Table \ref{ap.smatebar}. 

To realize the splittings between the irreducible representations using the auxiliary gauge symmetries alluded to in section \ref{sec:auxiliary}, we found the following construction to suffice.
{\em \ir{\ebar}{1} vacuum}
\begin{itemize}
\item \ir{X_i}{\Gamma^0}$_1$ is realized by the operator $\Delta^{\Ir{X_i}{\Gamma^0}}$.
\item \ir{X_i}{\Gamma^0}$_2$ is realized by the operators $\Delta^{\Ir{X_i}{\Gamma^0}} \Delta^{\Ir{e}{J_i}}$.
\item \ir{X_1}{\Gamma^2}$_1$ is realized by the operator $\Delta^{\Ir{X_1}{\Gamma^2}}$ with $\{ x_{i\sigma_1} = e, x_{-i\sigma_1} = i\sigma_2 \}. $
\item \ir{X_1}{\Gamma^2}$_2$ is realized by the operators $\Delta^{\Ir{X_1}{\Gamma^2}} \Delta^{\Ir{e}{J_1}}$ with $\{ x_{i\sigma_1} = e, x_{-i\sigma_1} = i\sigma_2 \}. $
\item \ir{X_2}{\Gamma^2}$_1$ is realized by the operator $\Delta^{\Ir{X_2}{\Gamma^2}}$ with $\{ x_{i\sigma_2} = e, x_{-i\sigma_2} = i\sigma_1 \}. $
\item \ir{X_2}{\Gamma^2}$_2$ is realized by the operators $\Delta^{\Ir{X_2}{\Gamma^2}} \Delta^{\Ir{e}{J_2}}$ with $\{ x_{i\sigma_2} = e, x_{-i\sigma_2} = i\sigma_1 \}. $
\item \ir{X_3}{\Gamma^2}$_1$ is realized by the operator $\Delta^{\Ir{X_3}{\Gamma^2}}$ with $\{ x_{i\sigma_3} = e, x_{-i\sigma_3} = i\sigma_1 \}. $
\item \ir{X_3}{\Gamma^2}$_2$ is realized by the operators $\Delta^{\Ir{X_3}{\Gamma^2}} \Delta^{\Ir{e}{J_3}}$ with $\{ x_{i\sigma_3} = e, x_{-i\sigma_3} = i\sigma_1 \}. $
\end{itemize}
{\em \ir{X_1}{\Gamma^0} vacuum}
\begin{itemize}
\item \ir{X_i}{\Gamma^0}$_1$ is realized by the operator $\Delta^{\Ir{X_i}{\Gamma^0}}$.
\item \ir{X_1}{\Gamma^0}$_2$ is realized by the operators $\Delta^{\Ir{X_1}{\Gamma^0}} \Delta^{\Ir{e}{J_1}}$.
\item \ir{X_2}{\Gamma^2}$_1$ is realized by the operator $\Delta^{\Ir{X_2}{\Gamma^2}}$ with $\{ x_{i\sigma_2} = e, x_{-i\sigma_2} = i\sigma_1 \}. $
\item \ir{X_3}{\Gamma^2}$_1$ is realized by the operator $\Delta^{\Ir{X_3}{\Gamma^2}}$ with $\{ x_{i\sigma_3} = e, x_{-i\sigma_3} = i\sigma_1 \}. $
\end{itemize}
\begin{table}[t]
\begin{center}
\tiny
\tabcolsep 3.8pt
\begin{tabular}{@{}r|rrrrrrrrrrrrrrrrrrrr@{}}

   & \bsw \ir{e}{1}\esw & \bsw\ir{e}{J_1}\esw & \bsw\ir{e}{J_2}\esw & \bsw\ir{e}{J_3}\esw & \bsw \ir{\ebar}{1}\esw & \bsw\ir{\ebar}{J_1}\esw & \bsw\ir{\ebar}{J_2}\esw & \bsw\ir{\ebar}{J_3}\esw & \bsw\ir{X_1}{\Gamma^0}$_1$\esw & \bsw\ir{X_1}{\Gamma^0}$_2$\esw & \bsw\ir{X_1}{\Gamma^2}$_1$\esw & \bsw\ir{X_1}{\Gamma^2}$_2$\esw & \bsw\ir{X_2}{\Gamma^0}$_1$\esw & \bsw\ir{X_2}{\Gamma^0}$_2$\esw & \bsw\ir{X_2}{\Gamma^2}$_1$\esw & \bsw\ir{X_2}{\Gamma^2}$_2$\esw & \bsw\ir{X_3}{\Gamma^0}$_1$\esw & \bsw\ir{X_3}{\Gamma^0}$_2$\esw & \bsw\ir{X_3}{\Gamma^2}$_1$\esw & \bsw\ir{X_3}{\Gamma^2}$_2$\esw  \\
 \hline
\ir{e}{1}               & 1 & 1 & 1 & 1 & 1 & 1 & 1 & 1 & 2 & 2 & 2 & 2 & 2 & 2 & 2 & 2 & 2 & 2 & 2 & 2 \\
\ir{e}{J_1}             & 1 & 1 & 1 & 1 & 1 & 1 & 1 & 1 & 2 & 2 & 2 & 2 &-2 &-2 &-2 &-2 &-2 &-2 &-2 &-2 \\
\ir{e}{J_2}             & 1 & 1 & 1 & 1 & 1 & 1 & 1 & 1 &-2 &-2 &-2 &-2 & 2 & 2 & 2 & 2 &-2 &-2 &-2 &-2 \\
\ir{e}{J_3}             & 1 & 1 & 1 & 1 & 1 & 1 & 1 & 1 &-2 &-2 &-2 &-2 & 2 & 2 & 2 & 2 &-2 &-2 &-2 &-2 \\
\ir{\ebar}{1}			& 1 & 1 & 1 & 1 & 1 & 1 & 1 & 1 & 2 & 2 & 2 & 2 & 2 & 2 & 2 & 2 & 2 & 2 & 2 & 2 \\
\ir{\ebar}{J_1}			& 1 & 1 & 1 & 1 & 1 & 1 & 1 & 1 & 2 & 2 & 2 & 2 &-2 &-2 &-2 &-2 &-2 &-2 &-2 &-2 \\
\ir{\ebar}{J_2}			& 1 & 1 & 1 & 1 & 1 & 1 & 1 & 1 &-2 &-2 &-2 &-2 & 2 & 2 & 2 & 2 &-2 &-2 &-2 &-2 \\
\ir{\ebar}{J_3}			& 1 & 1 & 1 & 1 & 1 & 1 & 1 & 1 &-2 &-2 &-2 &-2 & 2 & 2 & 2 & 2 &-2 &-2 &-2 &-2 \\
\ir{X_1}{\Gamma^0}$_1$  & 2 & 2 &-2 &-2 & 2 & 2 &-2 &-2 & 4 & 4 &-4 &-4 & 4 &-4 & 4 &-4 & 4 &-4 & 4 &-4 \\
\ir{X_1}{\Gamma^0}$_2$  & 2 & 2 &-2 &-2 & 2 & 2 &-2 &-2 & 4 & 4 &-4 &-4 &-4 & 4 &-4 & 4 &-4 & 4 &-4 & 4 \\
\ir{X_1}{\Gamma^2}$_1$  & 2 & 2 &-2 &-2 & 2 & 2 &-2 &-2 &-4 &-4 & 4 & 4 & 4 &-4 & 4 &-4 &-4 & 4 &-4 & 4 \\ 
\ir{X_1}{\Gamma^2}$_2$  & 2 & 2 &-2 &-2 & 2 & 2 &-2 &-2 &-4 &-4 & 4 & 4 &-4 & 4 &-4 & 4 & 4 &-4 & 4 &-4 \\ 
\ir{X_2}{\Gamma^0}$_1$  & 2 &-2 & 2 &-2 & 2 &-2 & 2 &-2 & 4 &-4 & 4 &-4 & 4 & 4 &-4 &-4 & 4 &-4 &-4 & 4 \\
\ir{X_2}{\Gamma^0}$_2$  & 2 &-2 & 2 &-2 & 2 &-2 & 2 &-2 &-4 & 4 &-4 & 4 & 4 & 4 &-4 &-4 &-4 & 4 & 4 &-4 \\
\ir{X_2}{\Gamma^2}$_1$  & 2 &-2 & 2 &-2 & 2 &-2 & 2 &-2 & 4 &-4 & 4 &-4 &-4 &-4 & 4 & 4 &-4 & 4 & 4 &-4 \\
\ir{X_2}{\Gamma^2}$_2$  & 2 &-2 & 2 &-2 & 2 &-2 & 2 &-2 &-4 & 4 &-4 & 4 &-4 &-4 & 4 & 4 & 4 &-4 &-4 & 4 \\
\ir{X_3}{\Gamma^0}$_1$  & 2 &-2 &-2 & 2 & 2 &-2 &-2 & 2 & 4 &-4 &-4 & 4 & 4 &-4 &-4 & 4 & 4 & 4 &-4 &-4 \\
\ir{X_3}{\Gamma^0}$_2$  & 2 &-2 &-2 & 2 & 2 &-2 &-2 & 2 &-4 & 4 & 4 &-4 &-4 & 4 & 4 &-4 & 4 & 4 &-4 &-4 \\
\ir{X_3}{\Gamma^2}$_1$  & 2 &-2 &-2 & 2 & 2 &-2 &-2 & 2 & 4 &-4 &-4 & 4 &-4 & 4 & 4 &-4 &-4 &-4 & 4 & 4 \\
\ir{X_3}{\Gamma^2}$_2$  & 2 &-2 &-2 & 2 & 2 &-2 &-2 & 2 &-4 & 4 & 4 &-4 & 4 &-4 &-4 & 4 &-4 &-4 & 4 & 4 \\
\end{tabular}
\end{center}
\caption{The broken $S$-matrix as measured in the \ir{\ebar}{1} vacuum, where the cloumns and rows of zeroes corresponding to the confined fields are left out. Identifying identical columns and rows we obtain  the familiar $S$-matrix of the $D(\mathbb{Z}_2\otimes\mathbb{Z}_2)$ theory.\label{ap.smatebar}}
\end{table}

\begin{table}[h!]
\begin{center}
\tiny
\tabcolsep 3.8pt
\begin{tabular}{@{}c|c|rrrrrrrrrrrrrrrrr@{}}

 \bsw\;\;$\sU$\esw  &\bsw $D(\mathbb{Z}_2\otimes\mathbb{Z}_2)$ \esw& \bsw \ir{e}{1}\esw & \bsw\ir{e}{J_1}\esw & \bsw\ir{e}{J_2}\esw & \bsw\ir{e}{J_3}\esw & \bsw\ir{X_1}{\Gamma^0}$_1$\esw & \bsw\ir{X_1}{\Gamma^0}$_2$\esw & \bsw\ir{X_1}{\Gamma^2}$_1$\esw & \bsw\ir{X_1}{\Gamma^2}$_2$\esw & \bsw\ir{X_2}{\Gamma^0}$_1$\esw & \bsw\ir{X_2}{\Gamma^0}$_2$\esw & \bsw\ir{X_2}{\Gamma^2}$_1$\esw & \bsw\ir{X_2}{\Gamma^2}$_2$\esw & \bsw\ir{X_3}{\Gamma^0}$_1$\esw & \bsw\ir{X_3}{\Gamma^0}$_2$\esw & \bsw\ir{X_3}{\Gamma^2}$_1$\esw & \bsw\ir{X_3}{\Gamma^2}$_2$\esw  \\
 \hline
\ir{e}{1}       			& $(++,++)$& 1 & 1 & 1 & 1 & 1 & 1 & 1 & 1 & 1 & 1 & 1 & 1 & 1 & 1 & 1 & 1 \\
\ir{e}{J_1}     			& $(++,+-)$& 1 & 1 & 1 & 1  & 1 & 1 & 1 & 1 &-1 &-1 &-1 &-1 &-1 &-1 &-1 &-1 \\
\ir{e}{J_2} 			    & $(++,-+)$& 1 & 1 & 1 & 1  & -1 &-1 &-1 &-1 & 1 & 1 & 1 & 1 &-1 &-1 &-1 &-1 \\
\ir{e}{J_3}   			& $(++,--)$& 1 & 1 & 1 & 1  &-1 &-1 &-1 &-1 &-1 &-1 &-1 &-1 &1 &1 &1 &1 \\
\ir{X_1}{\Gamma^0}$_1$  & $(-+,++)$&1& 1 &-1 &-1  & 1  & 1  &-1  &-1  & 1  &-1  & 1  &-1  & 1  &-1  & 1  &-1  \\
\ir{X_1}{\Gamma^0}$_2$  & $(-+,+-)$&1& 1 &-1 &-1  & 1  & 1  &-1  &-1  &-1  & 1  &-1  & 1  &-1  & 1  &-1  & 1  \\
\ir{X_1}{\Gamma^2}$_1$  & $(-+,-+)$&1 & 1 &-1 &-1  &-1  & -1  & 1  & 1  & 1  &-1  & 1  &-1  &-1  & 1  &-1  & 1  \\ 
\ir{X_1}{\Gamma^2}$_2$  & $(-+,--)$&1 & 1 &-1 &-1 &-1  & -1  & 1  & 1  &-1  & 1  &-1  & 1  & 1  &-1  & 1  &-1  \\ 
\ir{X_2}{\Gamma^0}$_1$  & $(+-,++)$&1 &-1 & 1 &-1 & 1  &-1  & 1  &-1  & 1  & 1  &-1  &-1  & 1  &-1  &-1  & 1  \\
\ir{X_2}{\Gamma^0}$_2$  & $(+-,-+)$&1 &-1 & 1 &-1  & -1  &1  &-1  & 1  & 1  & 1  &-1  &-1  &-1  & 1  & 1  &-1  \\
\ir{X_2}{\Gamma^2}$_1$  & $(+-,+-)$&1 &-1 & 1 &-1  &1  & -1  & 1  &-1  &-1  &-1  & 1  & 1  &-1  & 1  & 1  &-1  \\
\ir{X_2}{\Gamma^2}$_2$  & $(+-,--)$&1 &-1 & 1 &-1  & -1  & 1  &-1  & 1  &-1  &-1  & 1  & 1  & 1  &-1  &-1  & 1  \\
\ir{X_3}{\Gamma^0}$_1$  & $(--,++)$&1 &-1 &-1 & 1 &1  & -1  &-1  & 1  & 1  &-1  &-1  & 1  & 1  & 1  &-1  &-1  \\
\ir{X_3}{\Gamma^0}$_2$  & $(--,--)$&1 &-1 &-1 & 1  & -1  & 1  & 1  &-1  &-1  & 1  & 1  &-1  & 1  & 1  &-1  &-1  \\
\ir{X_3}{\Gamma^2}$_1$  & $(--,+-)$&1 &-1 &-1 & 1  &1  &-1  &-1  & 1  &-1  & 1  & 1  &-1  &-1  &-1  & 1  & 1  \\
\ir{X_3}{\Gamma^2}$_2$  & $(--,-+)$&1 &-1 &-1 & 1 & -1  & 1  & 1  &-1  & 1  &-1  &-1  & 1  &-1  &-1  & 1  & 1  \\
\end{tabular}
\end{center}
\caption{The  $S$-matrix of the $ \sU =D(\mathbb{Z}_2\otimes\mathbb{Z}_2)$ theory (up to normalisation factor of $1/D_U = 1/4$) as obtained from the  broken $S$-matrix  measured in the \ir{\ebar}{1} vacuum, after leaving out the rows and columns with only zeroes of the confined fields and after identifying identical rows and columns. \label{ap.smatdz2xz2}}
\end{table} 
We see that in Table \ref{ap.smatebar} the columns (rows) for  the sectors \ir{e}{\alpha} and \ir{\ebar}{\alpha} for $\alpha=1,J_1,J_2,J_3$ are identical and thus that the corresponding fields have to be identified. This leaves us with 16 sectors for the broken $\sU$ theory. Summing the entries as prescribed by formula (\ref{eq:suv}) yields exactly the $S$-matrix of the $D(\mathbb{Z}_2\otimes\mathbb{Z}_2)$ theory, which is given in Table \ref{ap.smatdz2xz2}.

In Table \ref{ap.smatx1} we have listed the result for broken $S$-matrix in the \ir{X_1}{\Gamma^0} vacuum. here we have to identify the sectors \ir{e}{1}, \ir{\ebar}{1} and \ir{X_1}{\Gamma^0}$_1$,  the sectors \ir{e}{J_1}, \ir{\ebar}{J_1} and \ir{X_1}{\Gamma^0}$_2$,  the sectors \ir{X_2}{\Gamma^0}$_1$ and \ir{X_3}{\Gamma^0}$_1$, and  the sectors \ir{X_2}{\Gamma^2}$_1$ and \ir{X_3}{\Gamma^2}$_1$. These results are all fully consistent with the algebraic analysis presented in Section \ref{breakalg}.

Let us illustrate the method by calculating a few sample $S$-matrix elements in the \ir{X_1}{\Gamma^0} condensed vacuum. The $\sU$ theory should be $D(\ZZ_2)$; let us first calculate the $S_{\Ir{+}{+}\Ir{+}{+}}$ element, the \ir{+}{+} sector being the new vacuum
\begin{eqnarray*}
S_{\Ir{+}{+}\Ir{+}{+}} = {1 \over q}  \Big\{ \left< S_{\Ir{e}{1}\Ir{e}{1}} \right>_\Phi + \left< S_{\Ir{e}{1}\Ir{\ebar}{1}} \right>_\Phi + \left< S_{\Ir{e}{1}\Ir{X_1}{\Gamma^0}} \right>_\Phi + \\
 \left< S_{\Ir{\ebar}{1}\Ir{e}{1}} \right>_\Phi + \left< S_{\Ir{\ebar}{1}\Ir{\ebar}{1}} \right>_\Phi + \left< S_{\Ir{\ebar}{1}\Ir{X_1}{\Gamma^0}} \right>_\Phi + \\
  \left< S_{\Ir{X_1}{\Gamma^0}\Ir{e}{1}} \right>_\Phi +  \left< S_{\Ir{X_1}{\Gamma^0}\Ir{\ebar}{1}} \right>_\Phi +  \left< S_{\Ir{X_1}{\Gamma^0}\Ir{X_1}{\Gamma^0}} \right>_\Phi \Big\} = \\
  {1 \over 4}{1 \over 8} \left( 1+1+2+1+1+2+2+2+4 \right) = {1 \over 2},
\end{eqnarray*}
in agreement with Table \ref{ap.smatz2}. The contributions to the above matrix element would be equal if we had used the $S$-matrix elements as measured in the trivial vacuum.

\begin{table}[t]
\begin{center}
\tiny
\tabcolsep 3.8pt
\begin{tabular}{@{}r|rrrrrrrrrr@{}}
                     & \bsw\ir{e}{1}\esw & \bsw\ir{e}{J_1}\esw & \bsw\ir{\ebar}{1}\esw & \bsw\ir{\ebar}{J_1}\esw & \bsw\ir{X_1}{\Gamma^0}$_1$\esw & \bsw\ir{X_1}{\Gamma^0}$_2$\esw & \bsw\ir{X_2}{\Gamma^0}$_1$\esw & \bsw\ir{X_2}{\Gamma^2}$_1$\esw & \bsw\ir{X_3}{\Gamma^0}$_1$\esw & \bsw\ir{X_3}{\Gamma^2}$_1$\esw\\
                     \hline
\ir{e}{1}           		& 1 & 1 & 1 & 1 & 2 & 2 & 2 & 2 & 2 & 2 \\
\ir{e}{J_1}         		& 1 & 1 & 1 & 1 & 2 & 2 &-2 &-2 &-2 &-2 \\
\ir{\ebar}{1}			& 1 & 1 & 1 & 1 & 2 & 2 & 2 & 2 & 2 & 2 \\
\ir{\ebar}{J_1}			& 1 & 1 & 1 & 1 & 2 & 2 &-2 &-2 &-2 &-2 \\
\ir{X_1}{\Gamma^0}$_1$	& 2 & 2 & 2 & 2 & 4 & 4 & 4 & 4 & 4 & 4 \\
\ir{X_1}{\Gamma^0}$_2$	& 2 & 2 & 2 & 2 & 4 & 4 &-4 &-4 &-4 &-4 \\
\ir{X_2}{\Gamma^0}$_1$	& 2 &-2 & 2 &-2 & 4 &-4 & 4 &-4 & 4 &-4 \\
\ir{X_2}{\Gamma^2}$_1$	& 2 &-2 & 2 &-2 & 4 &-4 &-4 & 4 &-4 & 4 \\
\ir{X_3}{\Gamma^0}$_1$	& 2 &-2 & 2 &-2 & 4 &-4 & 4 &-4 & 4 &-4 \\
\ir{X_3}{\Gamma^2}$_1$	& 2 &-2 & 2 &-2 & 4 &-4 &-4 & 4 &-4 & 4
\end{tabular}
\end{center}
\caption{The broken $S$-matrix for the $D(\dbar)$ theory as measured in the \ir{X_1}{\Gamma^0} vacuum.\label{ap.smatx1}}
\end{table}

\begin{table}[h!]
\begin{center}
\tiny
\tabcolsep 3.8pt
\begin{tabular}{@{}c|c|rrrrr@{}}
 \bsw\;\;$\sU$\esw  &\bsw $D(\mathbb{Z}_2)$ \esw & \bsw\ir{e}{1}\esw & \bsw\ir{e}{J_1}\esw & \bsw\ir{X_2}{\Gamma^0}$_1$\esw & \bsw\ir{X_2}{\Gamma^2}$_1$\esw\\
\hline
\ir{e}{1} & \ir{+}{+} & 1 & 1 & 1 & 1 \\
\ir{e}{J_1}&\ir{+}{-} & 1 & 1 & -1 & -1 \\
\ir{X_2}{\Gamma^0}$_1$ & \ir{-}{+} & 1 & -1 & 1 & -1 \\
\ir{X_2}{\Gamma^2}$_1$ & \ir{-}{-} & 1 & -1 & -1 & 1
\end{tabular}
\end{center}
\caption{The  modular $S$-matrix for the $D(\ZZ_2)$ theory (up to normalisation factor $1/D_U=1/2$)\label{ap.smatz2}}
\end{table}

To appreciate the importance of the measurements in the broken vacuum, consider the matrix element $S_{\Ir{-}{+}\Ir{-}{-}}$. The parents of the \ir{-}{+} sector are \ir{X_2}{\Gamma^0}$_1$ and \ir{X_3}{\Gamma^0}$_1$ and those of the \ir{-}{-} are \ir{X_2}{\Gamma^2}$_1$ and \ir{X_3}{\Gamma^2}$_1$.
\begin{eqnarray*}
S_{\Ir{-}{+}\Ir{-}{-}} = {1 \over q} \Big\{ \left< S_{\Ir{X_2}{\Gamma^0}_1\Ir{X_2}{\Gamma^2}_1} \right>_\Phi +\left< S_{\Ir{X_2}{\Gamma^0}_1\Ir{X_3}{\Gamma^2}_1} \right>_\Phi + \\
\left< S_{\Ir{X_3}{\Gamma^0}_1\Ir{X_2}{\Gamma^2}_1} \right>_\Phi + \left< S_{\Ir{X_3}{\Gamma^0}_1\Ir{X_3}{\Gamma^2}_1} \right>_\Phi \Big\} = \\
 {1 \over 4} {1 \over 8} \left( (-4) + (-4) + (-4) + (-4) \right) = -{1 \over 2},
\end{eqnarray*}.
We see that after completing the calculation along this line  we obtain the $S$-matrix of the $D(\ZZ)$ theory, as given in Table \ref{ap.smatz2}. Note that if we had used the $S$-matrix of the unbroken theory, the $S_{\Ir{X_2}{\Gamma^0}\Ir{X_3}{\Gamma^2}}$ and $S_{\Ir{X_3}{\Gamma^0}\Ir{X_2}{\Gamma^2}}$ would have been zero.

\section{Conclusions and outlook}
In this article we have studied euclidean lattice models for Discrete Gauge Theories. We have introduced a set of multiparameter actions for these  theories that display a rich phase structure, and showed in particular that all the allowed  condensates of pure magnetic flux are realized in certain well anticipated regions of coupling constant space. The set of open string operators that we defined form a set of order parameters that allowed us to determine the  content of the condensate and to measure the topological symmetry breaking index $q$.

Once the condensate is identified, we have shown how to unambiguously reconstruct the $S$-matrix of the low-energy theory in a broken or unbroken phase  by measurements of the complete set of braided loop operators, using the anyonic loop operators we proposed in  earlier work \cite{Bais:2008xi}. Due to an auxiliary gauge symmetry these operators are particularly well suited to detect the nontrivial splittings of fields that correspond to fixed points under fusion with the condensate.  We  found that as expected the excitations that are confined in a broken vacuum give rise to rows and columns of zeroes in the broken $S$-matrix. Our work clearly demonstrates that the euclidean approach allows for a very straightforward method to completely determine the nature of the broken phase.

Our work showed that the reason the modular $S$-matrix changes in the broken phase is largely due to the contribution of the so-called vacuum exchange diagram. In an upcoming more theoretical paper \cite{BER2010} we will extend the approach used in this work, the use of observables and in particular the $S$-matrix to determine the phase structure of a TQFT to a far wider range of theories, in particular the $SU(N)_k$ TQFT arising from Chern-Simons actions.

It would be interesting to study different models exhibiting different topological phases by somehow formulating them in the euclidean 3-dimensional framework, to our knowledge such an approach is unfortunately not yet available for Chern Simons theories. One  expects that for Levin Wen models \cite{Levin:2003ws} our approach could be implemented though. Another path is to investigate the phase structure after adding dynamical matter fields that transform nontrivially. It is known that in such situations the Wilson type  criteria break down as the strings can break, this necessitates the development of different diagnostic tools \cite{1367-2630-13-2-025009,PhysRevLett.56.223}.

\vskip 4mm
The authors would like to thank Jan Smit, Joost Slingerland, Simon Trebst and Sebas Eli\"ens  for useful discussions. JCR is financially supported by a grant from FOM.

\appendix
\section{\label{appfusion}Fusion rules for $\dbar$ DGT}
\begin{tabular}{l @{$\times$} l @{ = } l l@{$\times$} l @{ = } l}
\ir{e}{J_i} & \ir{e}{J_i} & \ir{e}{1}  \\
\ir{e}{J_i} & \ir{e}{J_j} & \ir{e}{J_k} \\
\ir{e}{J_i} & \ir{e}{\chi} & \ir{e}{\chi} \\
\ir{e}{\chi} & \ir{e}{\chi} & \ir{e}{1} + $\sum$ \ir{e}{J_i} \\
\ir{\ebar}{1} & \ir{e}{J_i} & \ir{\ebar}{J_i}\\
\ir{\ebar}{1} & \ir{e}{\chi} & \ir{\ebar}{\chi} \\
\ir{e}{J_i} & \ir{X_i}{\Gamma^{0,2}} & \ir{X_i}{\Gamma^{0,2}}\\
\ir{e}{J_i} & \ir{X_j}{\Gamma^{0,2}} & \ir{X_j}{\Gamma^{2,0}}\\
\ir{e}{\chi} & \ir{X_i}{\Gamma^0} & \ir{X_i}{\Gamma^1} + \ir{X_i}{\Gamma^3}\\
\ir{\ebar}{1} & \ir{\ebar}{1} & \ir{e}{1} \\
\ir{\ebar}{1} & \ir{X_i}{\Gamma^{0,2}} & \ir{X_i}{\Gamma^{0,2}}\\
\ir{\ebar}{1} & \ir{X_i}{\Gamma^{1,3}} & \ir{X_i}{\Gamma^{3,1}} \\
\ir{e}{J_i} & \ir{X_i}{\Gamma^{1,3}} & \ir{X_i}{\Gamma^{1,3}}\\
\ir{e}{J_i} & \ir{X_j}{\Gamma^{1,3}} & \ir{X_j}{\Gamma^{3,1}} \\
\ir{e}{\chi} & \ir{X_i}{\Gamma^{1,3}} & \ir{X_i}{\Gamma^0} + \ir{X_i}{\Gamma^2}\\
\ir{X_i}{\Gamma^{0,2}} & \ir{X_i}{\Gamma^{0,2}}  & \ir{e}{1} + \ir{\ebar}{1} + \ir{e}{J_i} + \ir{\ebar}{J_i} \\
\ir{X_i}{\Gamma^0} & \ir{X_i}{\Gamma^2}  & \ir{e}{J_j} + \ir{\ebar}{J_j} + \ir{e}{J_k} + \ir{\ebar}{J_k} \\
\ir{X_i}{\Gamma^{0,2}} & \ir{X_j}{\Gamma^{0,2}} &\multicolumn{4}{l}{$\!\!\!\!\!$ \ir{X_i}{\Gamma^{2,0}} $\times$ \ir{X_j}{\Gamma^{0,2}} = \ir{X_k}{\Gamma^0} + \ir{X_k}{\Gamma^2}} \\
\ir{X_i}{\Gamma^{0,2}} & \ir{X_i}{\Gamma^{1,3}} &\multicolumn{4}{l}{$\!\!\!\!\!$ \ir{X_i}{\Gamma^{2,0}} $\times$ \ir{X_i}{\Gamma^{1,3}} = \ir{e}{\chi} + \ir{\ebar}{\chi}}\\
\ir{X_i}{\Gamma^{0,2}}&\ir{X_j}{\Gamma^{1,3}}&\multicolumn{4}{l}{$\!\!\!\!\!$ \ir{X_i}{\Gamma^{2,0}} $\times$ \ir{X_j}{\Gamma^{1,3}} = \ir{X_k}{\Gamma^1} + \ir{X_k}{\Gamma^3}}\\
\ir{X_i}{\Gamma^{1,3}} & \ir{X_i}{\Gamma^{1,3}}  & \ir{e}{1} + \ir{e}{J_i} + \ir{\ebar}{J_j} + \ir{\ebar}{J_k}\\
\ir{X_i}{\Gamma^{1}} & \ir{X_i}{\Gamma^{3}} &\ir{\ebar}{1} + \ir{\ebar}{J_i} + \ir{e}{J_j} + \ir{e}{J_k} \\
\ir{X_i}{\Gamma^{1,3}} & \ir{X_j}{\Gamma^{1,3}}  & \multicolumn{4}{l}{$\!\!\!\!\!$ \ir{X_i}{\Gamma^{3,1}} $\times$ \ir{X_j}{\Gamma^{1,3}} = \ir{X_k}{\Gamma^0} + \ir{X_k}{\Gamma^2}}
\end{tabular}

\vskip 4mm
\noindent
{\bf References}
\bibliographystyle{iopart-num-jcr}
\bibliography{dgt}

\end{document}

%% file: Diagrams.tex
\usepackage[all,knot]{xy}
\xyoption{all}
\xyoption{knot}

\newdir{|>}{%
!/4.5pt/@{|}*:(1,-.2)@^{>}*:(1,+.2)@_{>}}

\newcommand{\smalllabels}{
	\def\objectstyle{\scriptstyle}
	\def\labelstyle{\scriptstyle}
}

\newcommand{\unknot}[1]{
	\xygraph{
		!{0;/r2.0pc/:}
		!{\vcap-}
		!{\vcap}
	}
}

\newcommand{\qdim}[2]{
	\smalllabels
	\xy
		(-4,0)="L"+(-2,0)*{#1};
		(4,0)="R";
		(0,4)="T";
		(0,-4)="B";
		"L";"T"*\crv{"L"+(0,2)&"T"+(-2,0)};
		"T";"R"*\crv{"T"+(2,0)&"R"+(0,2)};
		"R";"B"*\crv{"R"+(0,-2)&"B"+(2,0)};
		"B";"L"*\crv{"B"+(-2,0)&"L"+(0,-2)};
		"L"+(.05,1)*\dir{#2};
	\endxy
}

\newcommand{\anyon}[1]{
	\smalllabels
	\xy
		(0,6)*{}="T";
		(0,-6)*{}="B";
		"B"*{};"T"**\dir{-}?(.6)*\dir{>} + (2,-3)*{#1};
	\endxy
}

\newcommand{\qdanyon}[1]{
	\smalllabels
	\xy
		(0,6)*{}="T";
		(0,-6)*{}="B";
		"B"*{};"T"**\dir{-}?(.6)*\dir{>} + (5.5,-3)*{#1};
	\endxy
}

\newcommand{\twistright}[1]{
	\smalllabels
	\xy 
	(0,-8)*{}="T";
	(0,8)*{}="B";
	(0,-4)*{}="T'"+(-2,-2)*{#1};
	(0,6)*{}="B'";
	(3,0)*{}="MB";
	(7,0)*{}="LB";
	"T'";"T" **\dir{-};
	"B'";"B" **\dir{-}?(.4)*\dir{>};
	"T'";"LB" **\crv{(1,5) & (5,5)}; \POS?(.25)*{\hole}="2z";
	"LB"; "2z" **\crv{(8,-4) & (2,-4)};
	"2z"; "B'" **\crv{(0,2)};
\endxy 
}

\newcommand{\twistleft}[1]{
	\smalllabels
	\xy 
	(0,8)*{}="T";
	(0,-8)*{}="B";
	(0,5)*{}="T'";
	(0,-5)*{}="B'"+(-2,0)*{#1};
	(-3,0)*{}="MB";
	(-7,0)*{}="LB";
	"T'";"T" **\dir{-}?(.3)*\dir{>};
	"B";"B'" **\dir{-};
	"T'";"LB" **\crv{(-1,-3) & (-5,-4)}; \POS?(.25)*{\hole}="2z";
	"LB"; "2z" **\crv{(-8,6) & (-2,6)};
	"2z"; "B'" **\crv{(0,-3)};
\endxy 
}

\newcommand{\smatrix}[2]{
	\smalllabels
	\xy
		(-6,0)="L1";
		(-2,4)="T1";
		(2,0)="R1";
		(-2,-4)="B1"+(-3,-1)*{#1};
		(-2,0)="L2";
		(2,4)="T2";
		(6,0)="R2";
		(2,-4)="B2"+(3,-1)*{#2};
		"R1";"B1"**\crv{"R1"+(0,-2)&"B1"+(2,0)}?(.6)*{\hole}="lower";
		"B1";"L1"**\crv{"B1"+(-2,0)&"L1"+(0,-2)};
		"L1";"T1"**\crv{"L1"+(0,2)&"T1"+(-2,0)};
		"L2";"T2"**\crv{"L2"+(0,2)&"T2"+(-2,0)}?(.6)*{\hole}="higher";
		"T2";"R2"**\crv{"T2"+(2,0)&"R2"+(0,2)};
		"R2";"B2"**\crv{"R2"+(0,-2)&"B2"+(2,0)};
		"L2";"lower"**\crv{"L2"+(0,-2)"};
		"R1";"higher"**\crv{"R1"+(0,2)"};
		"B2";"lower"**\crv{"B2"+(-1,0)};
		"T1";"higher"**\crv{"T1"+(1,0)};
		"L2"+(0,-2);"L2"+(0,2)**\dir{}?(.7)*\dir{>};
		"R1"+(0,-2);"R1"+(0,2)**\dir{}?(.7)*\dir{>};
	\endxy
}

\newcommand{\qdsmatrix}[2]{
	\smalllabels
	\xy
		(-6,0)="L1";
		(-2,4)="T1";
		(2,0)="R1";
		(-2,-4)="B1"+(-3,-3)*{#1};
		(-2,0)="L2";
		(2,4)="T2";
		(6,0)="R2";
		(2,-4)="B2"+(3,-3)*{#2};
		"R1";"B1"**\crv{"R1"+(0,-2)&"B1"+(2,0)}?(.6)*{\hole}="lower";
		"B1";"L1"**\crv{"B1"+(-2,0)&"L1"+(0,-2)};
		"L1";"T1"**\crv{"L1"+(0,2)&"T1"+(-2,0)};
		"L2";"T2"**\crv{"L2"+(0,2)&"T2"+(-2,0)}?(.6)*{\hole}="higher";
		"T2";"R2"**\crv{"T2"+(2,0)&"R2"+(0,2)};
		"R2";"B2"**\crv{"R2"+(0,-2)&"B2"+(2,0)};
		"L2";"lower"**\crv{"L2"+(0,-2)"};
		"R1";"higher"**\crv{"R1"+(0,2)"};
		"B2";"lower"**\crv{"B2"+(-1,0)};
		"T1";"higher"**\crv{"T1"+(1,0)};
		"L2"+(0,-2);"L2"+(0,2)**\dir{}?(.7)*\dir{>};
		"R1"+(0,-2);"R1"+(0,2)**\dir{}?(.7)*\dir{>};
	\endxy
}

\newcommand{\qdcloseddoubleloop}[3]{
	\smalllabels
	\xy
	(6,-10)="B2";
	(6,10)="T2";	
	(0,-10)="B"+(-5,1)*{#3};
	(0,-6)="B'";
	(0,10)="T";
	(0,2)="T'";
	(0,-4)="M";
	(0,0)="M'";
	(-4,-3)="l"+(-5,0)*{#2};
	(4,-3)="r";
	(0,-5)="b";
	(-4,3)="ll"+(-5,0)*{#1};
	(4,3)="rr";
	(0,1)="bb";
	"B";"B'"**\dir{-};\vcap[-6];
	(0,5)*{\vcap[6]};
	"B2";"T2"**\dir{-};
	"T'";"T"**\dir{-}?(.4)*{\hole}="tt";"T'";"T"**\dir{}?(.9)*\dir{>};
	"M";"M'"**\dir{-}?(.7)*{\hole}="t";
	"l";"b"**\crv{"l"+(0,-1,5)&"b"+(-2,0)};
	"r";"b"**\crv{"r"+(0,-1,5)&"b"+(2,0)};
	"l";"t"**\crv{"l"+(0,2)};
	"r";"t"**\crv{"r"+(0,2)};
	"b"+(-1,.3);"b"+(1,.3)**\dir{}?(.7)*\dir{>};
	"ll";"bb"**\crv{"ll"+(0,-1,5)&"bb"+(-2,0)};
	"rr";"bb"**\crv{"rr"+(0,-1,5)&"bb"+(2,0)};
	"ll";"tt"**\crv{"ll"+(0,2)};
	"rr";"tt"**\crv{"rr"+(0,2)};
	"bb"+(-1,.3);"bb"+(1,.3)**\dir{}?(.7)*\dir{>};
	\endxy
}

\newcommand{\scondensate}[2]{
	\smalllabels
	 \xy
		(-9,9)*{}="L1";
		(-3,9)*{}="L2";
		(-9,3)*{}="L3";
		(-3,3)*{}="L4";
		(-9,-3)*{}="L5";
		(-3,-3)*{}="L6";
		(-9,-9)*{}="L7";
		(-3,-9)*{}="L8";
		(9,9)*{}="R1";
		(3,9)*{}="R2";
		(9,3)*{}="R3";
		(3,3)*{}="R4";
		(9,-3)*{}="R5";
		(3,-3)*{}="R6";
		(9,-9)*{}="R7";
		(3,-9)*{}="R8";
		"L2"*{};"L1"**\crv{"L2"+(-1,5)&"L1"+(1,5)};
		"L7"**\dir{-};
		"L8"**\crv{"L7"+(1,-5)&"L8"+(-1,-5)};
		"R2"*{};"R1"**\crv{"R2"+(1,5)&"R1"+(-1,5)};
		"R7"**\dir{-};
		"R8"**\crv{"R7"+(-1,-5)&"R8"+(1,-5)};
		\vtwist~{"L4"}{"R4"}{"L6"}{"R6"};
		\vtwist~{"L6"}{"R6"}{"L8"}{"R8"};
		"L4"*{};"L2"**\dir{-}?(.5)*\dir{>} + (-3,0)*{#1};
		"R4"*{};"R2"**\dir{-}?(.5)*\dir{>} + (+3,0)*{#2};
		"R4"*{};"L2"**\dir{--}?(.7)*\dir{} + (1.5,2)*{};
	\endxy
}

